\documentclass[]{jfm}
\usepackage{natbib}
\usepackage{psfrag}
\usepackage{graphicx}
\usepackage{color}
\def\drawline#1#2{\raise 2.5pt\vbox{\hrule width #1pt height #2pt}}

\def\trian{\raise 1.25pt\hbox{$\scriptscriptstyle\triangle$}\nobreak\ }

\def\square{${\vcenter{\hrule height .4pt
\hbox{\vrule width .4pt height 3pt \kern 3pt
\vrule width .4pt}
\hrule height .4pt}}$\nobreak\ }
\def\plus{\raise 1.25pt \hbox{$\scriptscriptstyle +$}\nobreak\ }
\def\solidtrian{\raise 1.25pt
\hbox to 3bp{
\hfill}\nobreak\ }
\definecolor{orange}{cmyk}{0,.5,1,0}
\definecolor{purple}{rgb}{1,0,1}
\definecolor{grey}{rgb}{.7,.7,.7}
\def\solidtriandown{{\raise 1.25pt 
\hbox to 3bp{
 \hfill}\nobreak\ }}

\def\om3{\omega_3}
\def\om2{\omega_2}
\def\tup{\widetilde{u}^\prime}

\def\tup2{\widetilde{u}_2^\prime}

\def\tt{\widetilde{\theta}}

\def\u2a{{\langle u_2^2 \rangle}}
\def\aQ{{\langle Q \rangle}}

\def\up2{\widetilde{\langle u_2^{\prime 2} \rangle^{1/2}}}
\def\tu2p{u_2^{\prime +}}
\def\der#1#2{ {{\partial #1 } \over {\partial #2}} }
\def\ded#1#2{ {{d #1 } \over {d #2}} }
\def\dedq#1#2{ {{d^{2} #1 } \over {d #2}} }

\begin{document}

\title
[
TKE production and  flow structures 
]
{
Turbulent kinetic energy production and flow structures in flows past smooth and rough walls
}
\author[ P.~ Orlandi  \\]
%                AUTHORS
{
P.\ns  O\ls R\ls L\ls A\ls N\ls D\ls I 
}
\affiliation
{
Dipartimento di Ingegneria Meccanica e Aerospaziale\\
Universit\`a La Sapienza, Via Eudossiana 16, I-00184, Roma
}
\date{\today}
\maketitle

\begin{abstract}

Data available in literature from direct numerical simulations of
two-dimensional turbulent channels by \cite{lee_15}, 
\cite{Bernardini2014}, \cite{yamamoto_18} and \cite{orlandi2015}
in a large range of Reynolds number have been used to
find that $S^*$ the ratio between the eddy turnover time ($q^2/\epsilon$)
and the time scale of the mean deformation ($1/S$), scales
very well with the Reynolds number in the near-wall
region. The good scaling is due to the eddy turnover time,
although the turbulent kinetic energy and the rate of
isotropic dissipation show a Reynolds dependence near the wall.
$S^*$ is linked to the flow structures, as well as 
$-\aQ=\langle{s_{ij}s_{ji}}\rangle-\langle{\omega_i\omega_i/2}\rangle$
and also this quantity presents a good scaling. It
has been found that the maximum of turbulent kinetic energy
production $P_k$ occurs in the layer with $-\aQ\approx 0$
that is where the unstable sheet-like structures roll-up
to become rods. The decomposition of $P_k$ in the contribution
of elongational and compressive strain demonstrates that
the two contribution present a good scaling. The perfect scaling
however holds when the near-wall and the
outer structures are separated. The same statistics
have been evaluated by direct simulations of turbulent channels
with different type of corrugations on both walls. The flow
physics in the layer near the plane of the crests is strongly
linked to the shape of the surface and it has been demonstrated
that the $u_2$ (normal to the wall) fluctuations are responsible
for the modification of the flow structures, for 
the increase of the resistance and of the turbulent 
kinetic energy production. These simulations at intermediate
Reynolds number indicated that in the outer region the Townsend
similarity hypothesis holds.

\end{abstract}

\section{Introduction}

Turbulent flows near smooth walls are characterised by
flow structures of different size and intensity that have
been observed and described by very impressive flow
visualizations by \cite{Kline}. In laboratory experiments
it is rather difficult to measure any quantity, therefore
the deep understanding of the flow complexity can not be
achieved.  The evolution
of hardware and software necessary for the solution  of
the non-linear Navier-Stokes equations allowed to 
evaluate any flow variable and to increase the knowledge of
the physics of turbulent flows. The simulations were performed for a large
amount of turbulent flows and in particular for wall
bounded flows such as boundary layers, circular pipes and
two-dimensional channels. In this paper the study is  focused 
on flows in two-dimensional channels past
smooth and corrugated walls. The first Direct Numerical Simulation (DNS)
of a two-dimensional channel by \cite{Moin_82} can be considered a scientific
revolution, in fact after this publication a large number of scholars
and even some experimentalist directed
their research work on the use of numerical methods to produce and
analyse turbulent data to understand the complex physics
of wall bounded turbulent flows.
The  direct comparison between numerical and laboratory flow visualizations
in \cite{Moin_97} can be considered a proof that the Navier-Stokes equations are the
valid model to describe the evolution of turbulent flows.
After that simulation at low Reynolds number
($R_\tau=u_\tau h/\nu=180$ with $u_\tau$ the friction velocity
$h$ half channel height and $\nu$ the kinematic viscosity) 
there was a large effort to increase the Reynolds number.
The relevant efforts were done, among several groups
by \cite{JimenezHoyas2008} up to $R_\tau=2000$ by
\cite{Bernardini2014} up to $R_\tau=4000$, by \cite{lee_15}
up to $R_\tau=5200$ and recently by \cite{yamamoto_18}
up to $R_\tau=8000$. Some of the statistics in these papers together with
others at low Reynolds number  in \cite{orlandi2015} are
used in this study to calculate quantities linked to the flow structures  
and therefore to investigate the dependence on
the Reynolds number. Namely these are the shear
rate parameter $S^*=Sq^2/\epsilon$ defined as the ratio between 
the eddy turnover time $q^2/\epsilon$ 
( $q^2$ is twice the turbulent kinetic energy and
$\epsilon$ is the isotropic dissipation rate) and the time scale 
of the mean deformation $1/S=2/\ded{U}{x_2}$
($U$ stands for $\langle u_1 \rangle$). The shear parameter was used
by \cite{lee_90} to prove that the elongated wall structures,
observed by \cite{Kline}, were not generated
by the presence of the solid wall, instead 
by the mean shear rate $S$ if it was greater than a threshold value. 
As it was shown by \cite{orlandi2015} the profiles of
the shear parameter do not largely vary with the Reynolds
number in presence of smooth walls. This conclusion is
a first evidence that the laboratory experiments have limitations
to explain the complex physics of turbulent flows due to
the impossibility to measure $\epsilon$ in the whole channel
and in particular near the wall.  The
profiles of $\epsilon^+$ as well as those of $q^{2+}$
(the superscript $+$ indicates wall units) are Reynolds dependent. 
In this paper it is
shown that, near the walls,  the eddy turnover time as well as
the mean shear in wall units do not depend
on the Reynolds number, therefore the 
shear parameters can be considered a fundamental quantity to
characterise the energetic scales near smooth walls.

The production of turbulent energy $P_k=-\langle u_1u_2\rangle \ded{U}{x_2}$ 
(small letters indicate fluctuations)
is strictly linked to $S$.  When the near-wall 
and the outer turbulent structures are separated the production
does not depend on the Reynolds number. At low $Re$, on the
other hand, the maximum of $P_k^+$, from zero for the laminar
regime jumps to  a value $0.15$ at the transitional
Reynolds number. Hence it gradually grows with $R_\tau$ to saturate at
$500 < R_\tau$ at a value equal to $0.25$.
Due to the key role  of $S$ the one-dimensional statistics
profiles can be projected on the eigenvectors of the 
tensor $S_{ij}$. In this frame there is a negative compressive 
$S_\alpha$ and a positive extensional $S_\gamma$ strain.
The turbulent kinetic production 
in this local frame is $P_k=-(P_\alpha+P_\gamma)$
with the terms $P_\alpha=R_{\alpha \alphaį} S_\alpha$ and
$P_\gamma=R_{\gamma \gamma} S_\gamma$ 
greater than $P_k$. Their profiles allow
to understand that the compressive strain generates
more kinetic energy than that destroyed by extensional  one.
The projection of the statistics along the
eigenvectors of the
tensor $S_{ij}$ shows a decrease on the anisotropy
of the velocity and vorticity correlation and
can give insights on turbulence closures. 
The stresses in the spanwise direction, as it should be expected, do not
change in this new reference frame with $S_\beta=0$.

The production of turbulent kinetic energy
can also be expressed  in a different way.
(\cite{Orlandi2000} at Pg.211) with
$P_k=P_T+P_C$. This expression is derived by the Navier-Stokes equation in
rotational form. $P_C=\der{ U  \langle u_1 u_2\rangle }{ x_2}$
is related to the action of the large
eddies advecting the turbulence across the channel.
$P_T= U (\langle u_3 \omega_2\rangle 
-\langle u_2 \omega_3\rangle )$
is linked to the energy transfer from large to small eddies.

\cite{Orlandi2013} in turbulent wall bounded flows emphasised the role
of the wall-normal Reynolds stress and therefore the statistics linked
to the $u_2$ velocity fluctuations and  in particular those 
connected to the flow structures should be analysed.
This stress received little
attention, in particular because of  the difficulty to get
accurate measurements near the walls. It is worth 
to recall that only this stress appears in the mean
momentum equation and it is balanced by the mean pressure
$\langle{p}\rangle$. As it was discussed in that paper
as well as by \cite{Tsinober2002} at Pg.162 the topology
of flow structures can be described by $-Q=s_{ij}s_{ji}-\omega_i\omega_i/2$,
where regions with $Q<0$ are sheet-dominated and
regions with $Q>0$ are associated with tube-like
structures. In  homogeneous turbulence $\aQ=0$
In non-homogeneous turbulent flows
$\dedq{\langle u_2^2\rangle}{ x_2^2}=
-\aQ=\langle{s_{ij}s_{ji}}\rangle
-\langle{\omega_i\omega_i/2}\rangle$
accounts for the disequilibrium between
$\langle{s_{ij}s_{ji}}\rangle$ and  $\langle{\omega_i\omega_i/2}\rangle$.
Hence the term  $\dedq{\langle u_2^2\rangle}{ x_2^2}$
determines whether in a region there is a prevalence of
sheet-like or rod-like structures. The former
are inherently unsteady, and roll-up
producing turbulent kinetic energy. 
A detailed study on the difference between the shape of ribbon- and
rod-like structures requires appropriate eduction schemes
as those described by \cite{pirozzoli2010}.
The profiles of the turbulent kinetic energy production,
in their different form,
together with the profiles of $\dedq{ \langle u_2^2\rangle}{ x_2^2}$
shows that the maximum production occurs in the
layer separating sheet- and
tubular-dominated regions.  One of the goals in performing DNS of two-dimensional
turbulent channels at high Reynolds numbers was and still is to investigate
the Reynolds number dependence on the statistics. The aim of the present study 
is to see whether the above mentioned statistics , in wall units, related to the 
flow structures present a minor or almost a complete Reynolds number independence
in the near-wall region.

A rediscovered and improved version of the old Immersed Boundary Technique
used by \cite{Peskin1972} for bio-inspired flows was developed by
\cite{OrlandiLeonardi2006} to perform DNS of turbulent flows past 
rough walls. The method was validated in several papers
and the convincing proof of its accuracy was reported by
\cite{burattini2008} by comparing the statistics derived by the
numerical experiments with those measured 
in laboratory, The  experiments were designed with the aim to show
that true DNS can be accomplished by the immersed boundary technique inserted
in a second order finite difference method. In this paper 
the corrugations are located in both walls and
the solutions are obtained at intermediate values of the
$Re$ number, namely approximately at $R_\tau=200$
in presence of smooth walls. Longitudinal transverse and
three-dimensional elements are considered producing
a drag increase with the exception of a geometry similar to that 
investigated by \cite{Choi} producing drag reduction. 
Near rough walls the flow structures 
change leading  to different profiles of 
the turbulent statistics above mentioned as it is shown in this
paper.

\section{Results}

\subsection{Smooth wall}

\begin{figure}
\centering
\vskip 0.0cm
\hskip -1.8cm
\psfrag{ylab}{\large $ q^{2 +}$}
\psfrag{xlab}{\large $ $ }
\includegraphics[width=7.5cm]{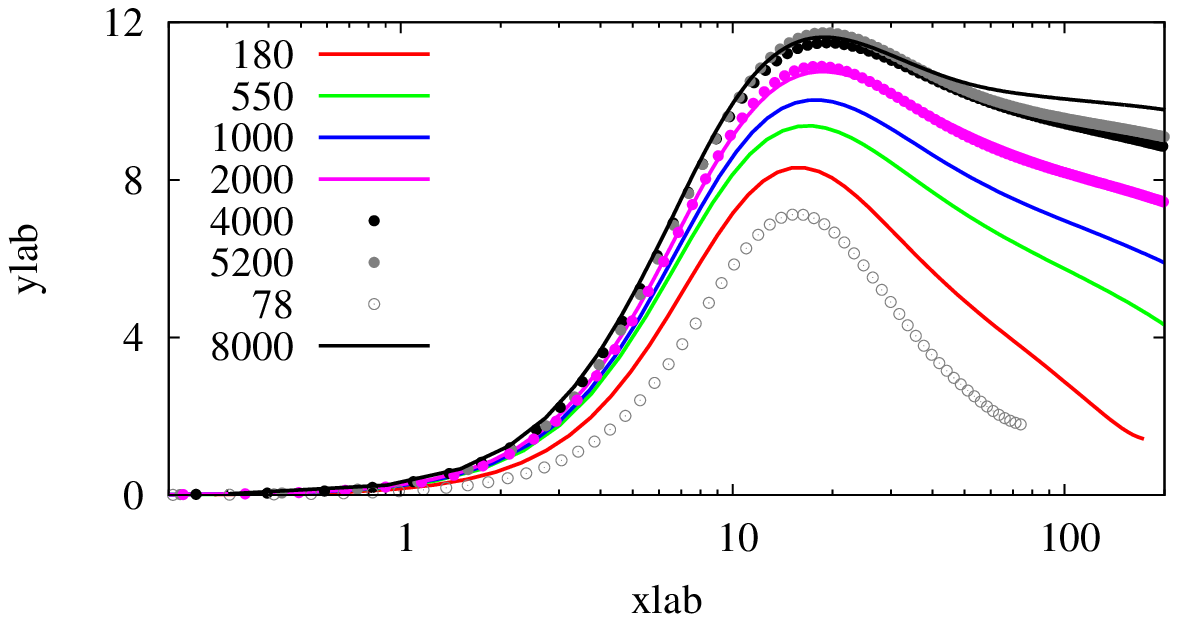}
\hskip -0.0cm
\psfrag{ylab}{\large $\epsilon^{ +}$ }
\psfrag{xlab}{\large $ $ }
\includegraphics[width=7.5cm]{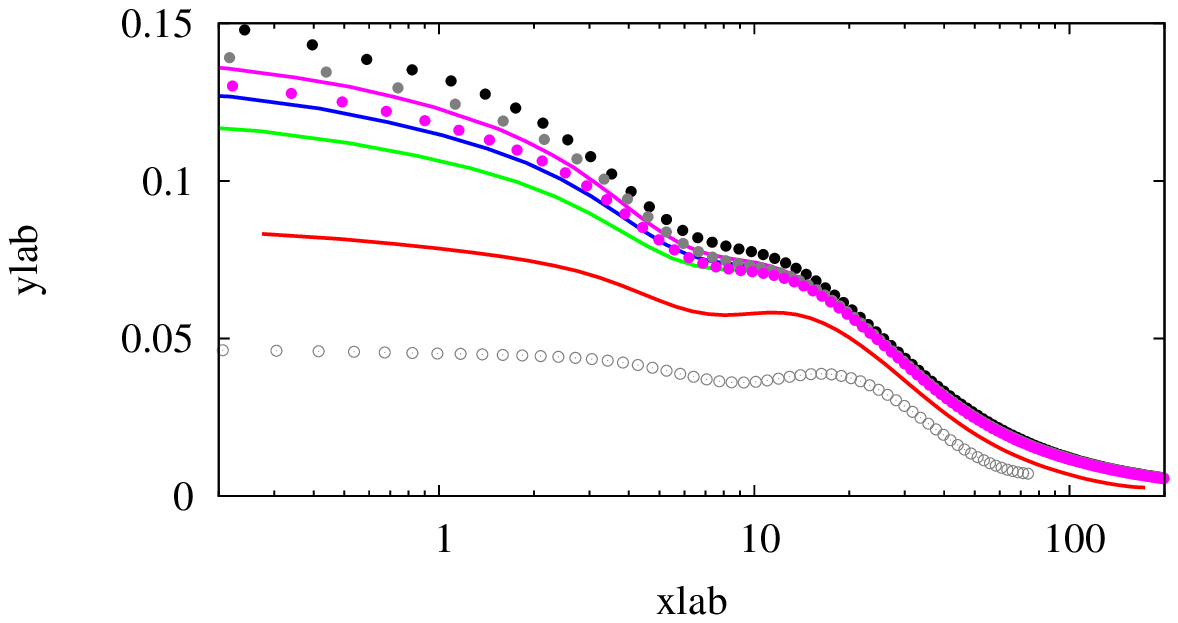}
\vskip -0.5cm \hskip 5cm a) \hskip 7cm b)
\vskip -.2cm
\hskip -1.8cm
\psfrag{ylab}{\large $ (q^2 /\epsilon)^{ +}$}
\psfrag{xlab}{\large $y^+ $ }
\includegraphics[width=7.5cm]{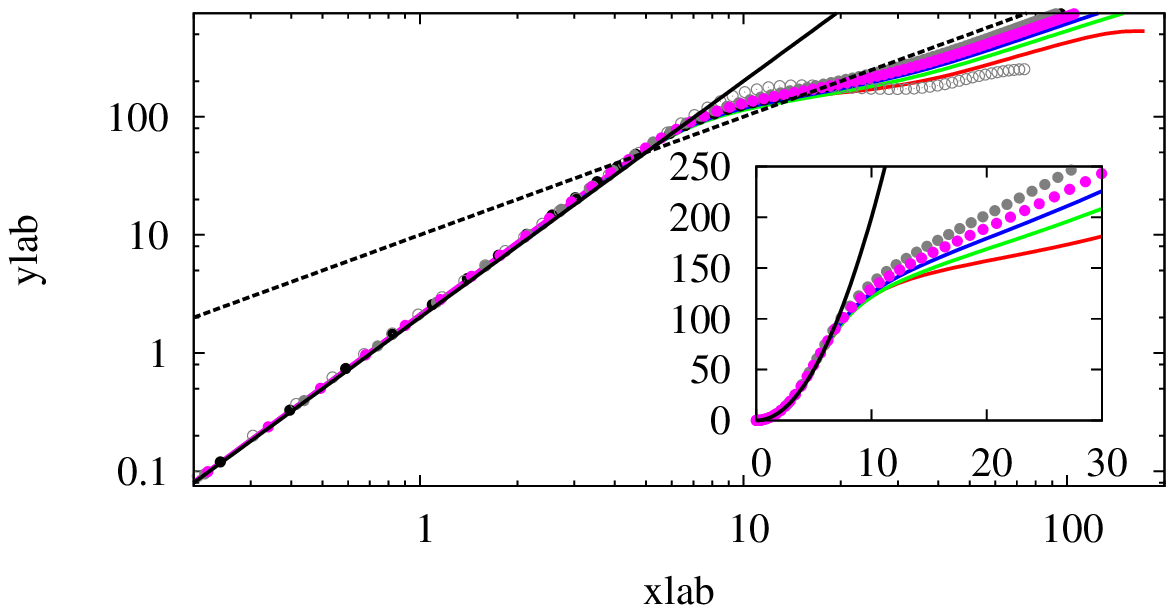}
\hskip -0.0cm
\psfrag{ylab}{\large $ Sq^2/\epsilon$}
\psfrag{xlab}{\large $y^+ $ }
\includegraphics[width=7.5cm]{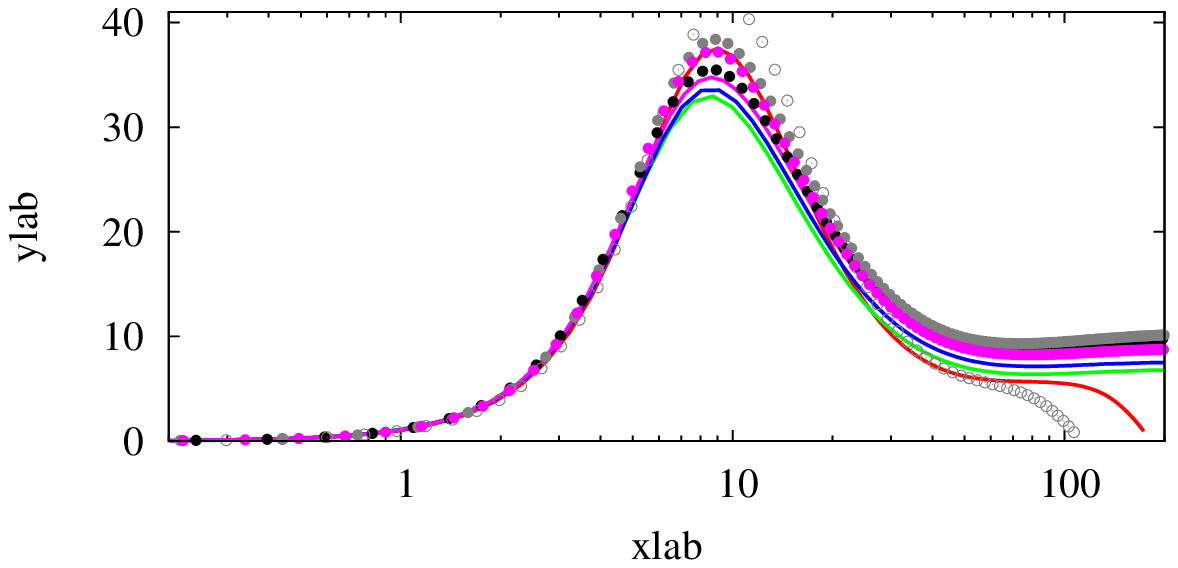}
\vskip -0.5cm \hskip 5cm c) \hskip 7cm d)
\vskip -.2cm
\hskip -1.8cm
\psfrag{ylab}{\large $ (q^2 /D_k)^{ +}$}
\psfrag{xlab}{\large $y^+ $ }
\includegraphics[width=7.5cm]{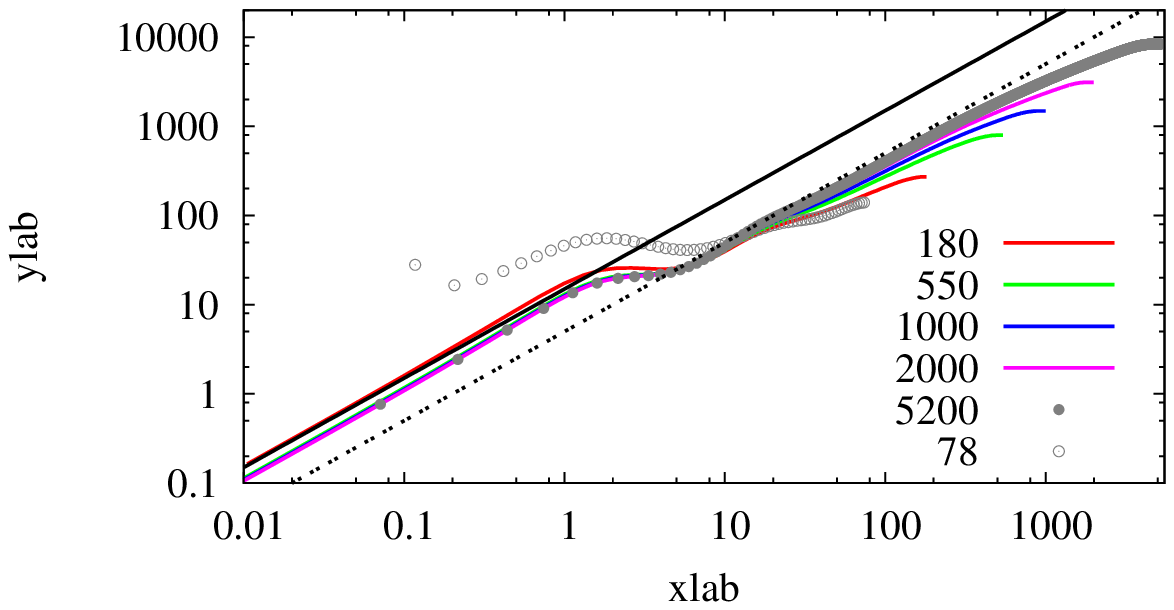}
\hskip -0.0cm
\psfrag{ylab}{\large $ $}
\psfrag{xlab}{\large $y^+ $ }
\includegraphics[width=7.5cm]{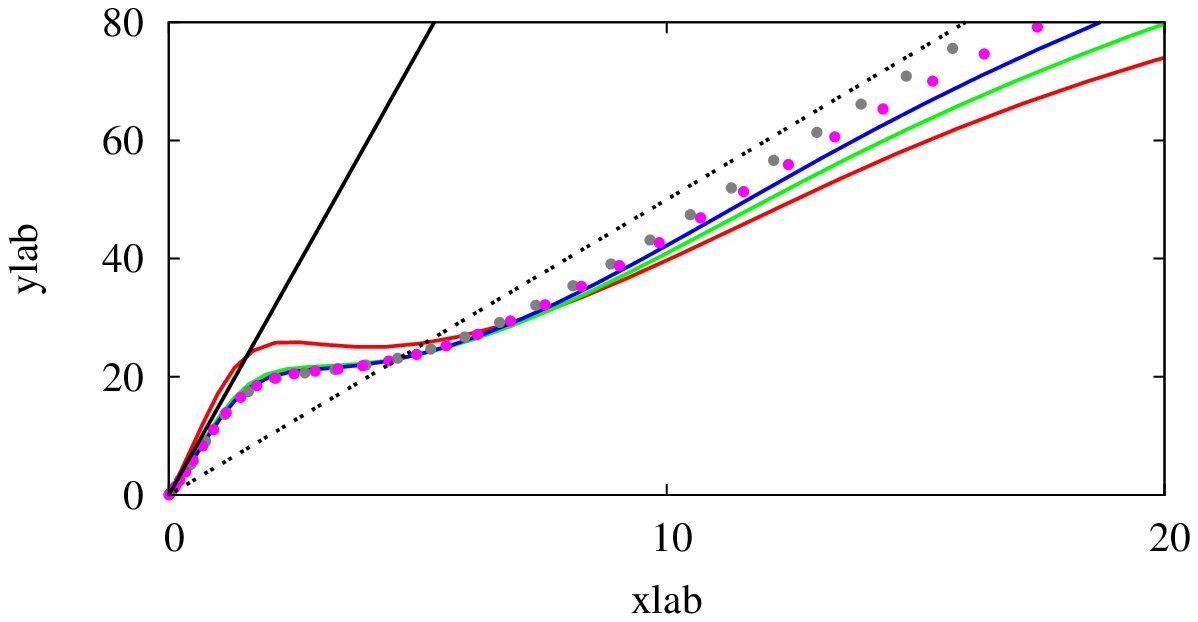}
\vskip -0.5cm \hskip 5cm e) \hskip 7cm f)
\caption{Profiles in wall units of:
a) Turbulent kinetic energy, 
b) rate of dissipation,
c) eddy turnover time evaluated with $\epsilon$
 (solid line $2 x^2$ dashed $10 x$), 
d) shear parameter,
e),f) eddy turnover time evaluated with $D_k$
 (solid line $15 x$ dashed $5 x$), 
for channel flows past smooth walls, the data
are from the references given in the text. The Reynolds
number are in the insets
 of a); in b), c), d), e), f) the profiles
at $R_\tau=8000$ could not be evaluated from the
\cite{yamamoto_18} data.
}
\label{fig1}
\end{figure}

In this section the data in the web of the DNS at high
Reynolds numbers by \cite{Bernardini2014}, by \cite{lee_15}
and by \cite{yamamoto_18} are used to investigate
the eventual Reynolds independence of 
the shear parameter and their components. The data
at low $Re$ are those used in \cite{orlandi2015}. The
shear parameter $S^*$ is one of the statistics linked
to certain kind of flow structures, in fact if $S^*$ is greater than
a threshold value, approximately $5$, very elongated anisotropic
longitudinal structures form. It has been observed that in the
near-wall region $S^*$ is high and in the outer is small,
consequently the flow structures are more intense near the wall
than those in the outer region.
The profiles of $q^{2 +}$ in figure \ref{fig1}a
and of $\epsilon^+$ in figure \ref{fig1}b
show a large $Re$ dependence in the near-wall
region, that has been emphasised by plotting the
quantities only in the viscous and buffer regions.
These two figures show that  large variations
appear at intermediate $Re$ and that both quantities 
tend to a limit at high $Re$. Still it has not been 
established whether  a saturation  or
a logarithmic growth of the maximum of $q^{2 +}$ 
occurs for $Re \rightarrow \infty$ .
However,  figures \ref{fig1}a seems to  infer a saturation.
The value of the maximum of $q^{2 +}$ at $R_\tau=8000$ is twice
the value reached at the transitional  Reynolds number.
At this $Re$ there is no separation between outer
and near-wall structures. The peak is located
almost at the same distance, in wall units, as that
at a $Re$ number  hundred times greater.  
The rate of isotropic energy dissipation in
figure \ref{fig1}b shows a large Reynolds number dependence
in the region $y^+< 10$, being this quantity linked
to the small scales. As it is discussed later on,
for $y^+<10$, the full rate of energy dissipation 
$D_k=2\nu \langle u_i\nabla^2 u_i \rangle$,
in wall units, shows a reduced dependence with the Reynolds number.
Since $D_k=\nu \dedq{ \langle q^2\rangle}{ x^2_2}-\epsilon$
it can be concluded that the Reynolds dependence of $\epsilon^+$,
in the near-wall region, is due to the viscous diffusion of $q^{2+}$.
The Reynolds dependence for $y^+<10$, at very high $Re$, in the
outer region disappears by looking
at the profiles of the eddy turnover time, in wall units in figure \ref{fig1}c.
\noindent In the near-wall region the eddy turnover time is proportional
to $y^{2+}$ instead in the outer region is proportional to $y^+$.
Figure \ref{fig1}c shows a deviation from the linear behavior
higher smaller $Re$ is. In order to appreciate better the 
variations in the region of transition in the inset some of the
profiles are plotted in linear scale.
The eddy turnover time based on $D_k^+$ (figure \ref{fig1}e) instead 
of on $\epsilon^+$ shows a linear increase with $y^+$ also near the wall. 
Therefore it can be stressed
that this dimensionless eddy turnover time depicts an universal behavior
for the inner and outer structures. 
These have the same characteristics being generated 
by the strain $S$, are fast near the wall,  and slow in the outer region.
The transition between the two similar structures occurs in the region
with high growth of turbulent kinetic energy production. To demonstrate
the collapse in the near-wall region and the tendency to the saturation
in the buffer region figure \ref{fig1}f has been plotted
in linear scales and the data at $R_\tau=78$ were not considered.
At $Re$ numbers close to 
the transitional value ($R_\tau \approx 80$) the two kind of
structures are strongly
connected and hence the similarity disappears and the flow physics
is more complex. 
In the transitional regime the turbulence 
could play a large effect in mixing processes or in heat transfer.
From the data in figure 2 in \cite{orlandi2015} 
it has been evaluated that for $70 < R_\tau < 200$,
$K/(h U_b^2)$ ( $K=\int q^2 dy$ and $U_b$
the bulk velocity) decays with $R_\tau^{1/3}$
and that for $500 < R_\tau < 5200$ with $R_\tau^{1/6}$.
Therefore it can be inferred that the effects of Reynolds number
are high when the near-wall and
the outer structures have the same size, and
a strong  interconnection. When the near-wall structures are much smaller
and far apart from the outer ones the effect of the Reynolds number
is reduced. The profiles of $S^+$ for $y^+<200$ are not given being
superimposed each other with the exception of that at $R_\tau=78$.
Therefore the Reynolds independence of the shear parameter $S^*$ 
in figure \ref{fig1}d
is due to the universality of the eddy turnover time
in the near-wall region. It can be argued 
that the similarity in the mean  velocity profiles 
for wall bounded turbulent flows past smooth walls,
having well defined boundary conditions,
forces the similarity in the eddy turnover time 
of turbulent flows.

\begin{figure}
\centering
\vskip 0.0cm
\hskip -1.8cm
\psfrag{ylab} {\hskip -1.0cm
\large $(10^2\ded{ \langle{u_2^2}\rangle}{ x^2_{2}})^+$}
\psfrag{xlab}{ $ $}
\includegraphics[width=7.5cm]{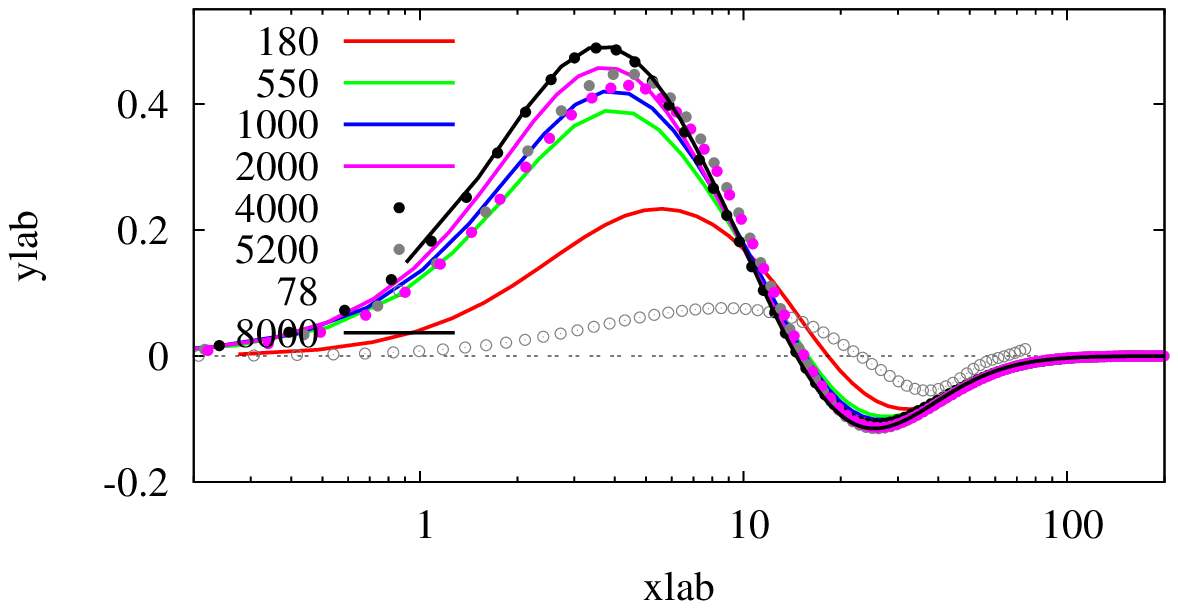}
\psfrag{ylab}{\large $ P_k^+$}
\psfrag{xlab}{\large $ $ }
\includegraphics[width=7.5cm]{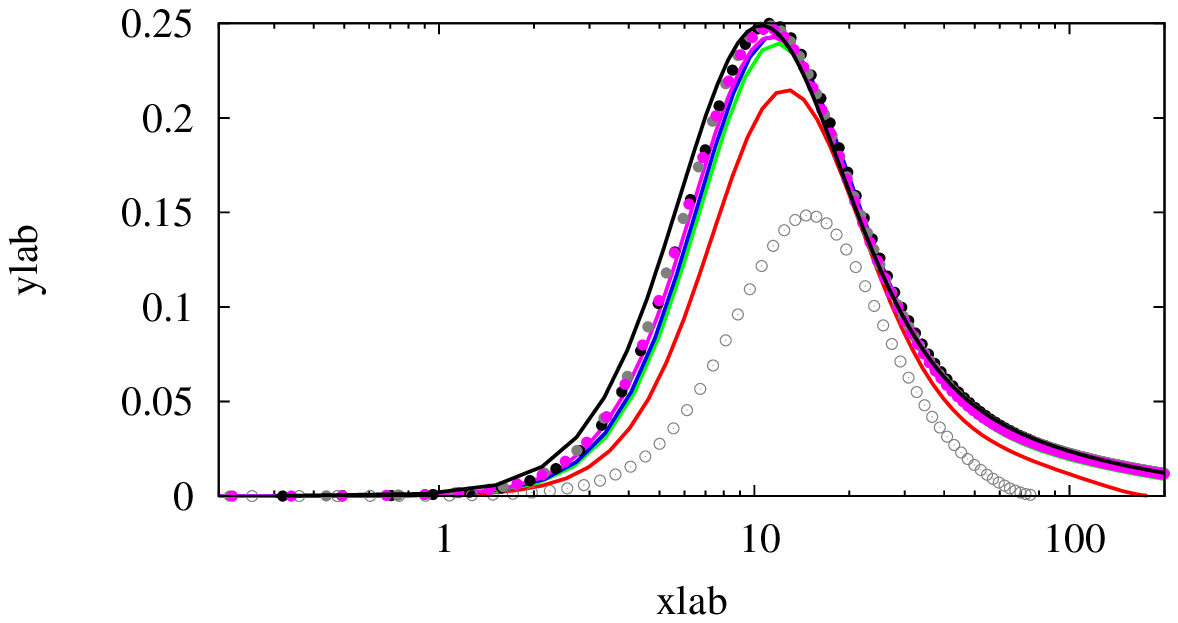}
\vskip -0.5cm \hskip 5cm a) \hskip 7cm b)
\vskip -0.2cm
\hskip -1.8cm
\psfrag{ylab}{\large $ D_k^+$}
\psfrag{xlab}{\large $ y^+ $ }
\includegraphics[width=7.5cm]{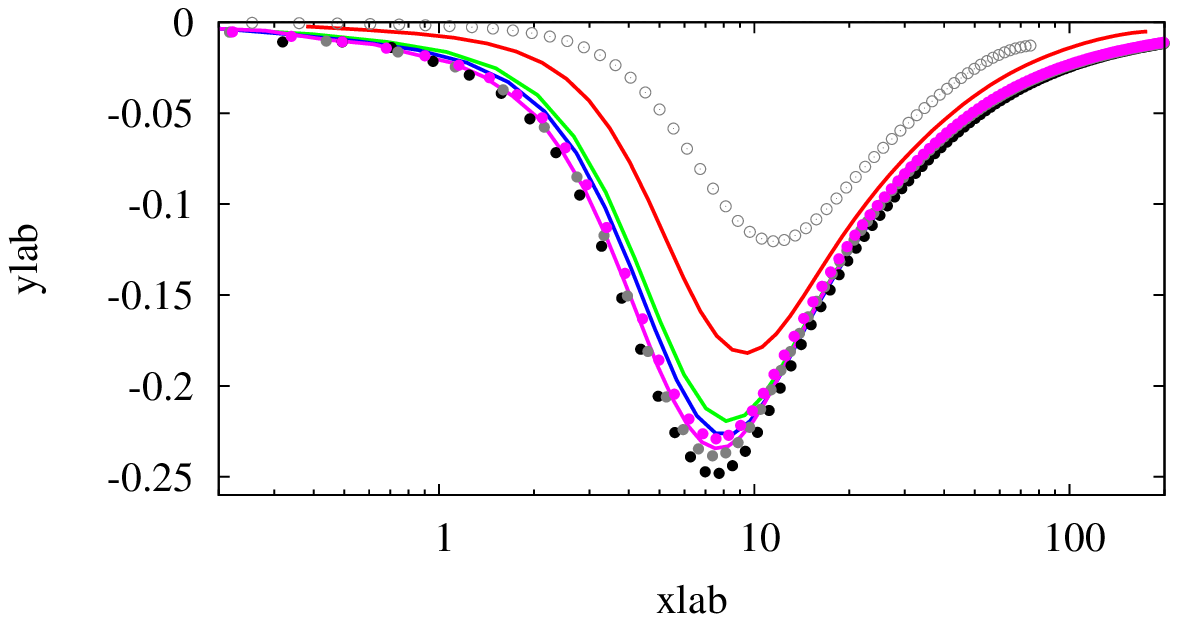}
\psfrag{ylab}{\large $ T_k^+$}
\psfrag{xlab}{\large $ y^+$ }
\includegraphics[width=7.5cm]{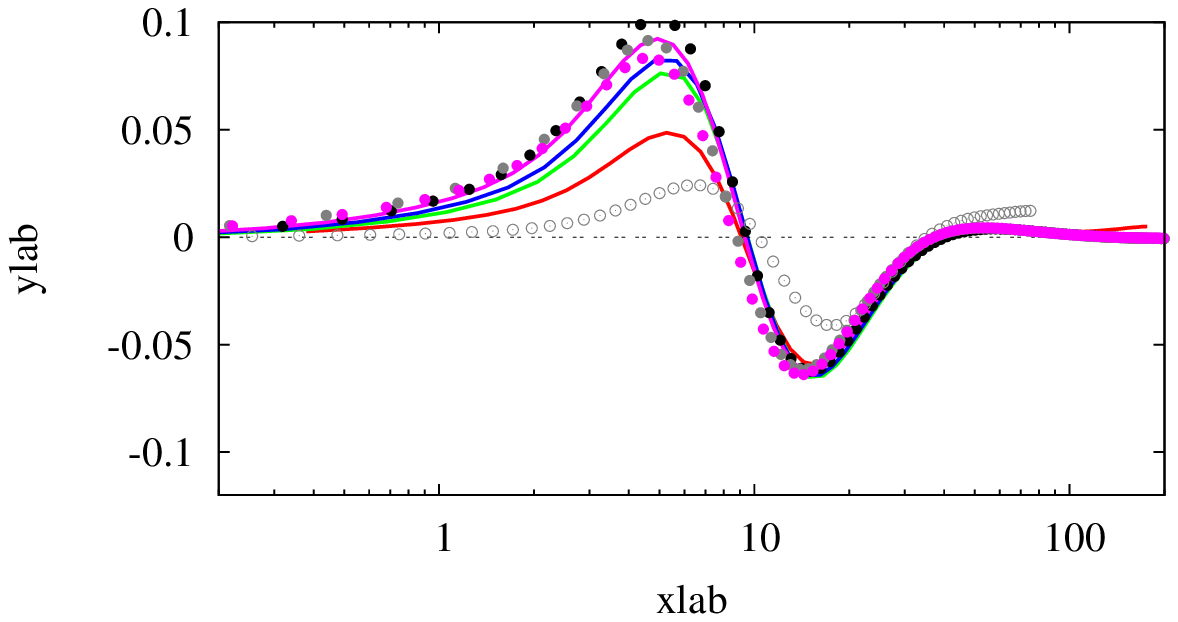}
\vskip -0.5cm \hskip 5cm c) \hskip 7cm d)
\caption{Profiles in wall units of:
a) $\dedq{ \u2a}{ x^2_2}$,
b) turbulent kinetic energy production,
c) full rate of dissipation ,
d) turbulent diffusion by non-linear terms,
the data
are from the references given in the text. The Reynolds
number are in the insets of a); in  c), d)
the profiles at $R_\tau=8000$ could not be evaluated from the
\cite{yamamoto_18} data.
}
\label{fig2}
\end{figure}

To understand in more detail the influence of the Reynolds
number on the flow structures in the near-wall region
it is worth to look at the profiles of
$\dedq{ \u2a}{ x^2_2}=-\aQ$.  As previously mentioned
this quantity is null in homogeneous turbulent flows,
for $-\aQ>0$ sheet-like structures prevail
on tubular like structures. In the near-wall region the sheets
produce and dissipate turbulent kinetic energy.
The profiles in wall units of $\dedq{ \u2a}{ x^2_2}$,
$P_k$ and  $D_k$ in figure \ref{fig2} show a similar dependence
upon the Reynolds number. Namely large variations 
for $70 < R_\tau < 200$ and small for $500 < R_\tau < 5200$.
In figure \ref{fig2} the data by \cite{yamamoto_18} at $R_\tau=8000$
are not reported, since the budgets profiles were not
given. Figure \ref{fig2}a for $500 < R_\tau < 5200$
depicts a good scaling for the sheet-like structures near the  
wall and even better for the tubular structures in the buffer region.
It can also been observed that the trend with $Re$ is not regular, in fact 
the peak at $R_\tau=5200$ (\cite{lee_15}) is smaller that that at $R_\tau=4000$
(\cite{Bernardini2014}) and that at $R_\tau=8000$ (\cite{yamamoto_18}).
The profiles for $70 < R_\tau < 200$ of $-\aQ$ largely depend
on the Reynolds number with the magnitude  
decreasing with $Re$ in both regions. By reducing the
Reynolds number the zero crossing point moves far from the wall. 
Figure \ref{fig2}b shows that the maximum energy
production is located, at low and high Reynolds numbers,
near the crossing point and it is slightly shifted in
the region where the sheet-like structures prevail.
In this location it may be inferred that the unstable
ribbon-like structures tend to roll-up to become
rod-like structures. When the Reynolds number increases
the saturation of the maximum, as well as of the
entire profiles up to $y^+\approx 200$ is evident, corroborating the saturation
of the maximum of the turbulent kinetic energy in
figure \ref{fig1}a. The total rate of dissipation 
in figure \ref{fig1}c behaves
similarly to the production, with the maximum located
in the region dominated by ribbon-like structures , therefore
during the roll-up of the unstable structures the
maximum of production and dissipation occur. The scaling
of $D_k^+$ at high $Re$ is rather good but not as good as
that of $P_k^+$ in figure \ref{fig2}b. This occurrence can be  explained
by considering that the production is directly linked to the mean shear, having a
perfect scaling with the Reynolds number. Since $P_k^+$ is balanced by 
$D_k^+$ and $T_k^+$ (turbulent kinetic energy diffusion)
and that the latter is smaller than
$P_k^+$  and $D_k^+$, the $Re$ dependence in 
$D_k^+$ should appear on the profiles of $T_k^+$. 
Indeed  the profiles of the turbulent diffusion,
in figure \ref{fig2}d, evaluated by including the small contribution 
of the pressure strain term, show a deterioration of the
wall scaling in the sheet dominated layer.  The 
transfer of energy from the region dominated by the tubular structures 
into the region dominated by the sheets depends on
the Reynolds number and this dependence should be expected
because of the influence of the viscosity on the roll-up of the 
ribbon-like structures. In figure \ref{fig2}d the perfect collapse 
of $T_k^+<0$, for the flows at $180<R_\tau$, suggests that the universal 
rod-like structures loose the same amount of energy independently from the 
value of $R_\tau$. The profiles in figure \ref{fig2}b-d
could be of interest to whom is interested to build low-Reynolds number 
RANS (Reynolds Averaged Navier-Stokes) closures at high
Reynolds number. In fact the
model of the rate of full dissipation should be  easier
growing, from zero at the wall, proportionally to $q^2$
in the viscous layer. 

\begin{figure}
\centering
\vskip 0.0cm
\hskip -1.8cm
\psfrag{ylab} { \large $ R_{\alpha \alpha}^+$}
\psfrag{xlab}{ $ $}
\includegraphics[width=7.5cm]{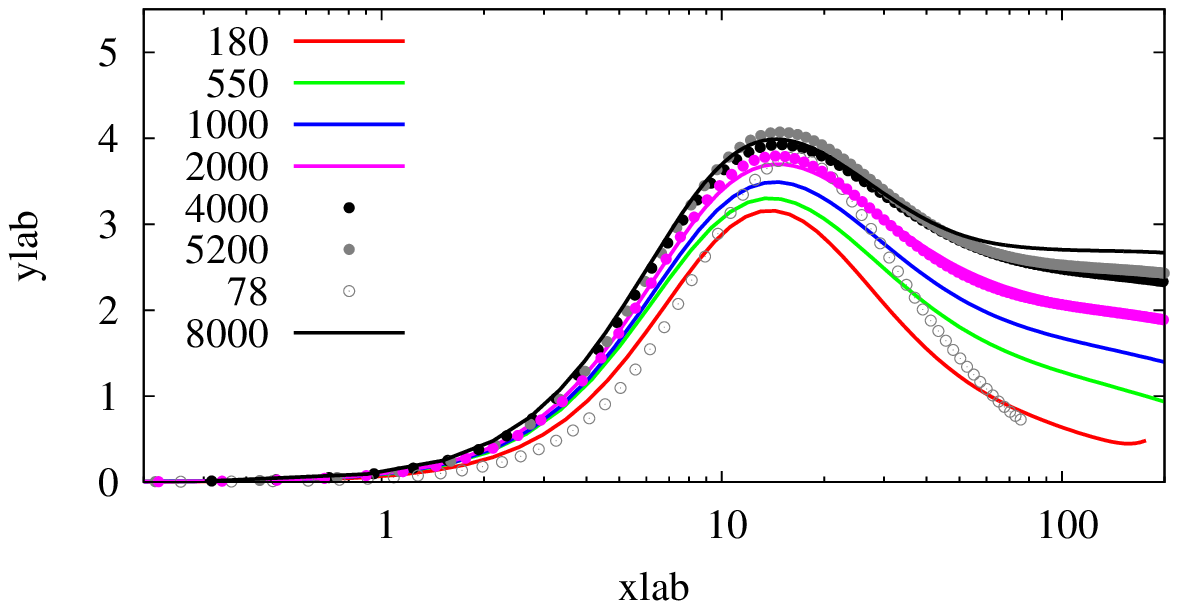}
\psfrag{ylab} {\large $ R_{\gamma \gamma}^+$}
\psfrag{xlab}{ $ $}
\includegraphics[width=7.5cm]{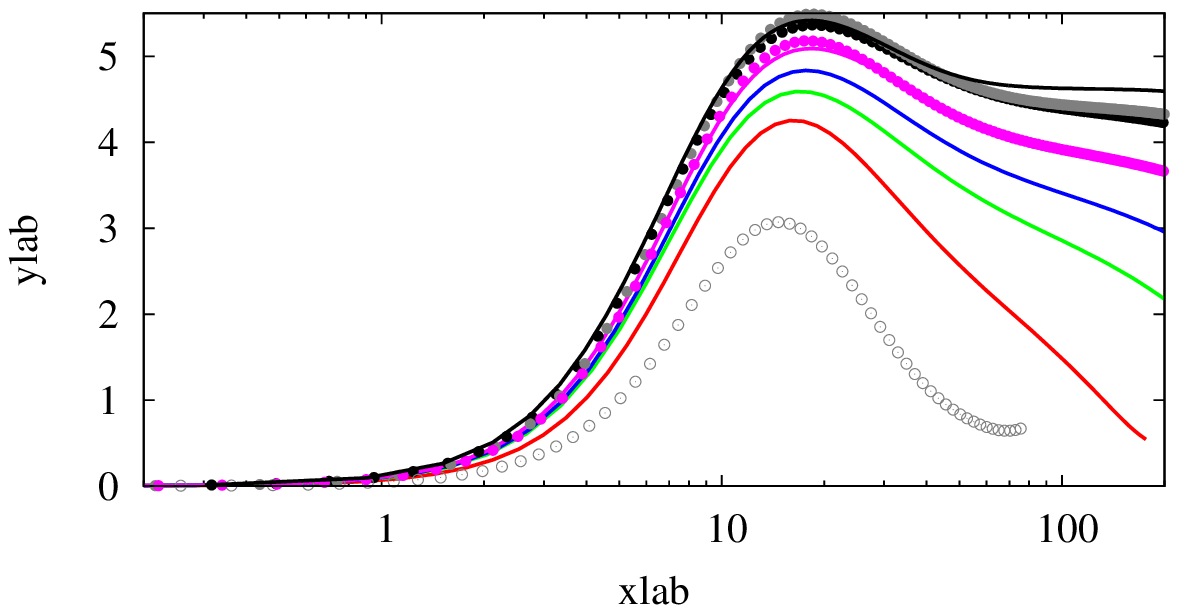}
\vskip -0.5cm \hskip 5cm a) \hskip 7cm b)
\vskip -0.2cm
\hskip -1.8cm
\psfrag{ylab}{\large $ P_\alpha^+$}
\psfrag{xlab}{\large $ y^+ $ }
\includegraphics[width=7.5cm]{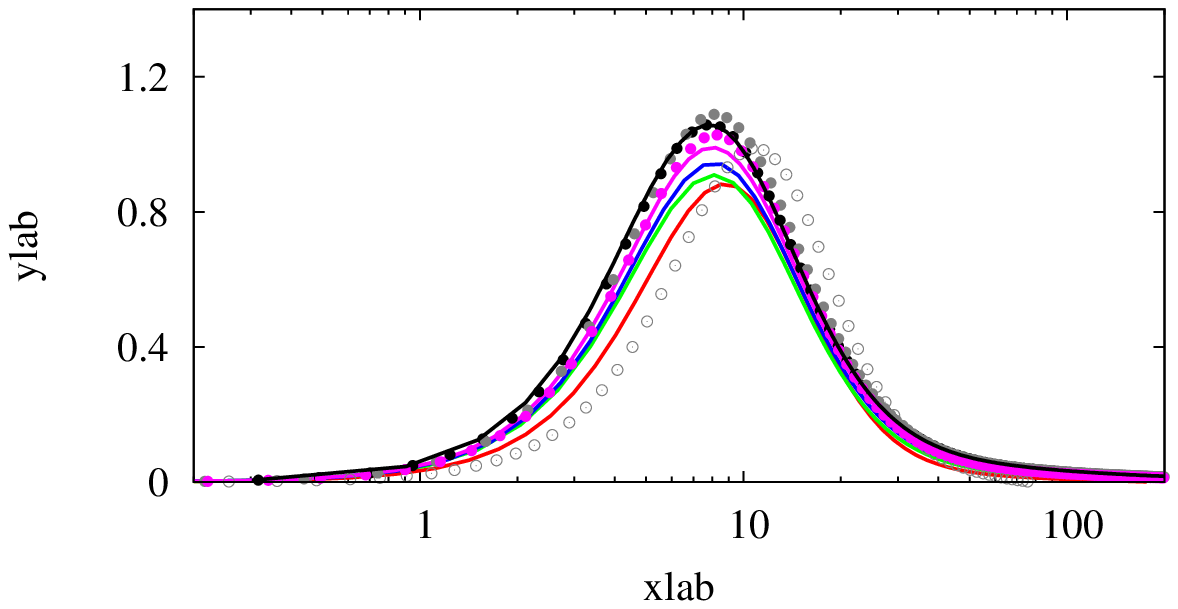}
\psfrag{ylab}{\large $ P_\gamma^+$}
\psfrag{xlab}{\large $ y^+$ }
\includegraphics[width=7.5cm]{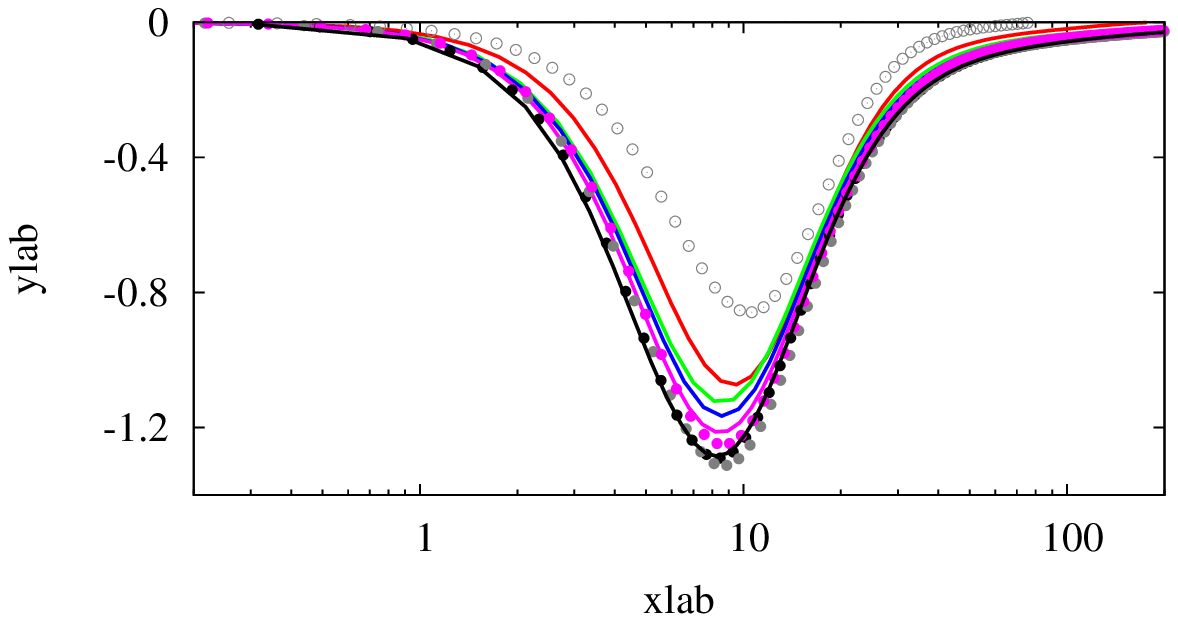}
\vskip -0.5cm \hskip 5cm c) \hskip 7cm d)
\caption{Profiles in wall units of the stress aligned with
a)  $S_\alpha$,
b)  $S_\gamma$;
of the turbulent kinetic energy aligned with 
c) $S_\alpha$,
d) $S_\gamma$,
the data are from the references given in the text. The Reynolds
number are in the insets of figure a).
}
\label{fig3}
\end{figure}

To get a different view of the contribution of the structures to the
turbulent kinetic energy production 
it is worth looking at the distribution of 
the normal stresses aligned with the eigenvectors of  
the strain tensor $S_{ij}$. The reason, as previously mentioned, is that the
good scaling of the production with the Reynolds number is
due  to its proportionality with the mean shear $S$.
Therefore the statistics aligned with the eigenvectors of $S_{ij}$
should be linked to  flow structures  different from those
visualised in the Cartesian reference frame. The
new reference frame is aligned with  a negative compressive
$S_\gamma$ and a positive extensional $S_\alpha$ strain.
In the Cartesian frame the near-wall inhomogeneity
is manifested by large differences in the
profiles of the normal stresses 
$R_{ii}=\langle u_i u_i\rangle$
with $R_{22}<R_{33}<< R_{11}$. Flow
visualizations in planes $x_1-x_3$ parallel to the wall,
show very elongated structures for $u_1$, while the other 
two fluctuating velocity components are concentrated in patches
of elliptical or circular shape. The contours of $u_2$
in several location depict the presence of an intense
negative patch surrounded by  two positive
patches of elliptical shape. In correspondence of the strong
$u_2<0$ (sweeps events) the positive elongated
streamwise structures form (\cite{Orlandi2016}). 
Several papers have been addressed to investigate this cycle of events, 
for instance that by \cite{jimenez_99}. The $u_1$ and $u_2$
are the fluctuations producing the active motion in turbulent
flows since their combination
interacts  directly  with the mean shear $S$ to produce new 
fluctuations.  The fluctuations $u_3$ in the spanwise direction can be considered
as an inactive motion, and these are concentrated in positive and
negative patches. The structures therefore are not well defined as
those of the other two velocity components.
These structures can be considered inactive
also because the profiles of the relative stress $R_{33}$ 
coincide with $R_{\beta \beta}$ aligned with $S_{\beta}=0$.
The vertical profiles of $R_{\beta \beta}^+$ are not reported, on the
other hand figure \ref{fig3}a and figure \ref{fig3}b
shows that $R_{\alpha \alpha}^+$ and $R_{\gamma \gamma}^+$ do not differ 
in shape, and that those aligned with $S_\gamma$ are greater than 
those aligned with $S_\alpha$. In each component a strong
Reynolds dependence, similar to that depicted in 
figure \ref{fig1}a for $q^{2 +}$ emerges. The similarity of
the profiles of $R_{\alpha \alpha}^+$ and of
$R_{\gamma \gamma}^+$ suggests that the contours
of $u_\alpha u_\alpha$ and $u_\gamma u_\gamma$
in a plane $x_1-x_3$ parallel to the wall, 
should be similar implying that the structures 
aligned with $S_\alpha$ and $S_\gamma$ do not largely differ. 
This is shown later on
discussing the differences between flows past smooth 
walls and flows past corrugated walls.
To investigate which of the two kind of structures
plays a large role in the near-wall
turbulent kinetic production it is worth
to decompose the production $P_k$.
In this local frame $P_k=-(P_\alpha+P_\gamma)$
with $P_\alpha=R_{\alpha \alphaį} S_\alpha$ and
$P_\gamma=R_{\gamma \gamma} S_\gamma$. The two terms
are greater than $P_k$ and their profiles in figure \ref{fig3}c
and in figure \ref{fig3}d show that the compressive strain generates
more kinetic energy than that eliminated by the extensional  one.
In both terms there is a Reynolds dependence at
high and low $Re$ numbers, while the sum of the two
in figure \ref{fig2}b shows that it is almost
absent at high $Re$. It can be, therefore, concluded,
that the universality of the wall structures is
evident  only in some of the statistics.

\begin{figure}
\centering
\vskip 0.0cm
\hskip -1.8cm
\psfrag{ylab} { \large $ P_T^+$}
\psfrag{xlab}{ $y^+ $}
\includegraphics[width=7.5cm]{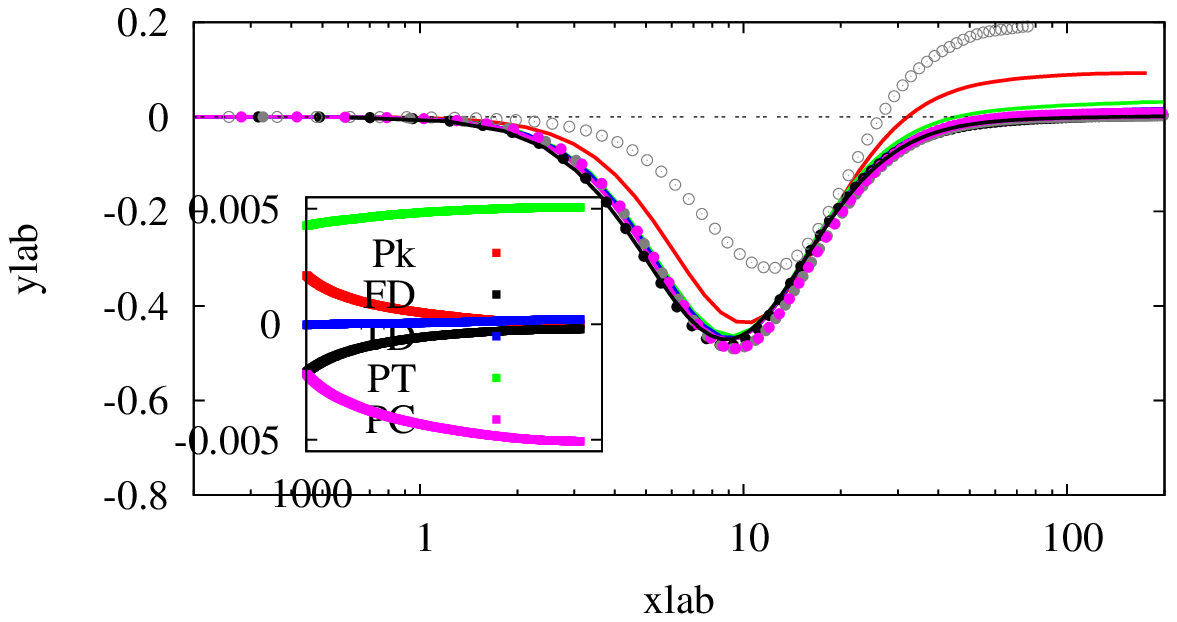}
\psfrag{ylab} { \large $ P_C^+$}
\psfrag{xlab}{ $ $}
\includegraphics[width=7.5cm]{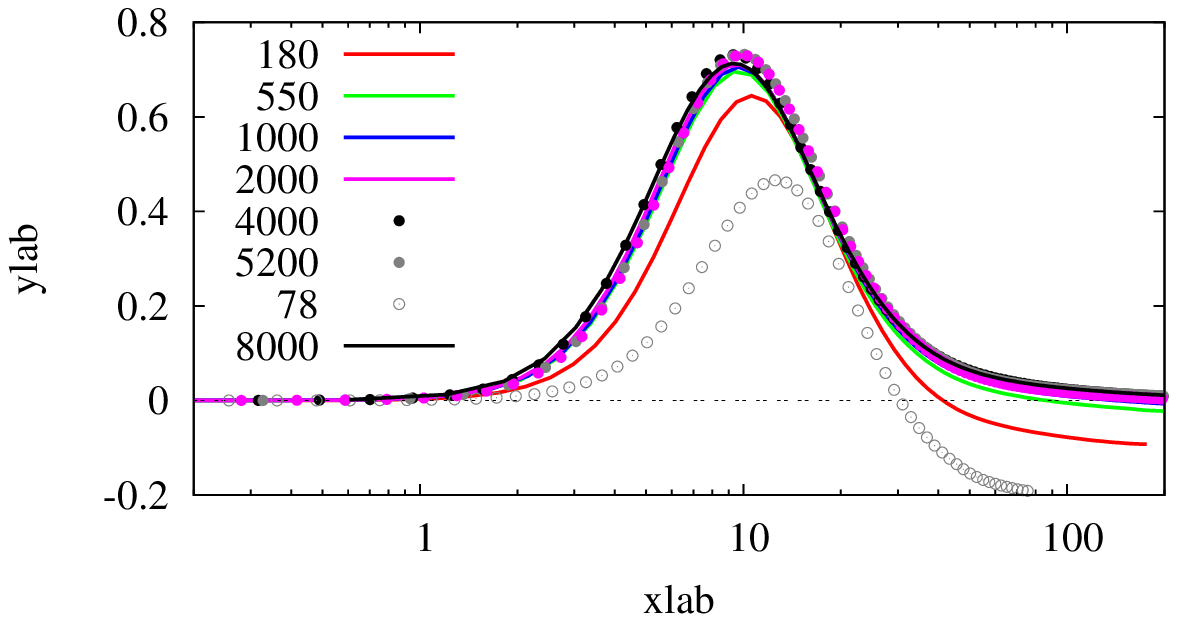}
\vskip -0.5cm \hskip 5cm a) \hskip 7cm b)
\caption{Profiles in wall units of:
a) $P_T$, 
b) $P_C$ as are defined in the text,
the data
are from the references given in the text. The Reynolds
number are in the insets of figure b).
}
\label{fig4}
\end{figure}

A different way to split $P_k$ was used 
by \cite{Orlandi2000} at Pg.211 to enlight 
the energy transfer from large to small eddies in the near-wall
region. The splitting was derived by the Navier-Stokes equation in
rotational form where the Lamb vector 
$\bf{\lambda}=-\bf{u}\times\bf{\omega}$
appears. This term is 
$P_T= U(\langle u_3 \omega_2\rangle 
-\langle u_2 \omega_3\rangle )$, and to get 
$P_k$ it should be added 
$P_C=\der{ U  \langle u_1 u_2\rangle }{ x_2}$
related to the action of the large
eddies advecting the turbulence across the channel.
The profiles of the vorticity velocity correlations
were not directly evaluated in the simulations here
used. However the identity $\langle u_3 \omega_2\rangle 
-\langle u_2 \omega_3\rangle
=\der{\langle u_2 u_1\rangle}{ x_2}$ allows to evaluate
$P_T$.  The two terms are plotted in figure \ref{fig4} with the characteristic
to have an universal behavior in the near-wall region for 
$500<R_\tau<8000$. The detailed analysis of the two terms
gives some insight on what occurs in the whole channel.
The expression of $P_C$ demonstrates that what is   
produced near the wall is transferred to the outer region.
In fact $P_C^+$ is negative in the outer region and in
magnitude higher smaller $Re$, being         
smaller than $P_T^+$ it follows that $P_k^+>0$.
\cite{Tsinober2002} at Pg.120 analysed the physical
aspects of the kinematic relationship previously mentioned,
by asserting "the component of the Lamb vectors
imply a statistical dependence by large scales 
($\bf{u}$) and  small scales ($\bf{\omega}$).
Without this dependence the mean flow does not
fill the turbulent part". Indeed figure \ref{fig4}a
shows that, for $100<y^+$, $P_T^+$ is negligible at high $R_\tau$,
however, the large contribution of $P_T^+$ infer the transfer of energy from large
to small scale, which is redistributed by $P_C^+$. The budget in \cite{lee_15}
at $R_\tau=5200$, in the outer region, shown in the inset
of figure \ref{fig4}a depicts the large contributions of
the $P_T^+$ and $P_C^+$ terms with respect to the turbulent
diffusion and to the full dissipation discussed respectively
in figure \ref{fig2}c and in figure \ref{fig2}d.
\cite{Tsinober2002} wrote 
" It is noteworthy that both correlation coefficients
$C_{u_3 \omega_2}=\frac{ (\langle u_3 \omega_2\rangle }
{(\langle u_3^{2}\rangle)^{1/2}
(\langle \omega_2^{ 2}\rangle)^{1/2}}$ and
$C_{u_2 \omega_3}=\frac{ (\langle u_2 \omega_3\rangle }
{(\langle u_2^2\rangle)^{1/2}
(\langle \omega_3^2\rangle)^{1/2}}$ 
(and many other statistical characteristics, e.g.
some, but not all, measures of anisotropy) are of order
$10^{-2}$ even at rather
small Reynolds numbers. Nevertheless, as we have seen, in view of the
dynamical importance of interaction between velocity and vorticity in 
turbulent shear flows such ‘small’ correlations by no means imply absence
of a dynamically important statistical dependence and a direct interaction
between large and small scales."  This is indeed true in the 
outer region where $S$ is rather small, however large enough to
give a $S^*\approx 10$ sufficient to create large
elongated structures in the outer region. Near the
wall the two velocity vorticity correlations are quite large producing
large negative values of $P_T^+$. This should be investigated in
more detail by looking at the joint pdf between the
velocity and vorticity components.

The easiest way to change the velocity fluctuations at the boundary
consists on the modification of the shape of the surface  with the
result to  produce large
differences among the statistics profiles in the near-wall region.
Therefore there is a  large probability that the universality with
the Reynolds number, described in this section, is not any longer
valid.  In the next section the behaviours of the 
quantities, here considered, are discussed 
for flows past surfaces leading to an increase or to a reduction of
the drag with respect to that in presence of smooth walls. 
Having several realisations the joint pdf between the
velocity components generating the turbulent stress
together with flow visualizations allow  to understand in more 
details the differences between smooth and rough walls.

\subsection{Rough walls}

\subsubsection{ Numerical Procedure and validation }

The numerical methodology was described in several previous papers
in particular \cite{OrlandiLeonardi2006}, however
it is worth to shortly summarise the main features of the 
method, and to recall the validation based on the comparisons between 
the numerical results and the laboratory data available in the literature.

The non-dimensional Navier-Stokes and continuity equations for
incompressible flows are
\begin{equation}
\frac{\partial u_{i}}{\partial t} + \frac{\partial u_{i}u_{j}}{\partial
x_{j}} = -\frac{\partial p}{\partial x_{i}} + \frac{1}{Re} \;
\frac{\partial^{2} u_{i}}{\partial x_{j}^2 } + \Pi \delta_{1i},
\hspace*{0.25in} \frac{\partial u_{j}}{\partial x_{j}} = 0,
\label{eqNS}
\end{equation}

\noindent where $\Pi$ is the pressure gradient required to maintain a
constant flow rate, $u_{i}$ is the component of the velocity vector in
the $i$ direction and $p$ is the pressure.
The reference velocity is the centerline laminar Poiseuille
velocity profile $U_P$, and the reference length is  the
half channel height $h$ in presence of smooth walls.  
The Navier-Stokes equations 
have been discretized in an orthogonal coordinate system through a
staggered central second-order finite-difference approximation.
The discretization scheme of the
equations is reported in chapter 9 of \cite{Orlandi2000}.
To treat complex boundaries, \cite{OrlandiLeonardi2006} developed
an immersed boundary technique, whereby the mean 
pressure gradient to maintain a constant flow rate in channels 
with rough surfaces of any shape is enforced.  In the presence of
rough walls, after the discrete integration
of $RHS_1$ (right-hand-side in the $i=1$ direction)
in the whole computational domain, a correction is necessary to account for 
the metrics variations near the body. This procedure,
requires a number of operations  proportional to
the number of boundary points, and the
flow rate remains constant within round-off errors.
In principle, there is no big difference in treating
two- or three-dimensional geometries. However,
in the latter case, a greater memory occupancy is necessary
to define the nearest points to the wall surface.

In the present paper several types of corrugations have
been considered in a computational domain with size $L_1=8$ 
in the streamwise and $L_3=2\pi$ in the spanwise directions.
Differently than in previous simulation, where
one wall was corrugated and the other was smooth, here
both walls have the same corrugation. This set-up has the
advantage to investigate whether at the steady state 
a symmetrical solution is obtained.  This condition should 
require a large number of realisations collected by
simulations requiring a great CPU time.
As for the flows past smooth walls, considered in the previous
section, the symmetric boundaries for flow past rough walls
require a reduced number of realisations to  get converged
statistics.

\begin{figure}
\centering
\vskip 0.0cm
\hskip -2.0cm
\includegraphics[width=4.0cm]{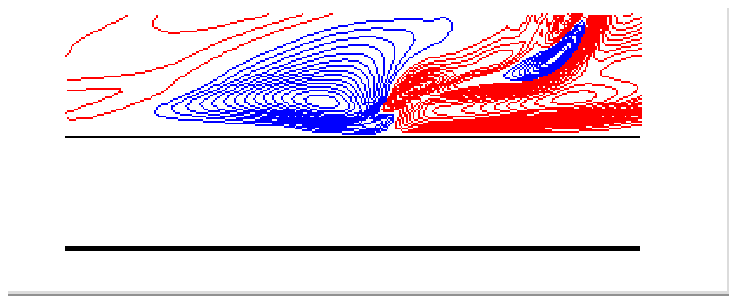}
\hskip -0.3cm
\includegraphics[width=4.0cm]{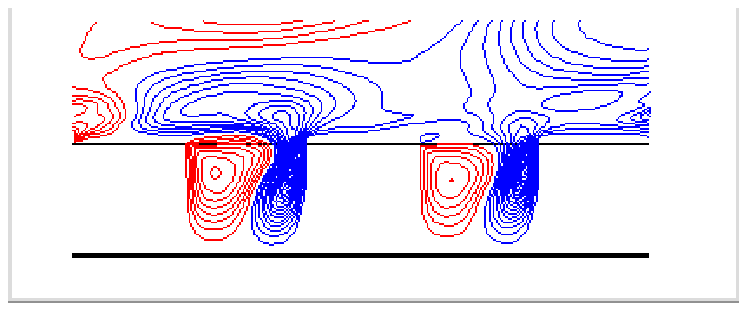}
\hskip -0.3cm
\includegraphics[width=4.0cm]{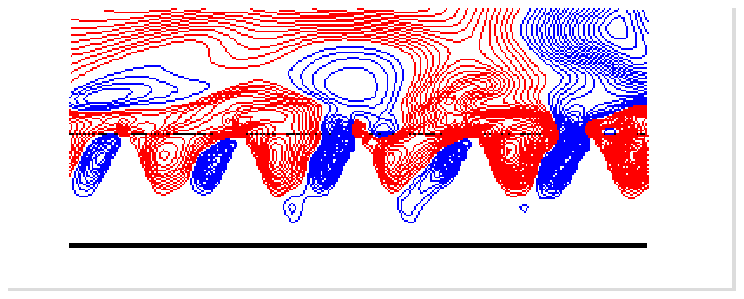}
\hskip -0.3cm
\includegraphics[width=4.0cm]{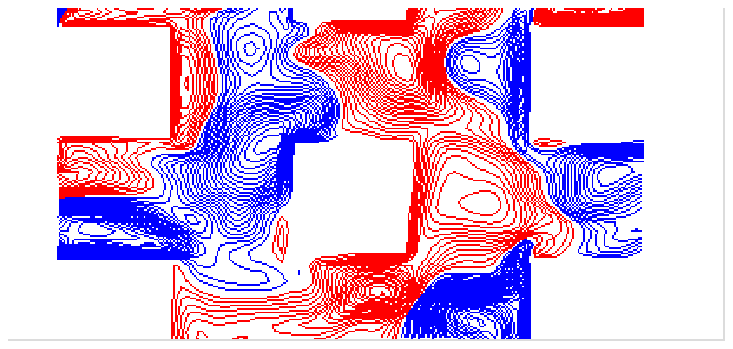}
\vskip -0.1cm \hskip 0cm a) \hskip 3.5cm b) \hskip 3.5cm c) \hskip 3.5cm d)
\vskip 0.5cm
\hskip -2.0cm
\includegraphics[width=4.0cm]{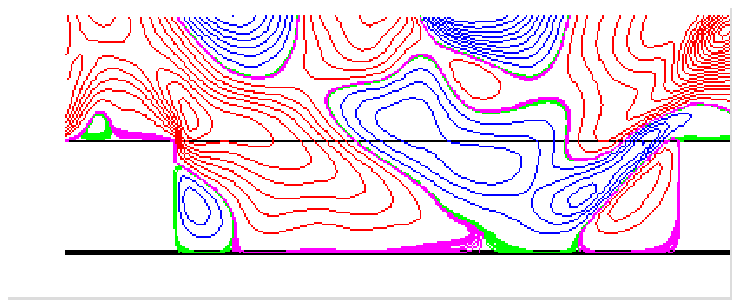}
\hskip -0.3cm
\includegraphics[width=4.0cm]{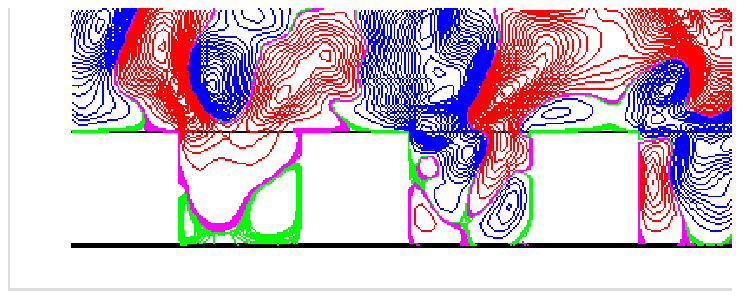}
\hskip -0.3cm
\includegraphics[width=4.0cm]{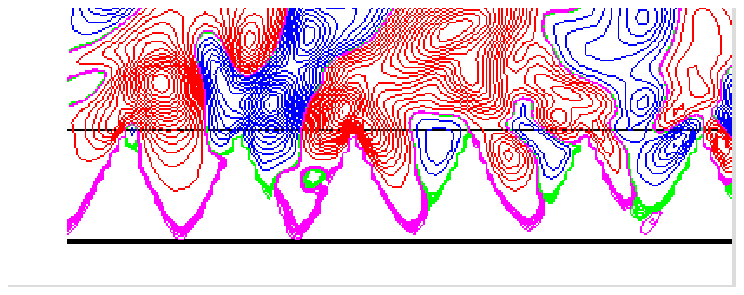}
\hskip -0.3cm
\includegraphics[width=4.0cm]{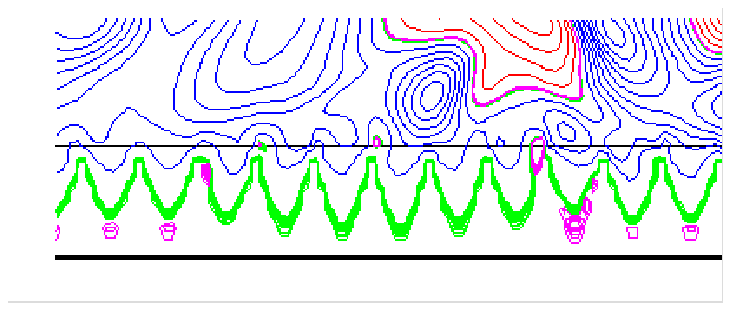}
\vskip -0.2cm \hskip 0cm e) \hskip 3.5cm f) \hskip 3.5cm g) \hskip 3.5cm h)
\caption{contour plots of $u_2$: 
a) $SM$, b) $TS$, c) $TT$ in planes $x_1-x_2$,
d) $CS$ plane $x_1-x_3$ at $x_2=-1.03$,
e) $LLS $, f) $LS$ , g) $LT$, h) $LTS$, planes $x_3-x_2$,
blue and green negative, red and magenta positive
$\Delta=0.005$ for red and blue 
$\Delta=0.0005$ for green and magenta.
}
\label{fig5}
\end{figure}

The validation of the immersed boundary technique was
presented in  \cite{OrlandiLeonardiAntonia2006} by a comparison of 
the pressure distribution on the rod-shaped elements 
with the measurements  by \cite{FMF}.
These authors studied the boundary layer over
circular rods, fixed to the wall transversely to the flow,
for several values of $w/k$ ($w$ is the streamwise separation
between two consecutive rods of height $k$).
The numerical validation was performed  for values of $w/k=3,7$ and $15$. 
It is important to point out that circular rods are
appropriate for numerical validation of the immersed boundary
method, owing to the variation of the metric along the
circle.  The numerical simulations were performed,
at $Re=U_P h/\nu=4200$, and the pressure distributions
around the circular rod were compared with those measured.
The good agreement reported in \cite{OrlandiLeonardiAntonia2006}
implies that
the numerical method is accurate and can be used to reproduce the flow
past any type of surface. From the physical point of view, the
agreement between low $Re$ simulations
and high $Re$ experiments ( \cite{FMF})
implies a similarity between the near-wall region of
boundary layers and channel flows. In addition
it can be asserted that, as in fully rough flows,
(\cite{NikuradseEN}) a Reynolds number independence for the
friction factor does exist.
The capability of the immersed boundary technique to
treat rough surfaces was further demonstrated by a comparison
with the experimental results of  \cite{burattini2008} 
for a flow past transverse square bars  with $w/k=3$.  

\subsubsection{ Global results   }

Several corrugations have been located below the plane
of the crest at $x_2\pm 1$, that coincides with the walls
of channels with  smooth walls ($SM $). The shape of the 
corrugation are given in figure \ref{fig5} by plotting
the contours of the $u_2$ velocity component
in the planes more appropriate to see the walls of the corrugations. 
These images demonstrate that the immersed boundary technique
accurately reproduces the flow around the corrugations. The velocity component $u_2$
coincides with the fluctuating component
being, for each realisation, the average in the homogeneous directions equal to
zero. In several previous papers it was
stressed the importance of the $u_2$ fluctuations and the relative statistics. 
The relevant papers are reported in \cite{Orlandi2013}. For the
smooth channel the $u_2$ contours, in a small region,
in figure \ref{fig5}a depict the sweep and the 
ejection events one after the other.
These are the events contributing to increase the drag
of turbulent flows with respect to that of laminar flows and to produce
turbulent kinetic energy. The recirculating motion
in figure \ref{fig5}b within the  cavities of the transverse square bars
configuration ($TS$), for this spanwise section, connect the
negative regions of $u_2$ inside with those of the same
sign above. However, in a different spanwise section, it has been
observed a connection  between positive values. The global results leads 
to a relative high value of $\langle u_2^2\rangle_W$ 
at the plane of the crests.
Triangular transverse bars, one attached to the 
other ($TT $), generate a more intense recirculating motion  
(figure \ref{fig5}c)), producing  large effects on the overlying turbulent
flows. A spanwise coherence of the recirculating motion inside the corrugations
is observed, that disappears at a distance $y=0.2$ from the plane of the crests.
The capability  of the numerical method to
describe the complex flow inside the three-dimensional
staggered cubes ($CS$) can be appreciated by the contours 
of $u_2$ in figure \ref{fig5}d in a $x_1-x_3$ plane at $x_2=-1.03$ 
The velocity disturbances ejected from three-dimensional
corrugations are large, and, therefore, large effects on the
overlying turbulent flow are produced. The motion inside the
longitudinal corrugation can be visualised by $u_2$
contours in $x_3-x_2$ planes; in these circumstances
the motion is rather weak, therefore contours with
$\Delta=0.0005$ are depicted in figures \ref{fig5}e-h
in green for negative and in magenta for positive $u_2$.
These images confirm that the immersed boundary technique
reproduces the complexity of the secondary motion, namely
for the corrugation $LLS $ (figure \ref{fig5}e) with $w/k=3$
($w$ is the distance between two square bars), and
$LS $ with $w/k=1$. Triangular bars ($LT$) with $s/k=1$
($s$ is the width of the base of the triangle)
in figure \ref{fig5}g show disturbances similar 
to those in figure \ref{fig5}f for $LS$.
On the other hand for the triangular bars with $s/k=0.5$
($LTS$) the recirculating motion in figure \ref{fig5}h is very weak, and,
as a consequence, the activity of the overlying
flow decreases, leading to a reduction of turbulent
kinetic energy and of the drag.

%{\tiny
\begin{table}
 \centering
 \begin{tabular*}{1.00\textwidth}
{@{\extracolsep{\fill}}ccccccccccc}
  \hline
   Flow  & $l$& $N_1 $ & $N_3 $  & $H_{fl}$ & $R_\tau$ & $10 U_W$ 
& $10^3 <u_2^2>_W$ 
& $10^2 u_\tau$ & $10^3 \tau_W$ & $10^3 <u_1u_2>_W$\\
  \hline
SM  &0 & 800 & 128 &2.00 &204.2& 0.0  & 0.0  & 4.1678 &17.362&0.    \\
CS  &1 & 800 & 512 &2.295&372.1& 1.397& 3.311& 7.5939 &15.279&34.985\\
TT  &2 & 800 & 128 &2.195&313.3& 1.048& 1.399& 6.3942 &20.384&16.882\\
TS  &3 & 800 & 128 &2.190&238.4& 0.370& 0.157& 4.8649 &20.000& 1.615\\
LS  &4 & 256 & 512 &2.200&228.8& 1.361& 0.528& 4.6699 &13.580& 6.245\\
LLS &5 & 256 & 512 &2.323&217.2& 3.817& 0.638& 4.4325 & 4.706&12.206\\
LT  &6 & 256 & 512 &2.195&205.7& 2.691& 0.338& 4.1981 & 9.777& 6.280 \\
LTS &7 & 256 & 512 &2.189&166.8& 2.143& 0.048& 3.4040 & 9.332& 1.253\\
  \hline
 \end{tabular*}
\caption{Values of some of the global quantities for the simulations
at $Re=4900$, the non-uniform grid $x_2$ is the same in all cases with
$N2=257$ points.
}
\label{table1}
\end{table}

Some of the global results and the resolution for the
cases depicted in in figure \ref{fig5} are reported in the
table \ref{table1}. The resolution in the streamwise and 
spanwise directions are different for transverse, 
longitudinal and three-dimensional corrugations. The
resolution in $x_1$ for transverse corrugations is dictated 
in order to have $20$ grid points
to describe the square and triangular cavities. For the longitudinal
corrugations $16$ grid points are used for the solid bars for the $LS$ 
and $LLS$ cases.
For all cases $20$ grid points in the direction $x_2$ have been
used to get the flow-fields depicted in figure \ref{fig5}.
The values of $U_W$ (the mean streamwise velocity at the plane of the crests)
in the table shows that, 
$U_W$   for the longitudinal bars are greater than
those for the transverse corrugations, implying a decrease
of $\tau_W=\nu\der{U}{x_2}|_W$. Therefore it should be expected a large
drag reduction due to this slip condition. As it has been
demonstrated by \cite{arenasleonardi2018}, if at the plane of
the crests the $u_2$ can be, ideally, set equal to zero for
any kind of corrugations a strong drag reduction is achieved. For
the surfaces here considered the largest reduction should be 
for the $LLS $ configuration. However, in the real flow, the 
$u_2$ fluctuations are large as it can be inferred by the values 
of $\langle u_2^2 \rangle_W$ in table \ref{table1}. The $u_2$ 
at the plane of the crests generates a turbulent
stress $\langle u_2 u_1 \rangle_W$ which can be considered as 
a "form" drag due to the corrugation of the  surfaces, contributing to 
the total  resistance $\tau_T=\tau_W+\langle u_2 u_1 \rangle_W$.
The friction velocities $u_{\tau l}=\sqrt{\tau_{T l}R_{V l}}$ 
($l$ is an index of the geometry of the surface) 
show that only for the surface $LTS$ there
is a drag reduction with respect to that in presence of smooth walls ($SM$).
In this expression $R_{V l}$ is given by the ratio
of $H_{fl}$ with respect to that of the channel with smooth walls ($H$).
$H_{fl}$  is the ratio between the volume occupied  by the fluid and the
area in the homogeneous directions ($L_1L_3$).

\subsubsection{ Viscous and turbulent  stresses   }

\begin{figure}
\centering
\vskip 0.0cm
\hskip -1.8cm
\psfrag{ylab} { \large $ \tau$}
\psfrag{xlab}{ $y $}
\includegraphics[width=7.5cm]{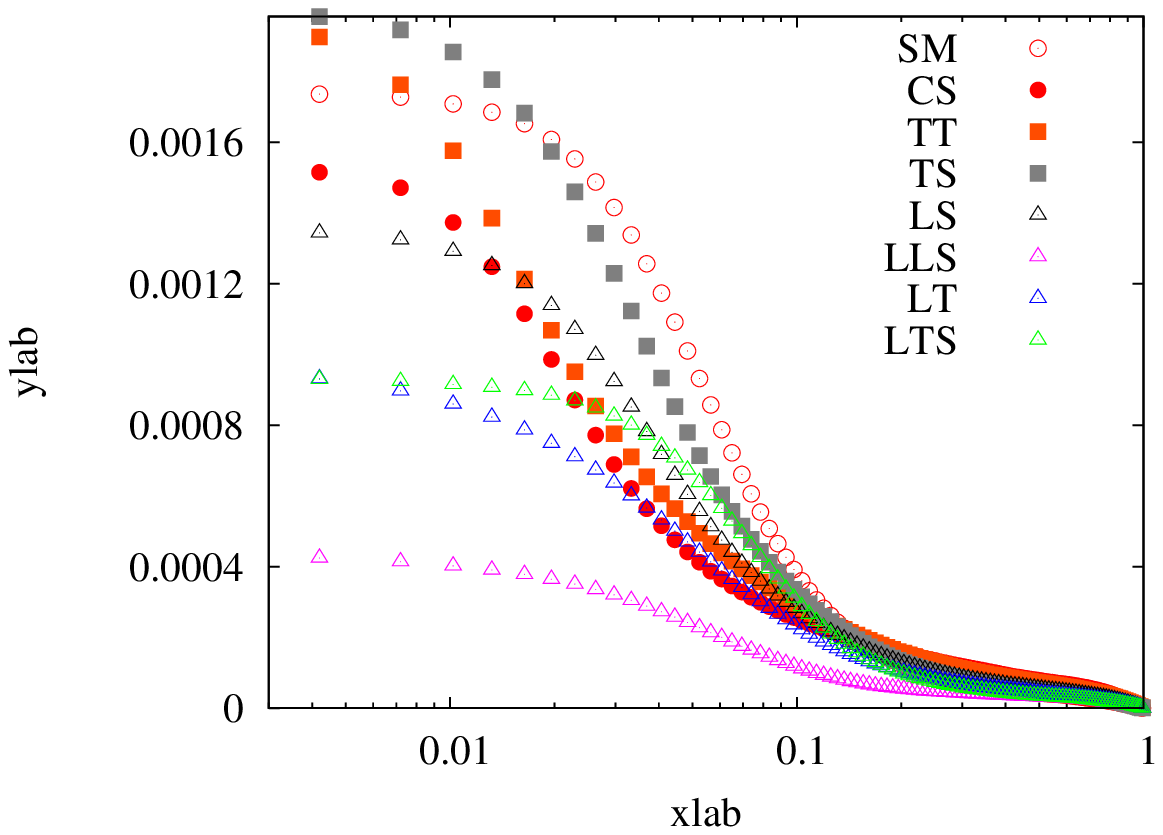}
\psfrag{ylab} { \large $ -\langle u_2 u_1 \rangle$}
\psfrag{xlab}{ $y $}
\includegraphics[width=7.5cm]{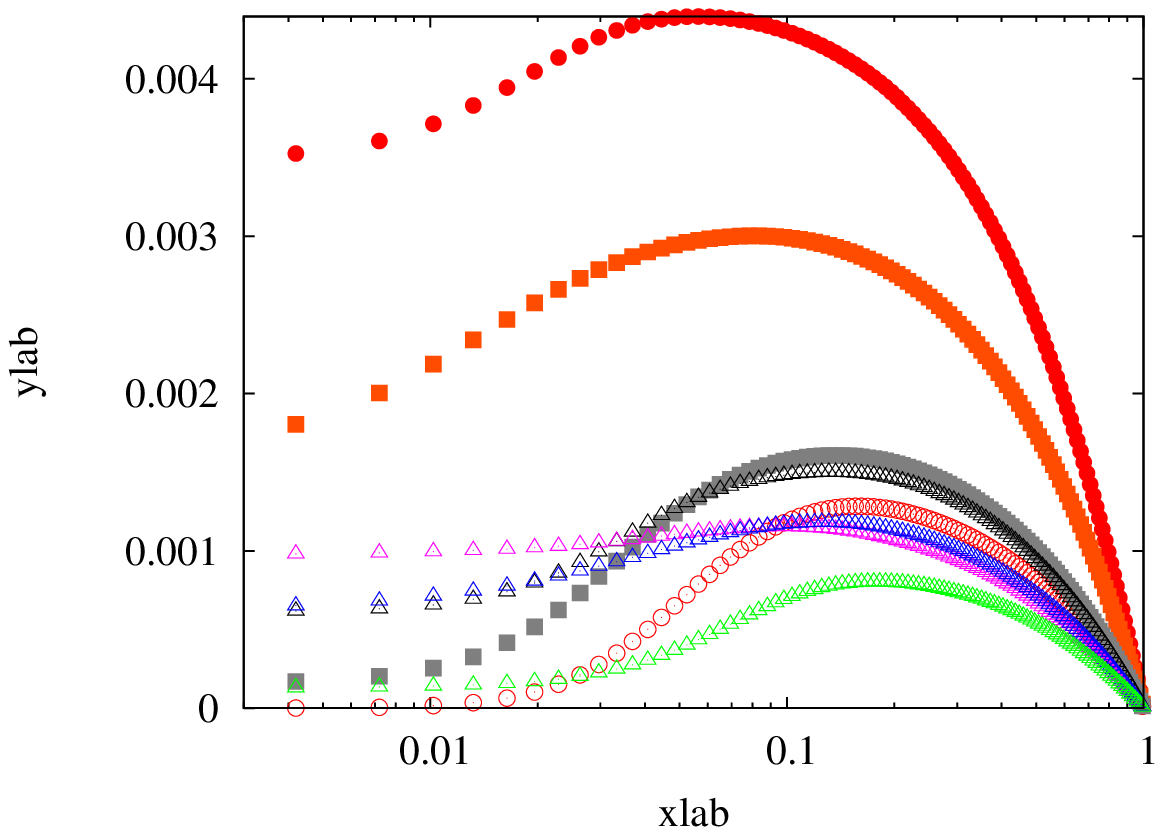}
\vskip -0.5cm \hskip 5cm a) \hskip 7cm b)
\vskip -.2cm
\hskip -1.8cm
\psfrag{ylab} { \large $ \langle u_2^2 \rangle^+$}
\psfrag{xlab}{ $y^+ $}
\includegraphics[width=7.5cm]{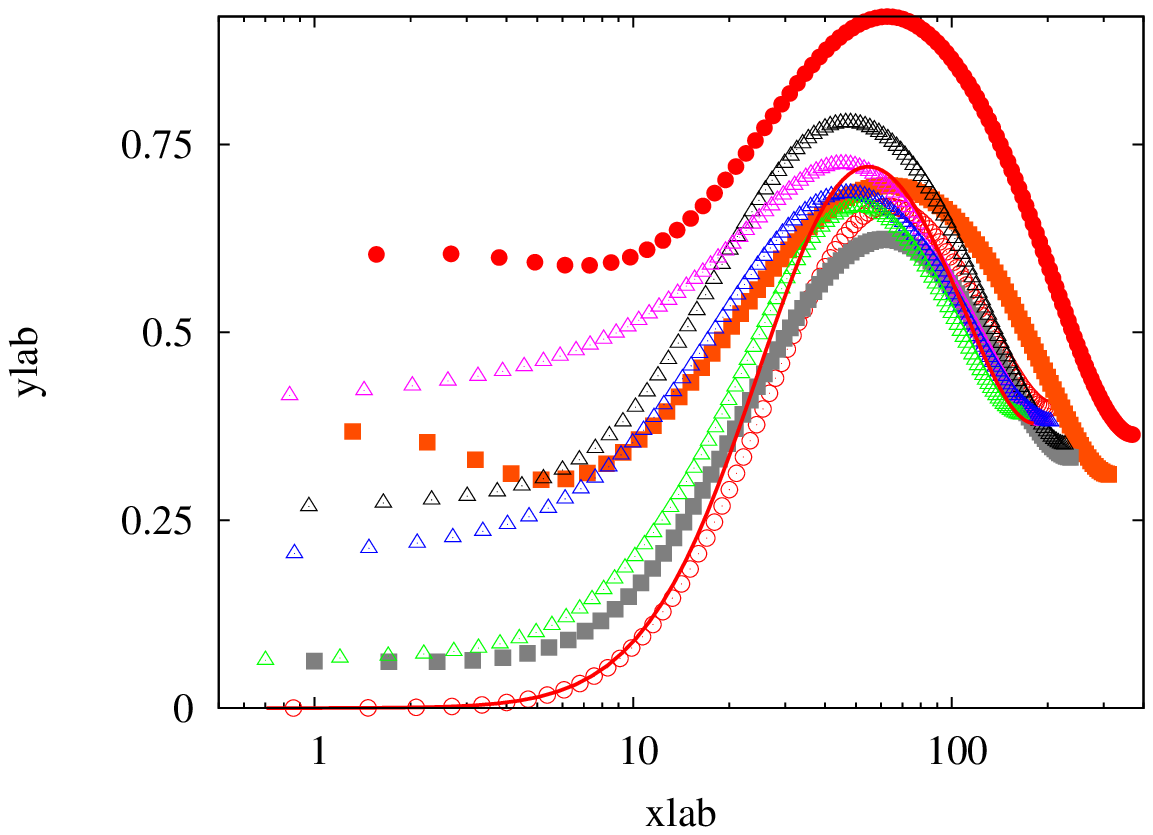}
\psfrag{ylab} { \large $ U^+-U_W^+$}
\psfrag{xlab}{ $y^+ $}
\includegraphics[width=7.5cm]{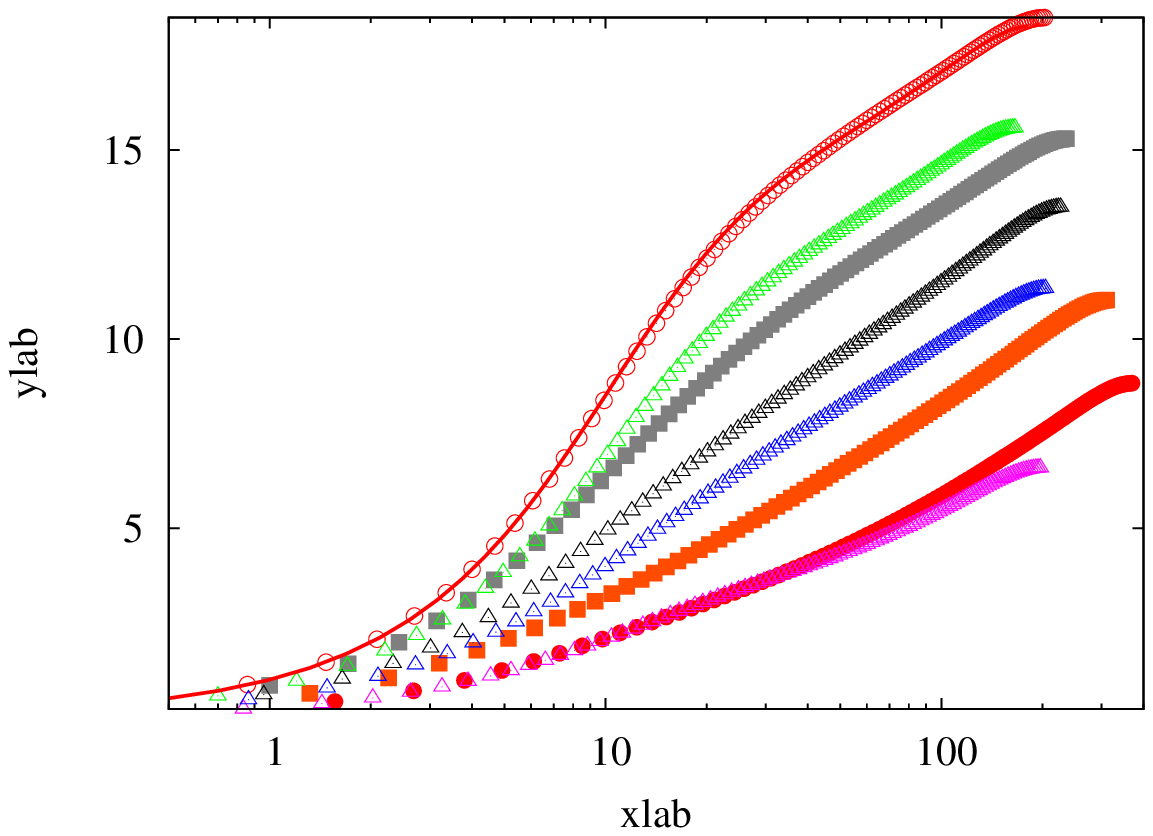}
\vskip -0.5cm \hskip 5cm c) \hskip 7cm d)
\caption{Profiles of:
a) viscous stress,
b) turbulent stress versus the distance from the 
plane of the crests, in computational units; 
c) the normal to the wall stress,
d) the mean velocity subtracted to the velocity at the
plane of the crests $U_W$,
in c) and d) the statistics and
the distance are in wall units.
The flows listed in the insets of a) corresponds to those
in table \ref{table1}.
}
\label{fig6}
\end{figure}

From  the global results it follows that the statistics of large
interest are the viscous $\tau=\nu\der{U}{x_2}$ in figure \ref{fig6}a and 
the turbulent $-\langle u_2 u_1 \rangle$ in figure \ref{fig6}b
stresses. The figures are in semi-log form to emphasise the 
different behavior in the region near the
plane of the crests and therefore  to enlight the difference
with the well known profiles in presence of smooth walls. 
Figure \ref{fig6}a shows a viscous stress, at the plane of the crests,
for transverse grooves  higher than
that of smooth walls. This occurs despite the presence of
a $\widehat{ u_1} |_W \ne 0$ in the regions of the cavities.
The over-script $\widehat{ \cdot}$ indicates an  average in time, 
in $x_3$ for the transverse, in $x_1$ for the longitudinal corrugations
of the generic quantity $q(x_1,x_2,x_3,t)$.
A further  phase average over several elements allows to have
the distribution of $\widehat{ u_1} |_W$  along the cavity.
The distribution of $\widehat{ u_1} |_W$ and of $\widehat{ u_1}$
above the cavities varies with the type
of corrugations and thus allow to understand which part of the
cavity contributes more to the reduction of $\der{\widehat{ u_1} }{y} |_W$. 
\cite{Orlandi2016}, for the $TS$ and $TT$ surfaces, described in detail the reduction
of the viscous stress above the cavity region and the large increase
near the solid leading to the values of $\tau$ in figure \ref{fig6}a
higher than that of $SM$.
Similar distribution along each transverse cavity for $\widehat{ u_1 u_2} |_W$
demonstrate why, in figure \ref{fig6}b,
a small for $TS$, and a large for $TT$, values
are found.  The latter is due to the strong ejections flowing along the slopes of 
the triangular cavities (figure \ref{fig5}c). 
For $TT $ the profiles of the viscous and the turbulent
stress largely differ from those in presence of smooth walls. 
The  statistics profiles are those typical of $k$ type roughness.
Instead for the $TS$ surface the profiles are those
typical of "d" type roughness. The turbulent 
stress profile for the flow past staggered cubes ($CS$) 
is the largest among all the cases here studied with a maximum 
four times greater than that of smooth walls. Even in this flow the causes
of the increase are due to the flow ejections from 
the roughness layer, qualitatively
depicted  in figure \ref{fig5}d. The viscous stress
profiles of the longitudinal corrugations in figure \ref{fig6}a
are largely reduced with respect to that of the smooth
wall, with the smallest values for $LLS $ due to
the high $\widehat{ u_1} |_W$ generated at the wide interface of 
the cavity.  Figure \ref{fig5}e shows a rather high
$u_2$ inside the cavity leading to a turbulent stress
three times greater than the viscous stress 
at the plane of the crests. The final result leads to 
a friction velocity slightly higher than that of smooth
walls. The other two surfaces $LT$ and $LS$, despite the different
profiles of the two stresses, lead to  similar values
of $u_\tau$ in table \ref{table1}. 
The recirculating motion inside the $LS$ cavity (figure \ref{fig5}f)
is similar to that inside the $LT$ (figure \ref{fig5}g), 
therefore the turbulent stress at the plane of the crest in
figure \ref{fig6}b is the same. The wider solid surface of $LS$ 
is the reason why $u_\tau$ in table \ref{table1} is slightly 
greater than that for $LT$. Thin triangular cavities, as
those in figure \ref{fig5}h ($LTS$), give at the plane of the crests
the same value of the viscous stress of $LT$, on the
other hand the values of turbulent stress, in figure \ref{fig6}b, 
are drastically reduced in the whole channel leading to 
a sensible drag reduction. In fact for $LTS$ $u_\tau$ is $18\%$ 
smaller than for $SM$.

Our view is that the normal to the
wall stress is the fundamental statistics to characterise
wall bounded flows. The values at the plane of the crests
are linked to the shape of the surfaces. It should be a difficult
task to relate it to the kind of the surface, in fact a large number
of  geometrical parameters enter in the characterisation of a surface. 
For instance in figure \ref{fig6}c the profile of $\langle u_2^2 \rangle$
of $LTS$ do not differ from that of $TS$ being the surfaces completely different.
Despite the quantitative differences with the smooth wall
it is interesting to notice that, in a thin layer of few
wall units, the growth is similar to that for smooth walls,
with the exception of the surfaces with very strong ejections ($CS$ and
$TT$). \cite {Orlandi2013} by investigating the importance of
$\langle u_2^2 \rangle|_W$  in wall bounded flows 
observed that the roughness function $\Delta U^+$ evaluated
by the profiles of $U^+-U_W^+$ is proportional to 
$\langle u_2^2 \rangle^+|_W$. This behavior can be
appreciated in figure \ref{fig6}d where the downward shift of the 
$\log$ law is greater higher $\langle u_2^2 \rangle^+|_W$.
In figure \ref{fig6}d the results
by \cite{lee_15} are in  perfect agreement with the present one
corroborating the accuracy of the present numerics. The
differences, in figure \ref{fig6}c, between the present $SM$ and the
\cite{lee_15} profiles should be, in part, attributed
to the effect of the Reynolds number. In fact, in the previous
section, large differences  have been observed  at low $Re$,
here $R_\tau=204$ instead in  \cite{lee_15} $R_\tau=180$.

\begin{figure}
\centering
\vskip 0.0cm
\hskip -1.8cm
\psfrag{ylab}{\large $ q^{2 +}$}
\psfrag{xlab}{\large $ $ }
\includegraphics[width=7.5cm]{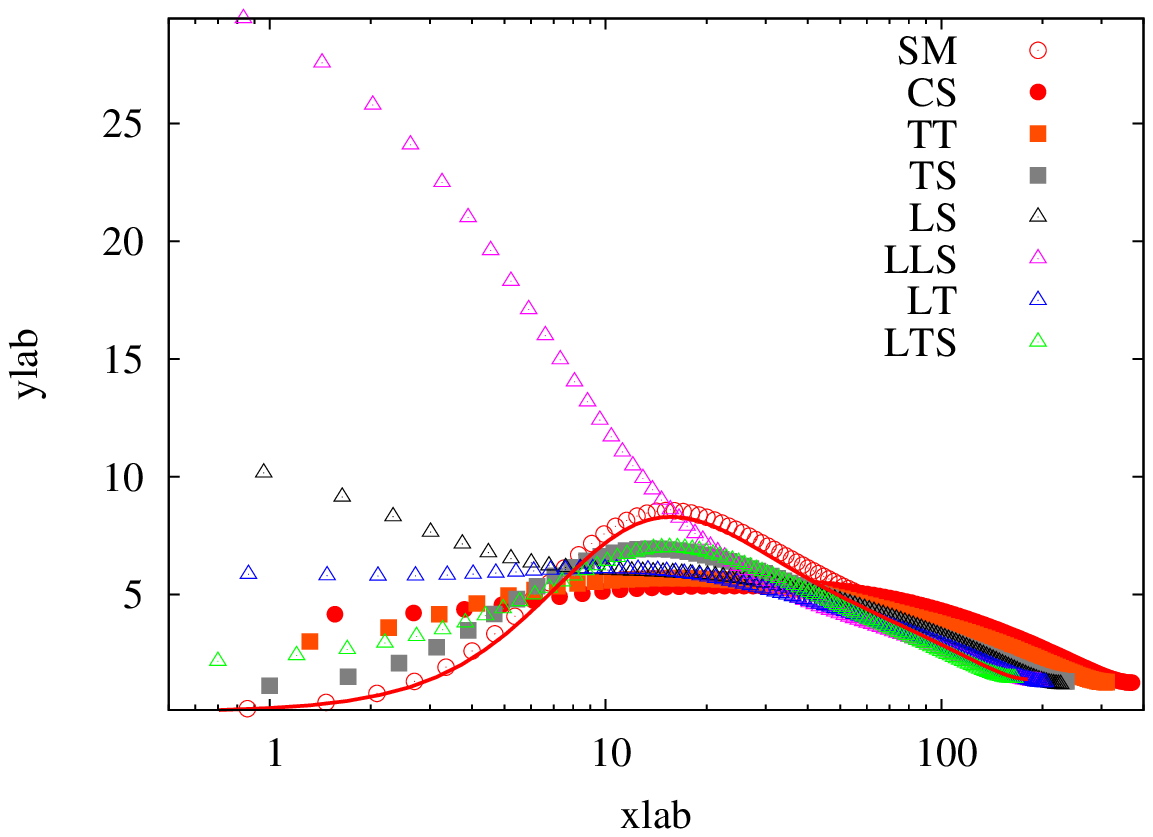}
\hskip -0.0cm
\psfrag{ylab}{\large $\epsilon^{ +}$ }
\psfrag{xlab}{\large $ $ }
\includegraphics[width=7.5cm]{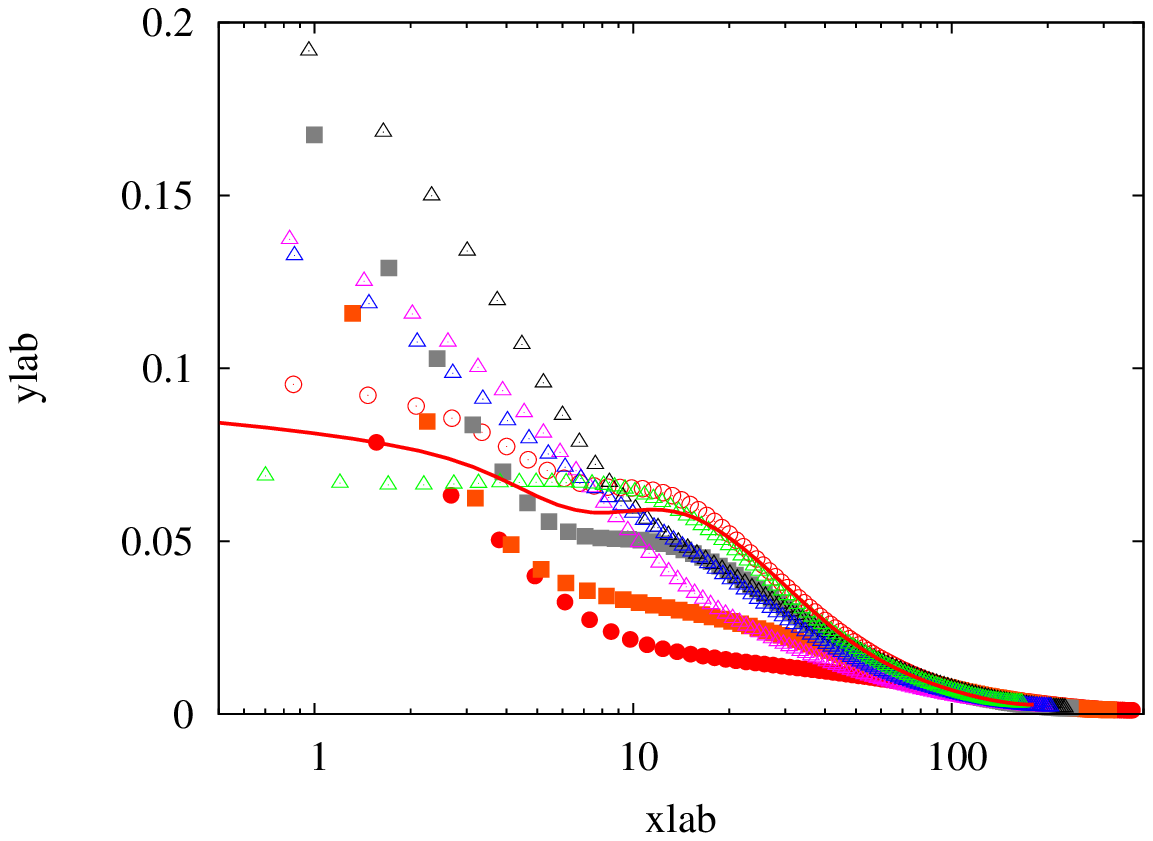}
\vskip -0.5cm \hskip 5cm a) \hskip 7cm b)
\vskip -.2cm
\hskip -1.8cm
\psfrag{ylab}{\large $ (q^{2 }/\epsilon)^{ +}$}
\psfrag{xlab}{\large $y^+ $ }
\includegraphics[width=7.5cm]{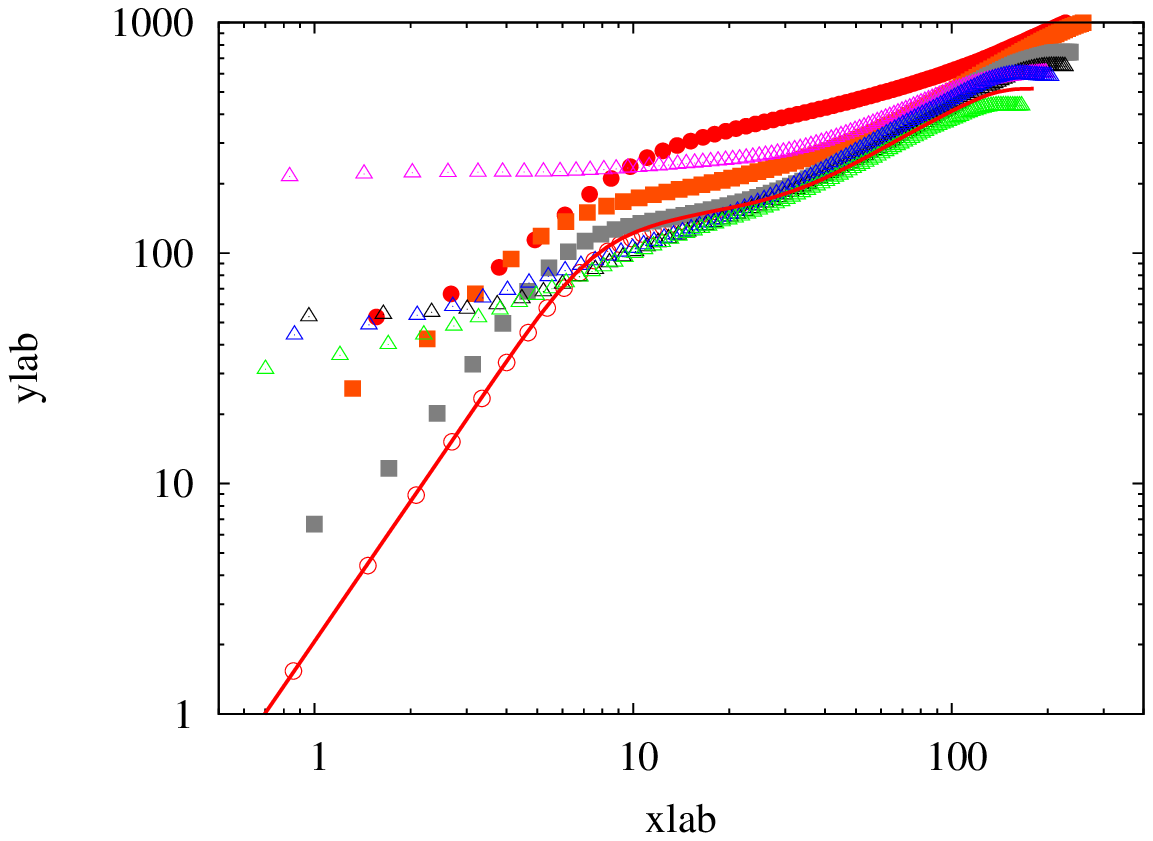}
\hskip -0.0cm
\psfrag{ylab}{\large $ Sq^{2 }/\epsilon$}
\psfrag{xlab}{\large $y^+ $ }
\includegraphics[width=7.5cm]{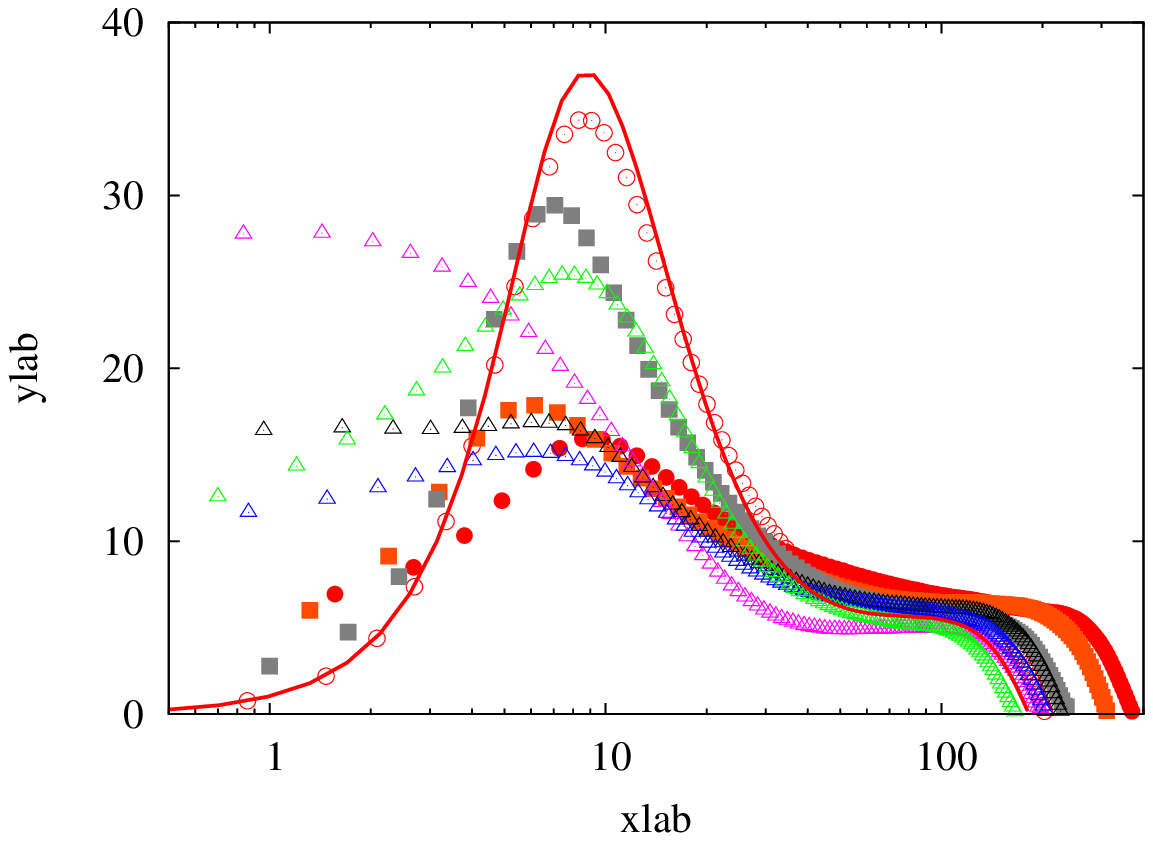}
\vskip -0.5cm \hskip 5cm c) \hskip 7cm d)
\caption{Profiles in wall units of:
a) turbulent kinetic energy, 
b) rate of isotropic dissipation,
c) eddy turnover time, 
d) shear parameter 
for the flows with rough surfaces listed in the inset in a),
compared with
those in presence of smooth walls  
(open circle present at $R_\tau=204$,
lines \cite{lee_15} at $R_\tau=180$).
}
\label{fig7}
\end{figure}

\subsubsection{ Shear parameter }

For flows past smooth walls 
despite the $Re$ variations in the near-wall region
of the turbulent kinetic energy and of the
rate of energy dissipation in figure \ref{fig1}c
there was a good scaling of the eddy turnover time. Figure \ref{fig7}a shows
unexpected behavior of $q^{2 +}$ depending on the type of roughness.
For instance it is rather difficult
to predict the large increase of $q^{2 +}$ for $LLS$
with respect to that for $CS$ being 
the differences between the $\langle u_2 u_2 \rangle^+$
in figure \ref{fig6}c rather small. The profiles
of each normal stress, not reported,
show that the growth of $q^{2 +}$ is due to the large
increase of $\langle u_1 u_1 \rangle^+$. The increase 
of $\langle u_3 u_3 \rangle^+$, instead, is moderate.
The message of figure \ref{fig7}a is that 
in the near-wall region the longitudinal grooves
generate values of $q^{2 +}$ greater than those for
transverse and three-dimensional corrugations, 
due to the large streamwise fluctuations inside
the longitudinal cavities. 
In figure \ref{fig7}b large variations in the
near-wall region of the profiles of the rate of dissipation 
do not have the same trend as those of $q^{2 +}$. 
The $LS$ and the $TS$ surfaces have a high rate of dissipation,
in the near-wall region, due to large amount of solid at the plane 
of the crests, generating high
$s_{12}^{2 +}$ contributing  more to  $\epsilon^+$
than the other fluctuating shears.
Only for the $LTS$ flow the small fluctuations
near the plane of the crests, and the small amount of solid 
give rise to a rate of dissipation smaller than that of the smooth
wall.
The profile of the eddy turnover time of the $TS$ surface,
in figure \ref{fig7}c,  is the only one close 
to that of smooth walls and the difference is mainly
due to the $\epsilon^+$ in figure \ref{fig7}b.
Interestingly figure \ref{fig7}c depicts a completely
different behavior for transverse and longitudinal
corrugations. In the latter $ q^{2 +}/\epsilon^{ +}$
remains constant while in the former it decays similarly
to the smooth walls. The shear parameter $S^*$
corroborates the similarity between the smooth and
the $TS$ surface, classified as "d" type 
roughness, with a weak drag increase with respect to $SM$. In
both surfaces as well as for $LTS$ with drag reduction
the maximum is located approximately at $y^+=10$.
Flow visualizations for $LTS$ depicts the
formation of  streaky structures similar to those of
the smooth channel. For the other longitudinal corrugations
the maximum of $S^*$, near the plane
of the crests, depend on the type of surface, indicating that the shape
of the surface dictates the structures formation. 
The values of $S^*$ suggest that for $LS$ 
the longitudinal structures are  coherent, these become more strong
and coherent for the $LLS$ surfaces. The low values $S^*$
indicate an isotropization of the structures, that was
investigated by \cite{OrlandiLeonardi2006} through the
profiles of the normal stresses.  In figure \ref{fig7}e 
the collapse of the $S^*$ profiles in the outer layer, despite the 
differences near the plane of the crests, is a first indication
of the validity of the Townsend similarity hypothesis (\cite{Townsend}).

\begin{figure}
\centering
\vskip 0.0cm
\hskip -1.8cm
\psfrag{ylab} {\hskip -1.0cm
\large $(10^2\dedq {\langle{u_2^2}\rangle}{ x^2_2})^+$}
\psfrag{xlab}{ $ y^+$}
\includegraphics[width=7.5cm]{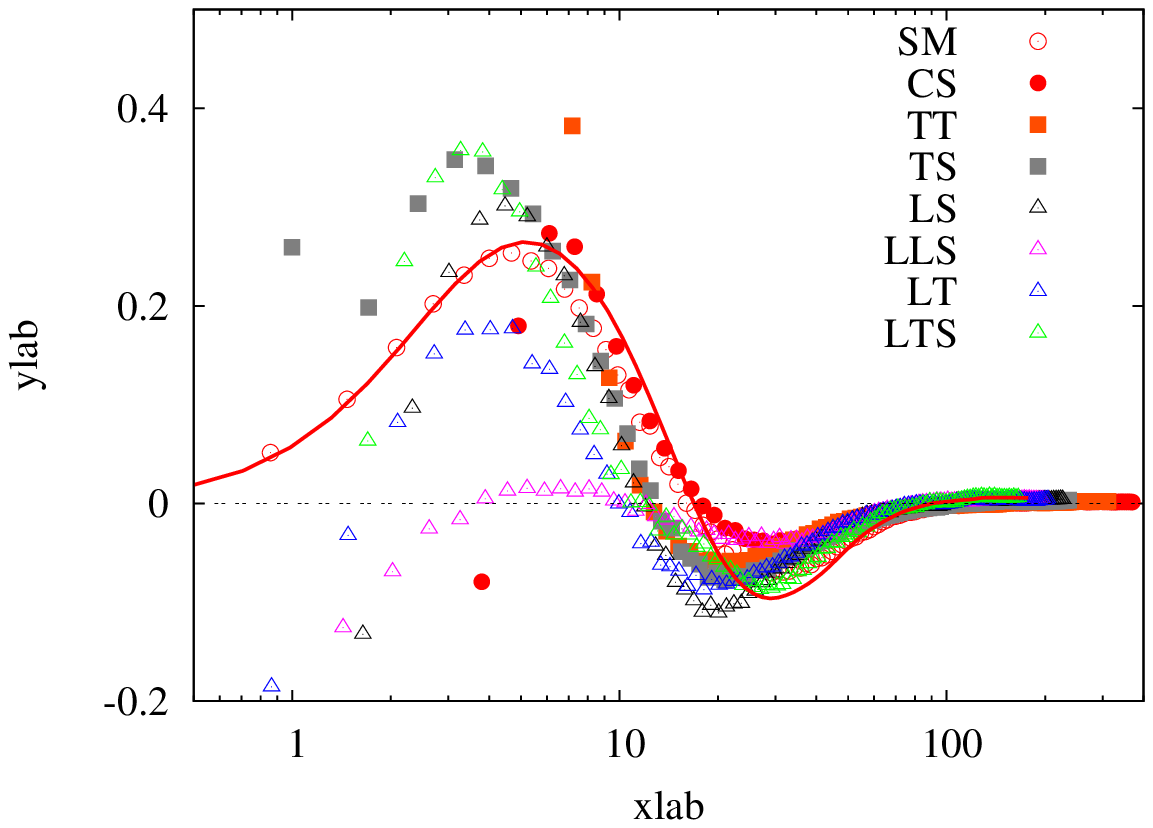}
\psfrag{ylab}{\large $\langle \omega_2^2 \rangle^+ $}
\psfrag{xlab}{\large $ y^+$ }
\includegraphics[width=7.5cm]{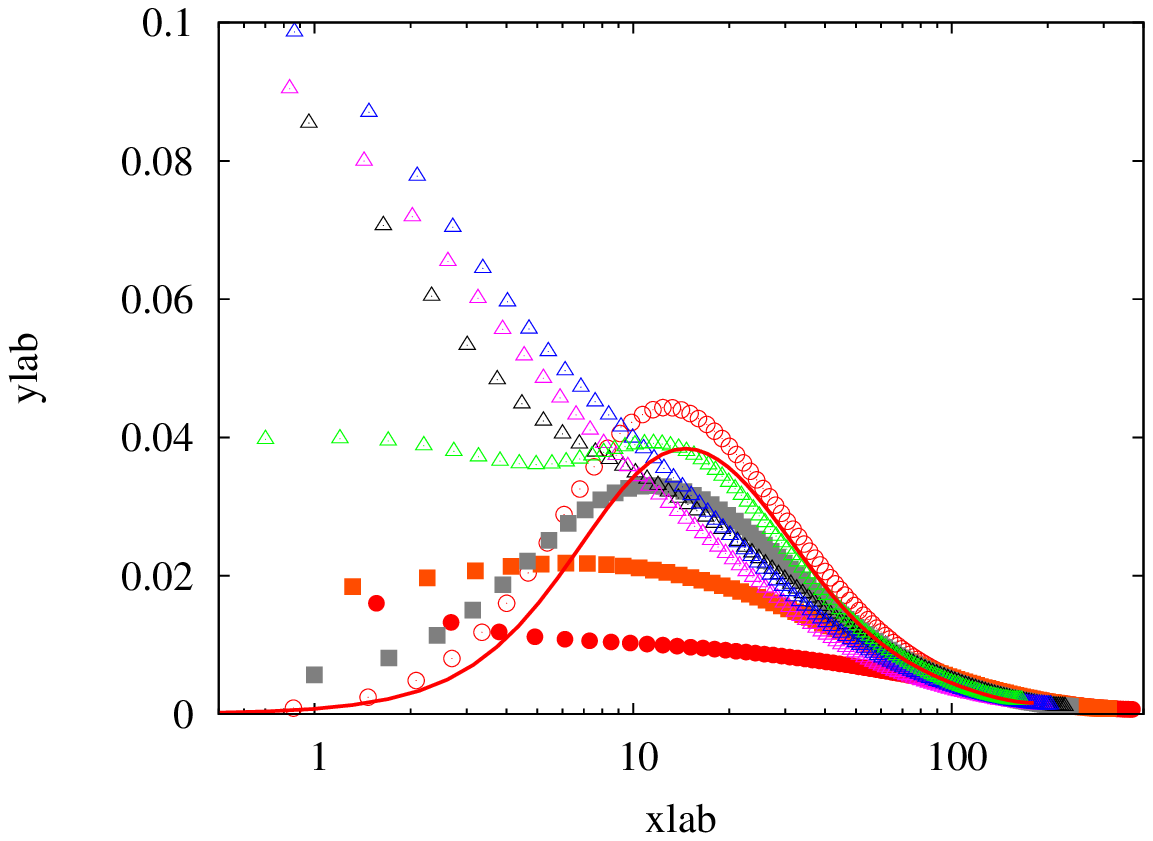}
\vskip -0.5cm \hskip 5cm a) \hskip 7cm b)
\caption{Profiles in wall units of:
a) $\dedq {\langle{u_2^2}\rangle}{ x^2_2}$,
b) $rms$ of normal to the wall vorticity component,
for the flows with rough surfaces listed in the inset of a),
compared with
those in presence of smooth walls  
(open circle present at $R_\tau=204$,
lines \cite{lee_15} at $R_\tau=180$).
}
\label{fig8}
\end{figure}

\subsubsection{ Structural statistics }

As for the smooth channel the analysis of the kind
of structures near the surfaces can be drawn by the profiles
of $(\dedq{ \langle{u_2^2}\rangle}{ x^2_2})^+$.
It is worth to recall that positive values indicate a layer
dominated by sheet-like and negative by 
rod-like structures. In figure \ref{fig8} as well as in figure \ref{fig7}b
differences can be noticed 
between the present $SM$ data and those at
$R_\tau=180$ in figure \ref{fig2}a. The reason should be ascribed in a large
measure to the different $R_\tau$ and in a reduced measure to 
the coarse grid, here used to have a smooth transition  of the 
resolution in the flow side with that required to reproduce the roughness
layer.  The resolution and the Reynolds number affect more the profiles of 
$\langle \omega_2^2 \rangle^+ $ in figure \ref{fig8}b and of $\epsilon^+$
in figure \ref{fig7}b.
Both figures \ref{fig8} show drastic
differences between smooth and rough walls in the inner
region that disappear in the outer region. For the longitudinal
corrugations, near the plane of the crests,
and, in particular, in contact with the solid 
tubular-like structures form, as it is qualitatively depicted
by the $u_2$ contours in figure \ref{fig5}. Even for the
transverse triangular bar ($TT$) as well as for the cubes ($CS$) there is
a tendency to the formation of tubular-like structures.  For
flow past the $TS$ surface, on the other hand, there is a prevalence
of the sheet-like structures even higher than that for
smooth walls ($SM$). This can be also
deduced by comparing figure \ref{fig5}c  and figure \ref{fig5}d.
Only the $LLS$ flow shows small variations of $-\aQ$,
and once more this occurrence is corroborated by the  smooth  contours in
figure \ref{fig5}e of $u_2$ lying in large structures. The intensification
of the contours of $u_2$ near the wedges in figure \ref{fig5}f and
figure \ref{fig5}g, relative to the longitudinal corrugations $LLS$
and $LT$, explain the negative values of $-\aQ$ near the plane of the crests
in figure \ref{fig8}a.
The locations where $-\aQ=0$ vary between
$y^+=10$ and $y^+=18$, in correspondence of this region the maximum
of turbulent kinetic energy production is located, as 
discussed later on.

Near smooth walls the elongated structures, the so called near-wall
streaks usually are characterised by regions of negative and positive
$u_1$. The same picture is obtained by contours of $\omega_2$, therefore
it is interesting to look at the effects of the shape  of the surfaces
on the profiles of $\langle \omega_2^2 \rangle^+$. 
In presence of smooth walls the streaky structures are very intense at
the distance where these are generated, accordingly the peak of 
$\langle \omega_2^2 \rangle^+$ is located at $y^+\approx 15$.
Figure \ref{fig8}b shows a 
different trend, near the plane of the crests, between the longitudinal
and the transverse corrugations. For the $TT$ and the $CS$ surfaces
the  small and constants values of $\langle \omega_2^2 \rangle^+$
suggest an isotropization of the small scales in the near
wall region. In the near-wall region the isotropization is 
further corroborated by the
profiles of the three vorticity $rms$, not shown. These profiles
do not depict the large differences among the three components
reported by \cite{kimetal87} for the channel with smooth walls. 
On the other hand, for the longitudinal grooves the anisotropy of the structures 
is appreciated by the growth of $\langle \omega_2^2 \rangle^+$ 
moving towards the plane of the crests.
The strong planar motion at the top of
the cavities, in particular for the $LS$ and $LT$
surfaces causes this growth. This motion is due to the the large $u_2$
fluctuations generated inside the longitudinal cavities
depicted in figure \ref{fig5}f and figure \ref{fig5}g.
For the drag reducing surface ($LTS$) these fluctuations reduce
in the near-wall region (figure \ref{fig5}h), therefore the strength of 
small and large scales reduce in accord to the decrease of
$q^{2+}$ and $\epsilon^+$ in figure \ref{fig7}. In figure \ref{fig8}b
$\langle \omega_2^2 \rangle^+$ for $LTS$ is smaller
than that of the other longitudinal corrugations.

\subsubsection{ Flow visualizations and statistics }

\begin{figure}
\centering
\vskip 0.0cm
\hskip -1.8cm
\psfrag{ylab} {\hskip -1.0cm
\large $ \langle{u_2^2}\rangle$}
\psfrag{xlab}{ $ y$}
\includegraphics[width=7.5cm]{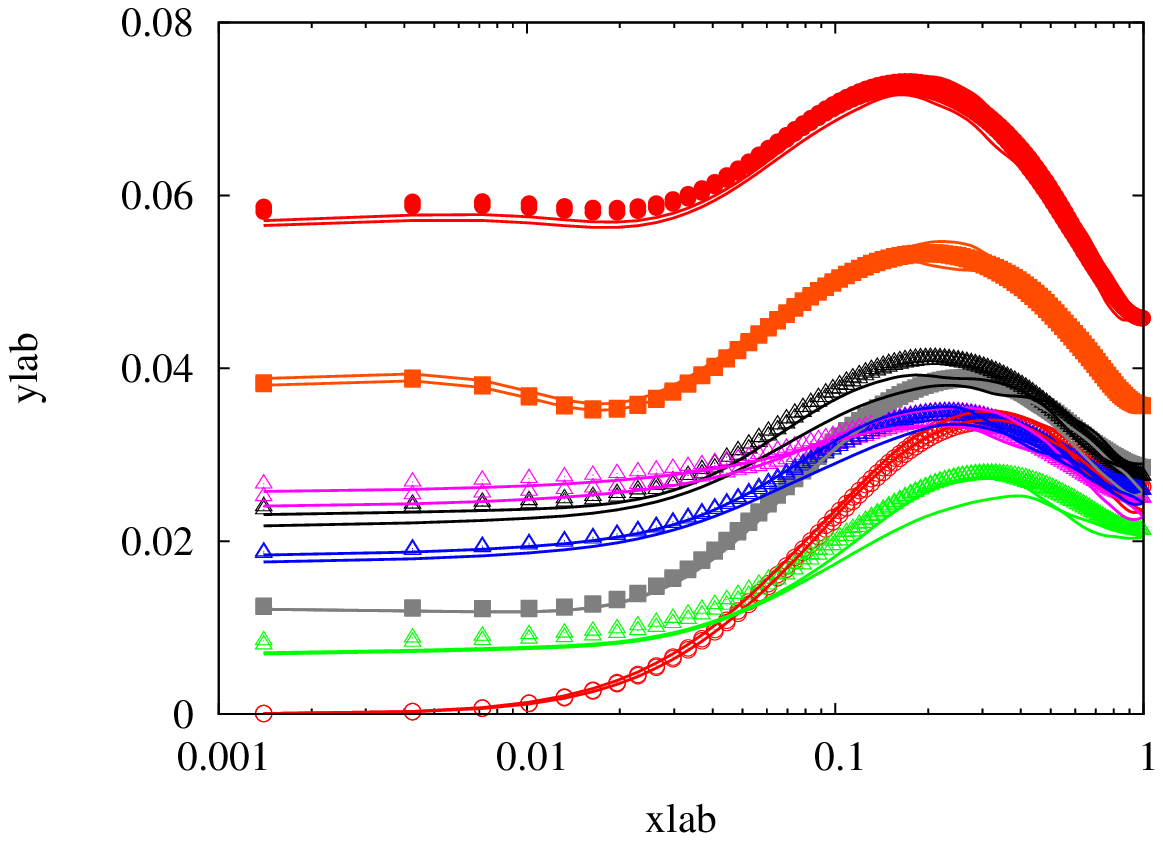}
\psfrag{ylab}{\large $\langle \omega_2^2 \rangle $}
\psfrag{xlab}{\large $ y$ }
\includegraphics[width=7.5cm]{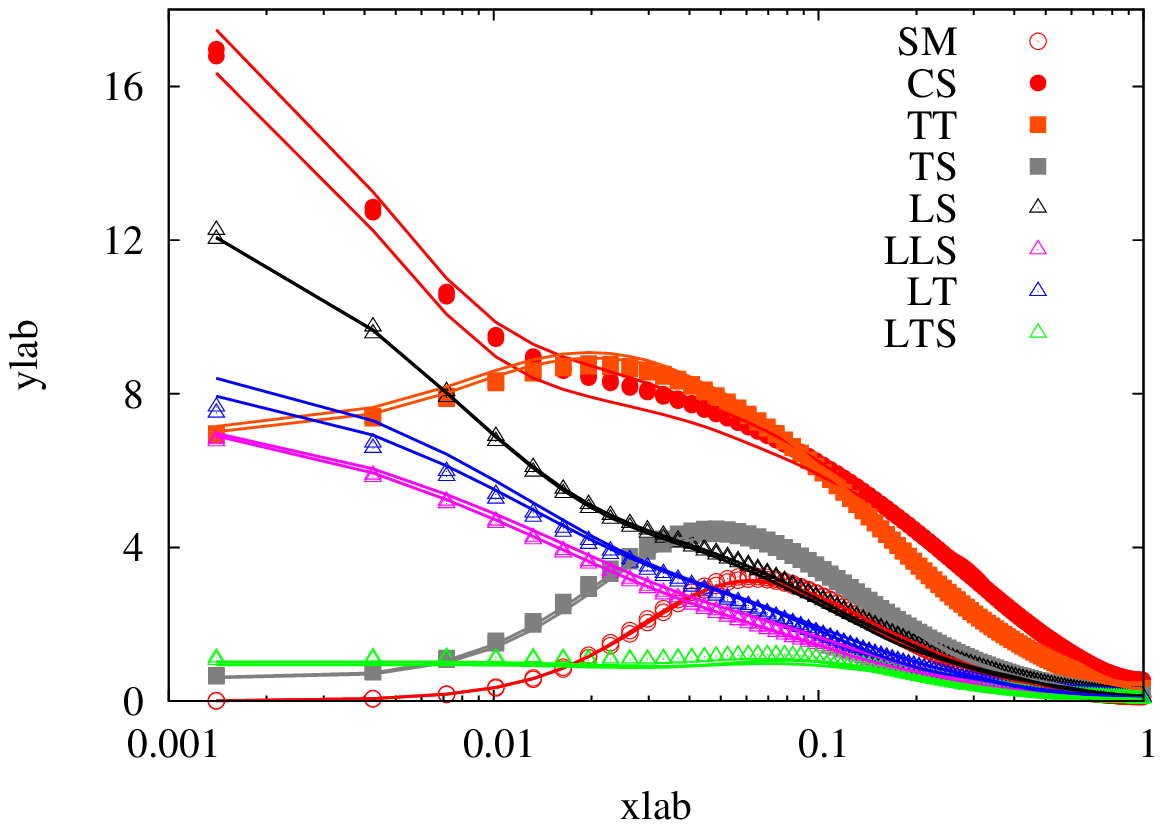}
\vskip -0.5cm \hskip 5cm a) \hskip 7cm b)
\vskip -0.1cm
\hskip -2.0cm
\includegraphics[width=4.0cm]{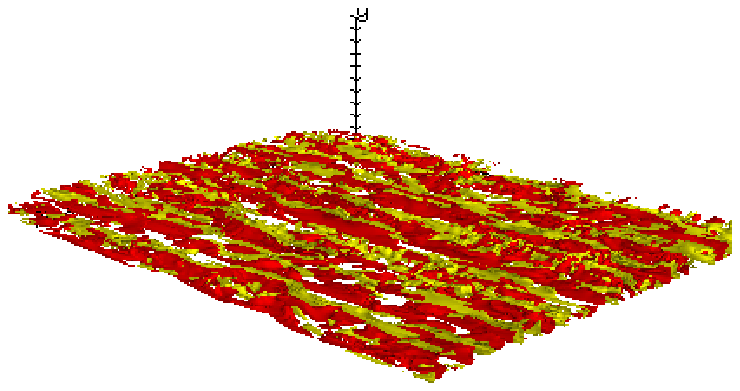}
\hskip -0.35cm
\includegraphics[width=4.0cm]{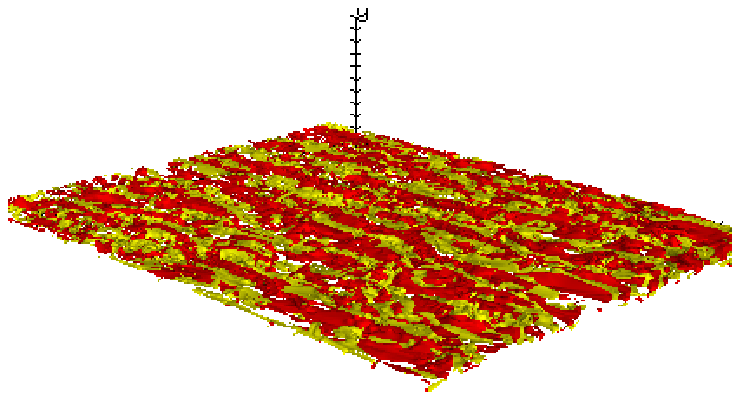}
\hskip -0.35cm
\includegraphics[width=4.0cm]{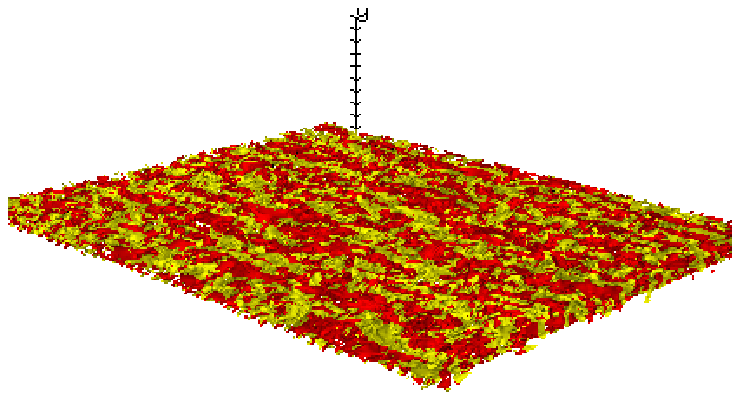}
\hskip -0.35cm
\includegraphics[width=4.0cm]{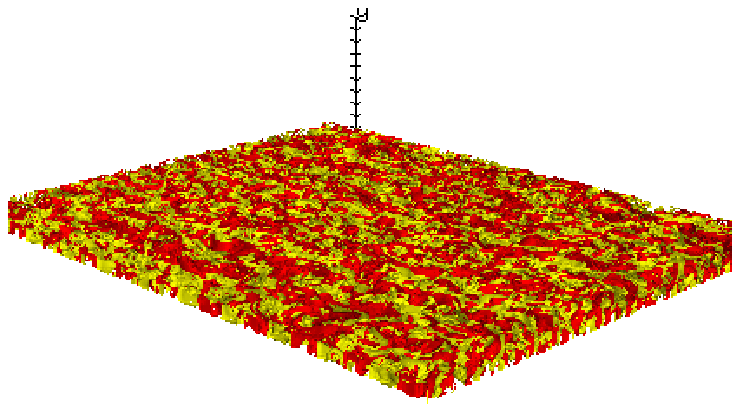}
\vskip -0.2cm \hskip 0cm c) \hskip 3cm d) \hskip 3cm e) \hskip 3cm f)
\vskip -0.1cm
\hskip -2.0cm
\includegraphics[width=4.0cm]{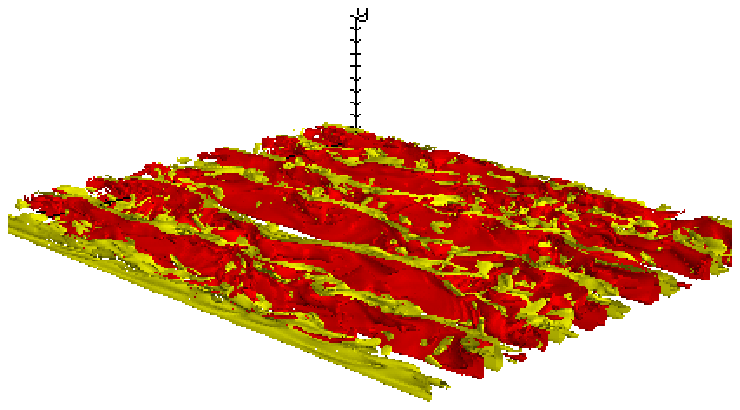}
\hskip -0.35cm
\includegraphics[width=4.0cm]{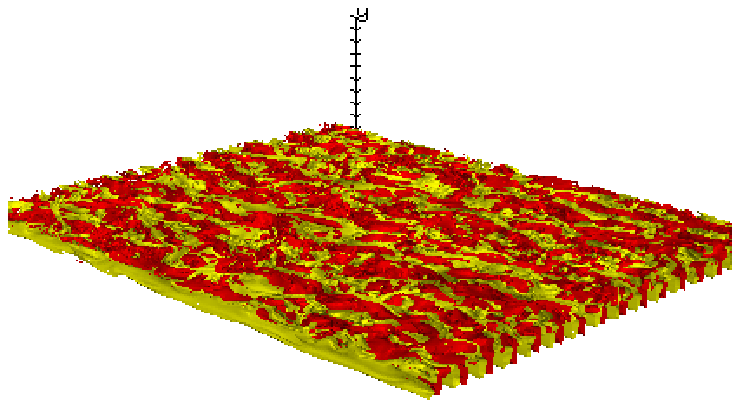}
\hskip -0.35cm
\includegraphics[width=4.0cm]{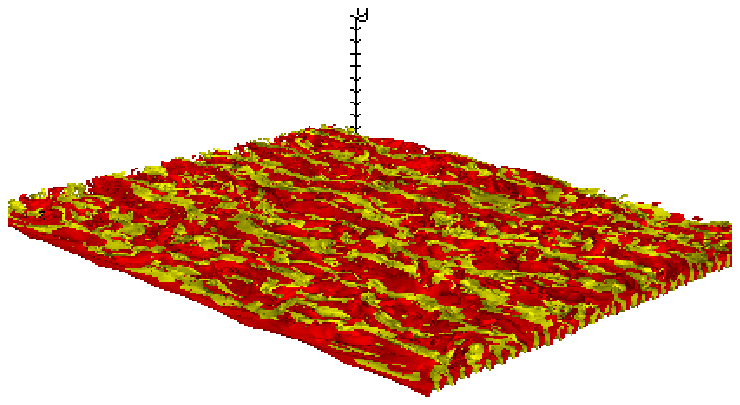}
\hskip -0.35cm
\includegraphics[width=4.0cm]{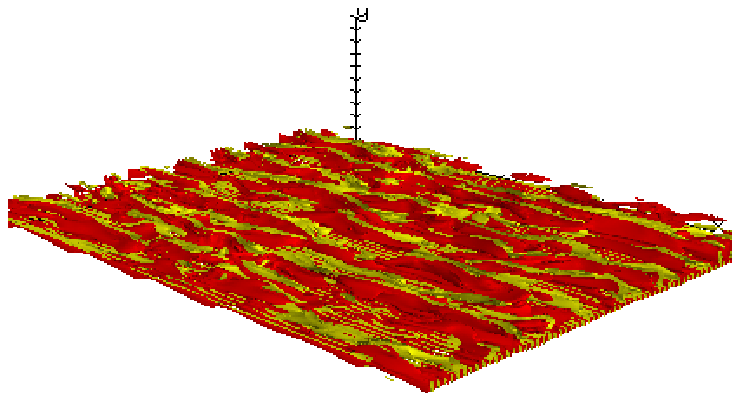}
\vskip -0.2cm \hskip 0cm g) \hskip 3cm h) \hskip 3cm i) \hskip 3cm l)
\caption{Profiles in computational  units of the $rms$ of the:
a) normal to the wall velocity,
b) normal to the wall vorticity components,
for the flows with rough surfaces listed in the inset of b),
symbols averages in time and in the homogeneous directions
$x_1$ and $x_3$, lines the same quantities averaged in $x_1$ and $x_3$
of the fields used to get the visualizations of $\omega_2$ 
(red  $\omega_2=+1$, yellow $\omega_2=-1$):
c) $SM$, d) $TS$, e) $TT$, f) $CS$ ,
g) $LLS$, h) $LS$, i) $LT$, l) $LTS$
}
\label{fig9}
\end{figure}

The surface contours of $\omega_2$, in the near-wall region, for the
different corrugations may be of help to explain
what has been previously discussed . The visualizations are
performed by taking only one realisation, from which the
$rms$ profiles of $\langle \omega_2^2 \rangle $  are calculated.
The  comparison between these profiles, indicated by lines, and
those calculated by taking several fields, indicated by symbols, demonstrates     
that the main features, previously described by converged statistics, 
are captured by one realisation.
This is shown in   figure \ref{fig9}a and in  figure \ref{fig9}b
through the profiles of $\langle u_2^2 \rangle$ and of
$\langle \omega_2^2 \rangle$ in computational units. These
profiles show, in figure \ref{fig9}a, that, $\langle u_2^2 \rangle $
is rather constant near the plane of the crests, and 
$\langle \omega_2^2 \rangle$ depends on the type of
corrugations. In some of the flows, and, in particular,
for those with a large resistance or those with large longitudinal
corrugations ($LLS$) $\langle \omega_2^2 \rangle$ decreases moving far from the 
plane of the crests. For transverse corrugations
$\langle \omega_2^2 \rangle$ increases as for 
smooth walls. It remains constant in a thick layer
for the drag reducing flow ($LTS$). The weak $u_2$ fluctuations
created by the $TS$ corrugation do not produce large 
differences in the near-wall streaks, as it can be appreciated by
comparing figure \ref{fig9}c  for $SM$ and figure \ref{fig9}d 
for $TS$. On the other
hand, the strong $u_2$ fluctuations emerging from the $CS$ and
the $TT$ surfaces break the streamwise coherence of the near
wall structures emphasised in figure \ref{fig9}e ($TT$) 
and in figure \ref{fig9}f ($CS$).
\noindent The $\omega_2$ contours are clustered in short regions.
The tendency towards the isotropization 
of  the small scales is also corroborated by
visualizations, not shown of $\omega_1$ and $\omega_3$.
The impact of the  geometry surfaces on the $\omega_2$ vorticity 
is depicted in figure \ref{fig9}g by the positive and
negative surface contours of $\omega_2$ attached to the corners 
of $LLS$, spanning the entire length in the streamwise direction. 
These vorticity layers are generated by  the strong
$\frac{\partial u_1}{\partial x_3}$ forming 
near the vertical walls inside the cavities.  A similar view 
is obtained in the $LS$ (figure \ref{fig9}h) and $LT$
(figure \ref{fig9}i) flows by the layers of $\omega_2$ 
generated near the cavities walls.
For $LT$  the $\omega_2$ layers are less intense than those for $LS$, 
accordingly to the profiles in figure \ref{fig9}b. In the drag reducing 
$LTS$ flow the weak motion near the plane of the crest
creates a more uniform flow and therefore the vorticity
structures in figure \ref{fig9}l are weak. 

\subsubsection{ Turbulent kinetic energy production}

\begin{figure}
\centering
\vskip 0.0cm
\hskip -1.8cm
\psfrag{ylab} { \large $ R_{\alpha \alpha}^+$}
\psfrag{xlab}{ $ $}
\includegraphics[width=7.5cm]{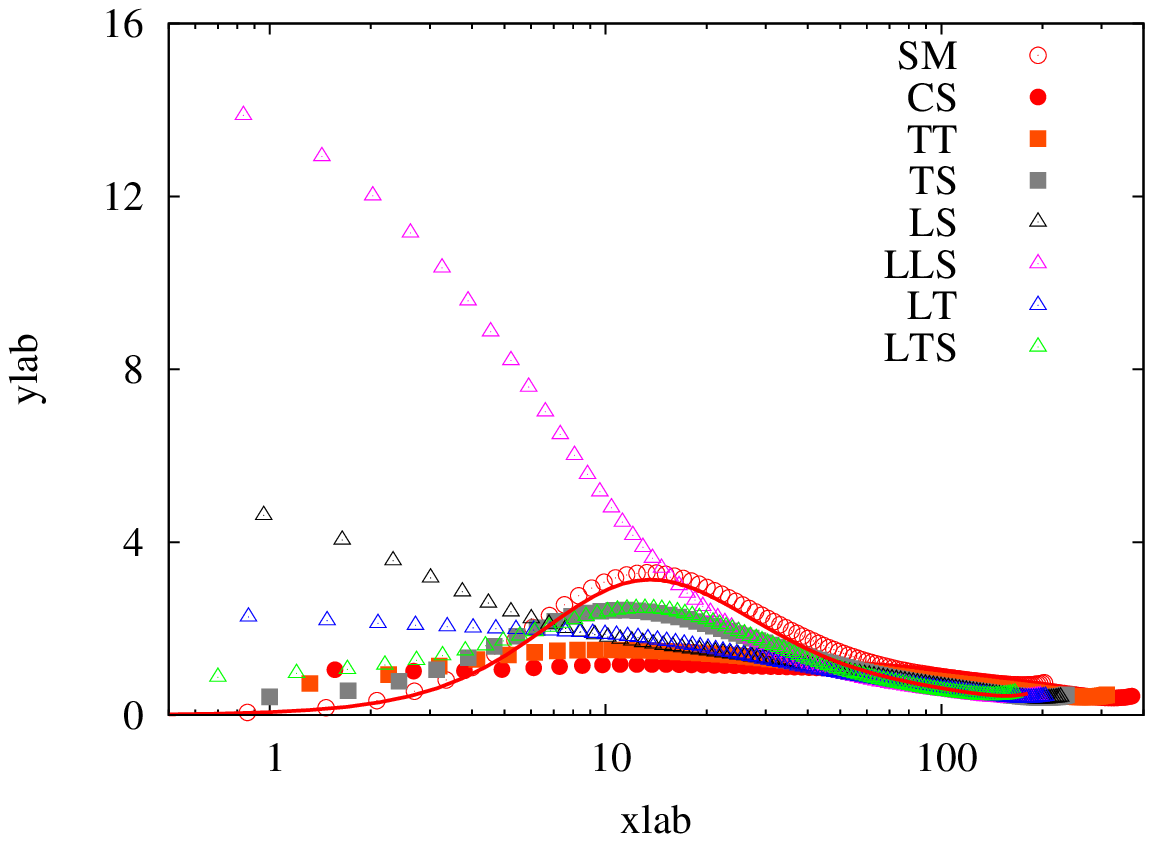}
\psfrag{ylab} {\large $ R_{\gamma \gamma}^+$}
\psfrag{xlab}{ $ $}
\includegraphics[width=7.5cm]{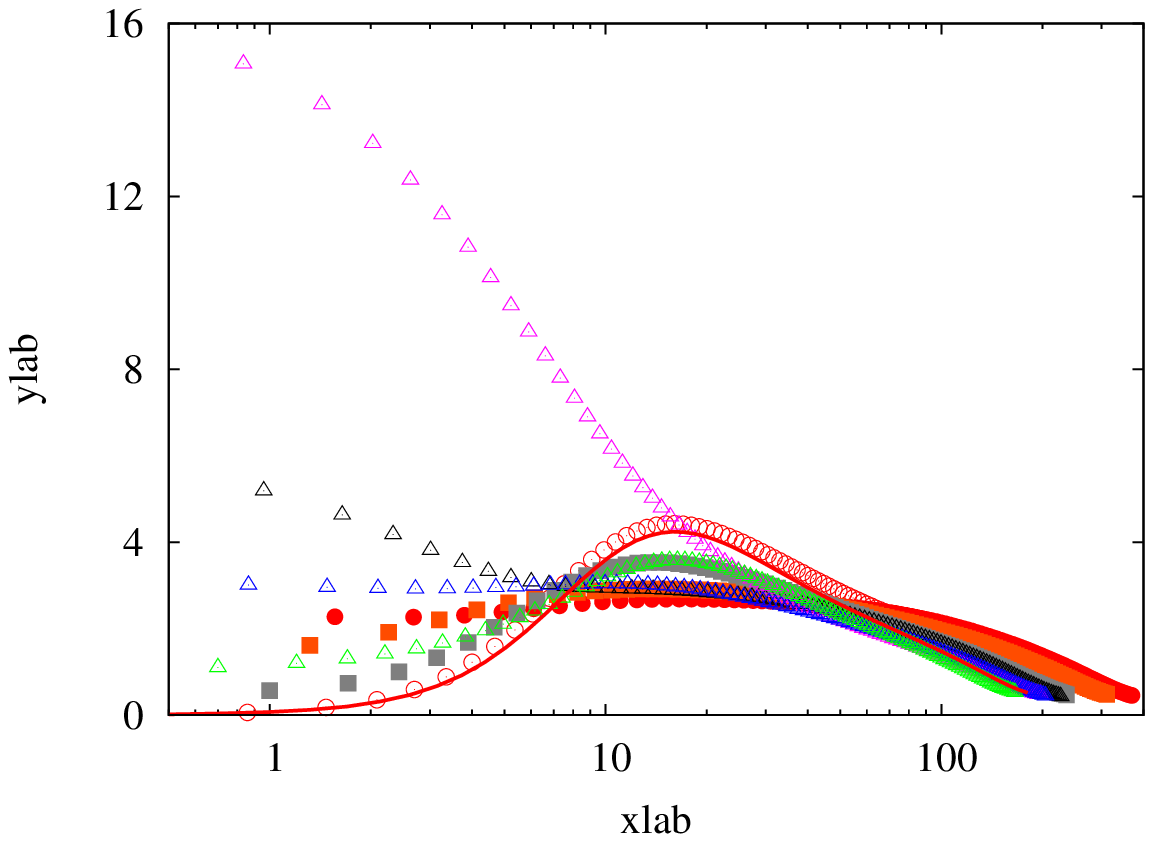}
\vskip -0.5cm \hskip 5cm a) \hskip 7cm b)
\vskip -0.2cm
\hskip -1.8cm
\psfrag{ylab}{\large $ P_\alpha^+$}
\psfrag{xlab}{\large $ $ }
\includegraphics[width=7.5cm]{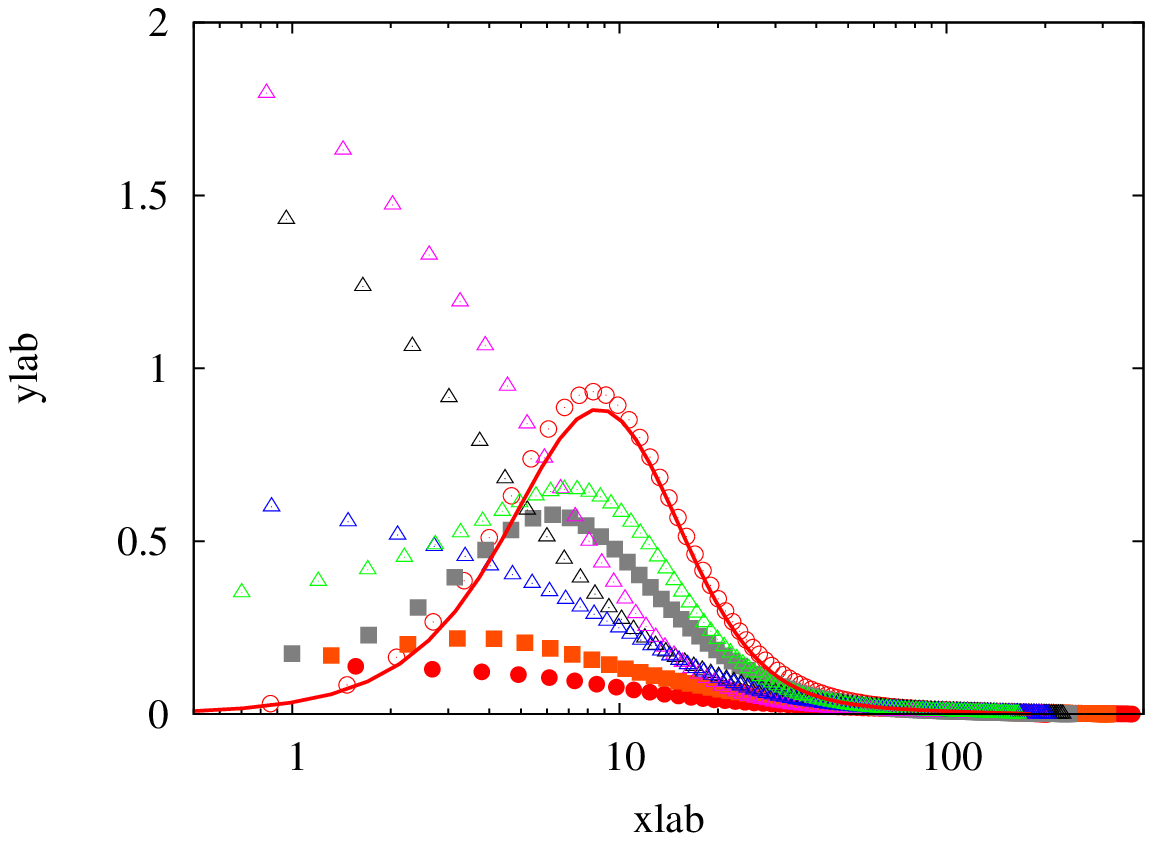}
\psfrag{ylab}{\large $ P_\gamma^+$}
\psfrag{xlab}{\large $ $ }
\includegraphics[width=7.5cm]{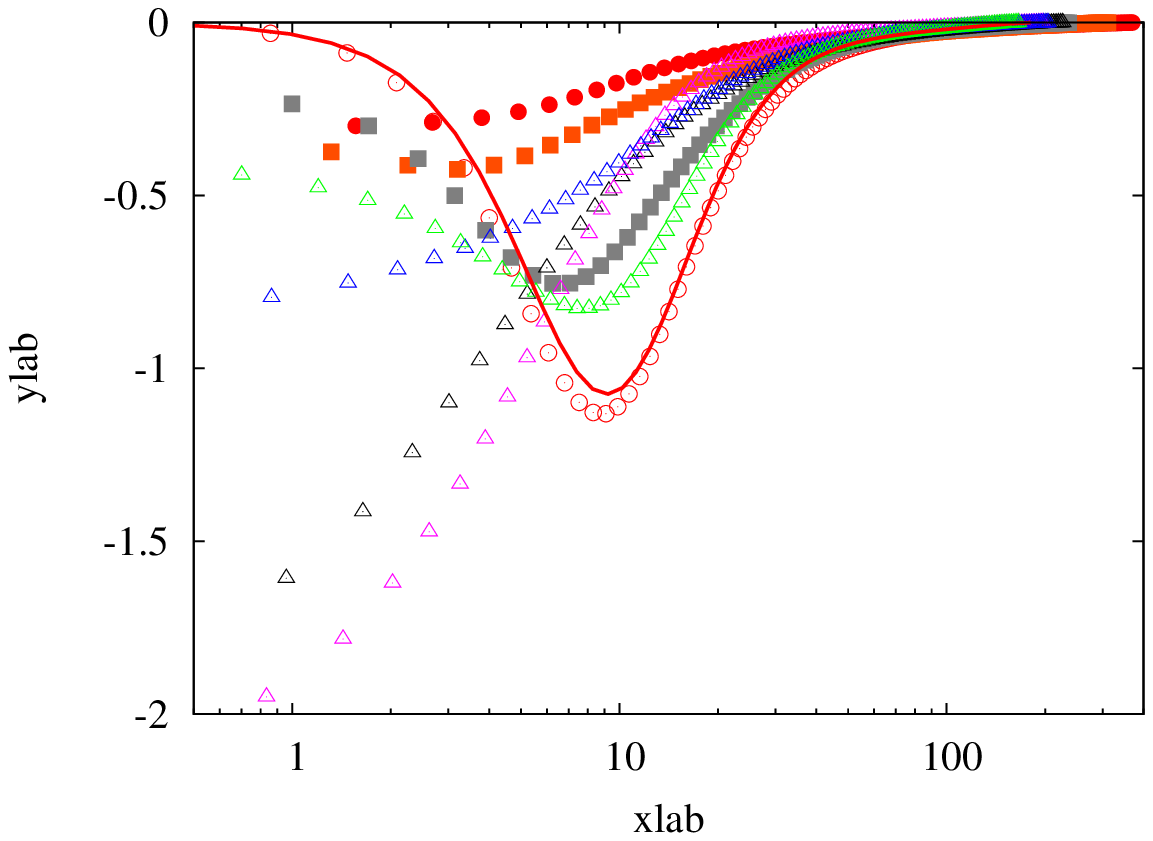}
\vskip -0.5cm \hskip 5cm c) \hskip 7cm d)
\vskip -0.2cm
\hskip -1.8cm
\psfrag{ylab}{\large $ P_k^+$}
\psfrag{xlab}{\large $ y^+ $ }
\includegraphics[width=7.5cm]{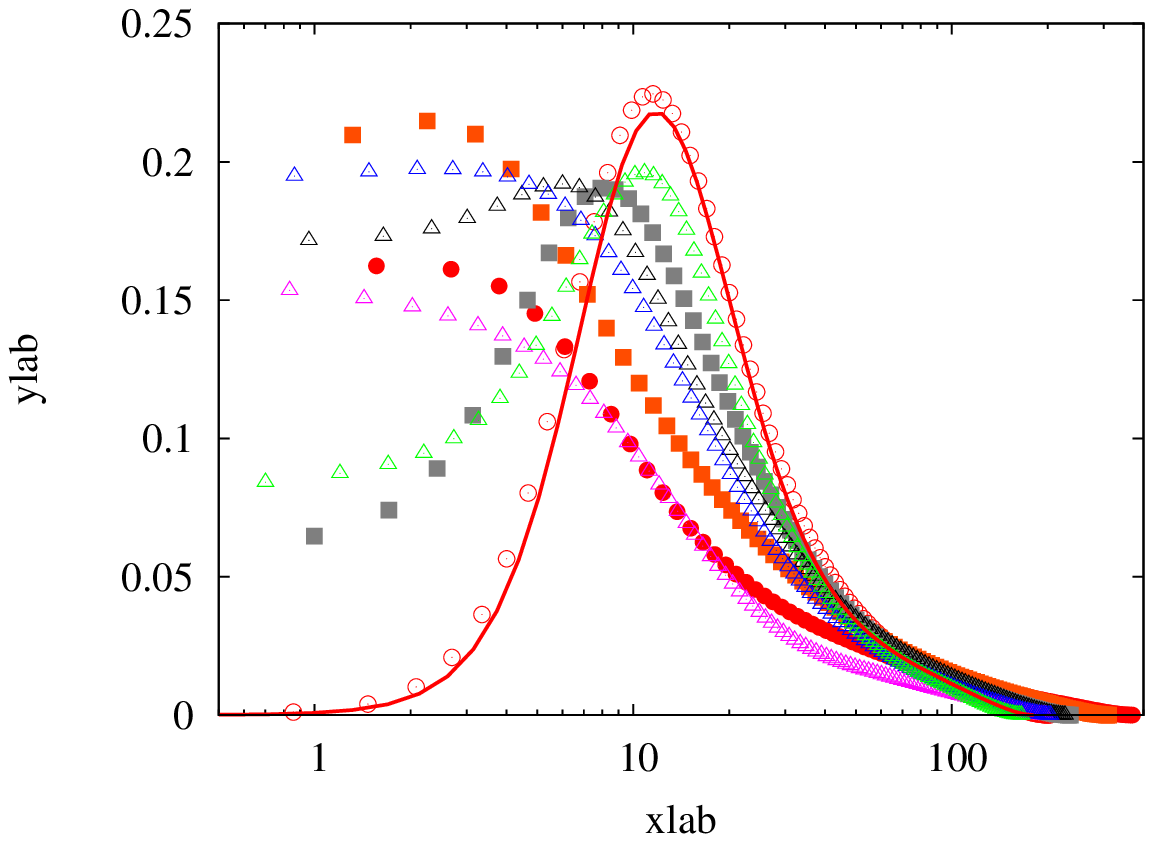}
%\psfrag{ylab}{\large $ \epsilon^+$}
\psfrag{ylab}{\large $ D_k^+$}
\psfrag{xlab}{\large $ y^+$ }
\includegraphics[width=7.5cm]{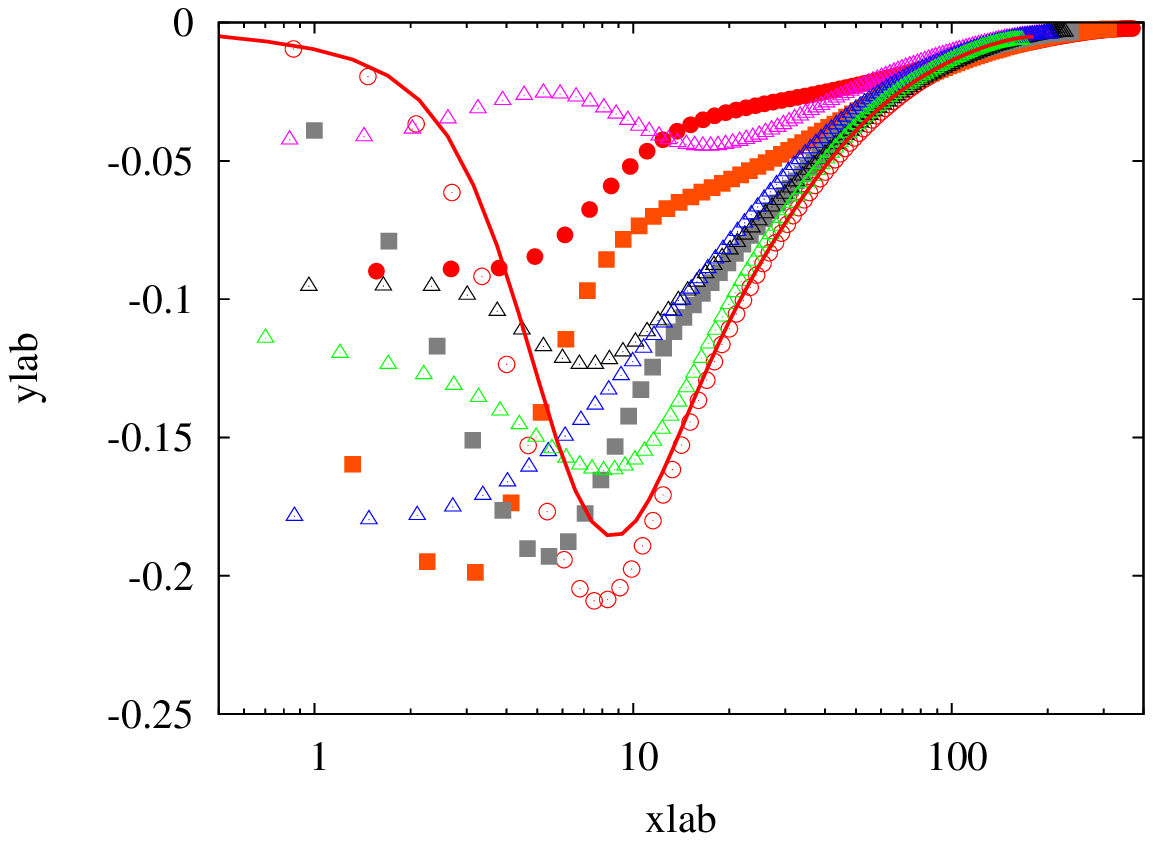}
\vskip -0.5cm \hskip 5cm e) \hskip 7cm f)
\caption{Profiles in wall units of the stress aligned with:
a) $S_\alpha$,
b) $S_\gamma$,
of the turbulent kinetic energy production aligned with
c) $S_\alpha$,
d) $S_\gamma$,
e) turbulent kinetic energy production,
f) full rate of dissipation $D_k^+$,
for the flows with rough surfaces listed in the inset of a),
compared with
those in presence of smooth walls  
(open circle present at $R_\tau=204$,
lines \cite{lee_15} at $R_\tau=180$).
}
\label{fig10}
\end{figure}

The large dependence of the statistics upon the shape 
of the corrugations should be also observed
in the components of the normal stresses aligned with the 
eigenvectors of the strain tensor $S_{ij}$. As 
for smooth walls, in this reference system the stress $R_{\gamma \gamma}$
aligned with the compressive $S_\gamma$
and the $R_{\alpha \alpha}$ aligned with the extensional 
$S_\alpha$ strain become of the same order. 
The stress in the spanwise direction ($R_{33}$), does
not change, and coincide with $R_{\beta \beta}$
aligned with $S_\beta=0$. For any surface figure \ref{fig10}a
and figure \ref{fig10}b show that the stress
aligned with $S_\gamma$ are greater than those
aligned with $S_\alpha$. For the longitudinal corrugations
$R_{\alpha \alpha}$ and $R_{\gamma \gamma}$ become
very large. This is mainly due to the growth of
$R_{11}$ in particular for the $LLS$ surface.
In this reference system no one component
has a constant trend near the plane of the crests
as that in figure \ref{fig9}a.
The  $R_{\alpha \alpha}$ and the $R_{\gamma \gamma}$ grow 
or decrease with  slopes that  depend on the shape of the surface. 
The slope is zero for the $LT$ surface. This
stress decomposition allows to split the 
turbulent kinetic production $P_k=-(P_\gamma+P\alpha)$;
in magnitude $P_\gamma$ in 
figure \ref{fig10}d, is greater than $P\alpha$, in
figure \ref{fig10}c. Near the walls the two component
have a similar trend with the highest  values for
the $LLS$ due to the strong fluctuations 
generated within the cavities. 
The increase of $R_{\alpha \alpha}$
and  $R_{\gamma \gamma}$ despite the reduction  of $S_\alpha$
and $S_\gamma$ leads to the increase of $P^+_\alpha$
and of $P_\gamma^+$.  For the drag reducing surface 
the components of $P_k$ decrease.  These are rather small
for the three dimensional corrugations.  The same trend should be
expected for $P_k$, on the other hand, figure \ref{fig10}e 
shows a different trend with maximum production
for $TT$ and a sensible reduction for $LLS$ and $LS$ corrugations.
In some of the flows the maximum production is located near the plane
of the crests, with the exception of the $LTS$ and $TS$
surfaces, which  do not differ
much with the $P_k$ profile in presence of smooth walls. 
The rate of isotropic  dissipation $\epsilon^+$ in the near
wall region, in figure \ref{fig7}b, is greater than the production 
$P_k^+$.           
The  trend of the maximum of $\epsilon^+$ are not similar to  those
of  the production $P_k^+$. Near smooth walls,
in figure \ref{fig2},  the same trend for the profiles
of $P_k^+$ and $D_k^+$ was observed and the difference
between the two was balanced by the turbulent
diffusion $T_k^+$ due to the non-linear terms. In figure \ref{fig10}f
$D_k^+$ has been plotted showing a complex
behavior. For the $CS$, $LT$, $LTS$ and $LS$ corrugations
a trend similar to that of $P_k^+$ is found while for the 
other flows large differences occurs. Therefore very large
differences should be expected in the profiles of $T_k^+$.

\begin{figure}
\centering
\hskip -2.0cm
\psfrag{ylab}{ \hskip -0.5cm $BUDG^+ $}
\psfrag{xlab}{\large $  $ }
\includegraphics[width=7.5cm]{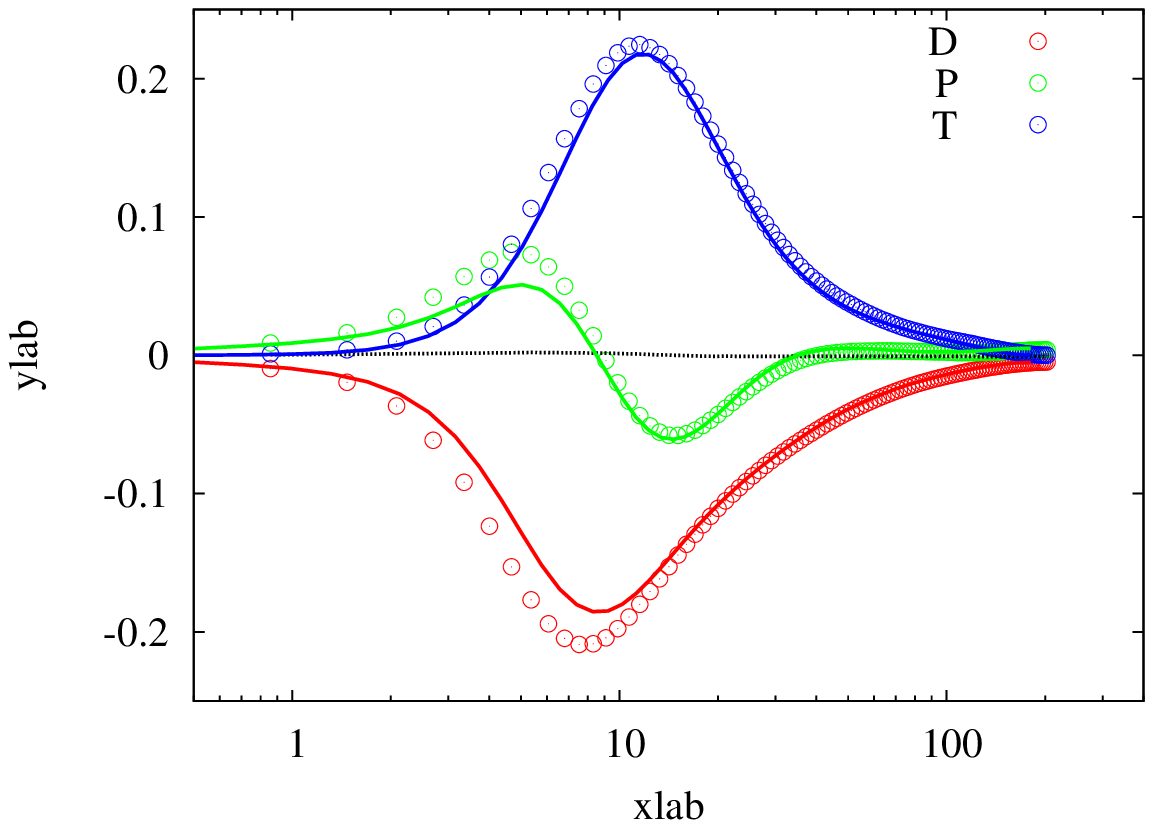}
\hskip -0.35cm
\psfrag{ylab}{\large $ $}
\psfrag{xlab}{\large $  $ }
\includegraphics[width=7.5cm]{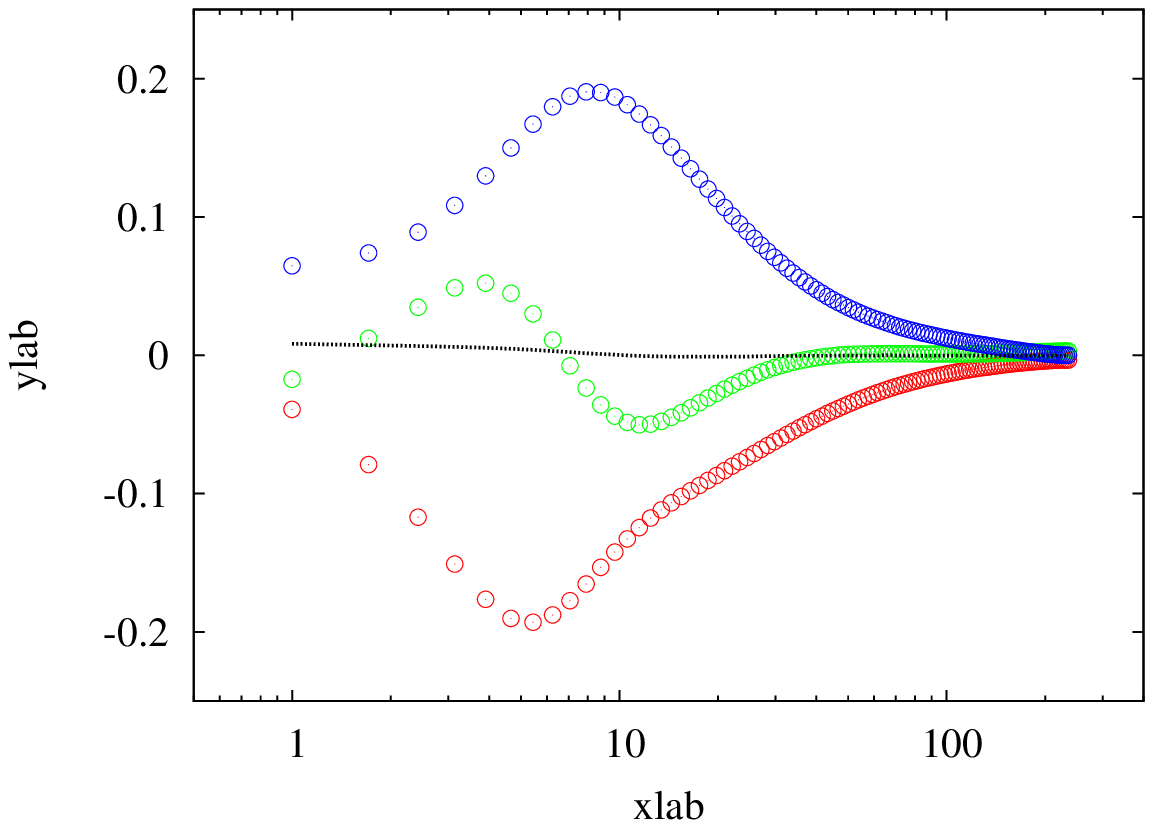}
\vskip -0.5cm \hskip 5cm a) \hskip 7cm b) 
\vskip -0.1cm
\hskip -2.0cm
\psfrag{ylab}{\large $ $}
\psfrag{xlab}{\large $  $ }
\includegraphics[width=7.5cm]{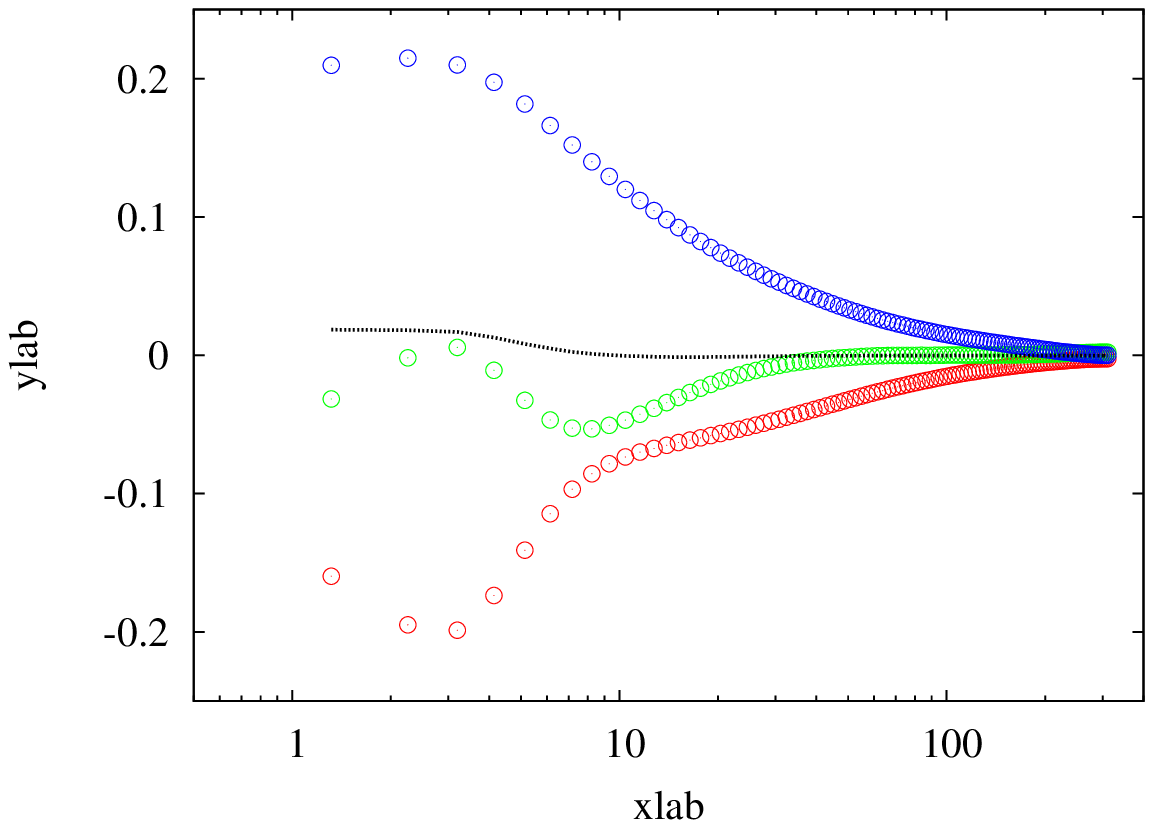}
\hskip -0.35cm
\psfrag{ylab}{\large $ $}
\psfrag{xlab}{\large $  $ }
\includegraphics[width=7.5cm]{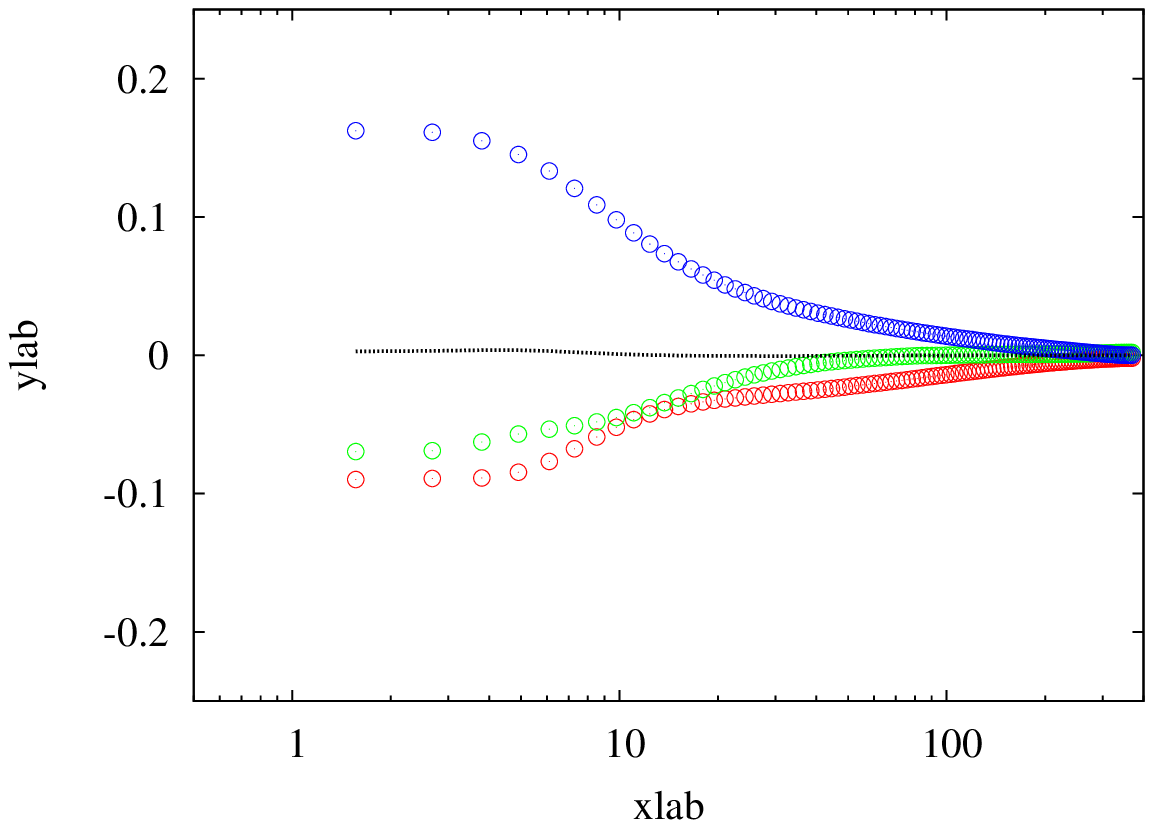}
\vskip -0.5cm  \hskip 5cm c) \hskip 7cm d)
\vskip -0.1cm
\hskip -2.0cm
\psfrag{ylab}{ \hskip -0.5cm $BUDG^+ $}
\psfrag{xlab}{\large $ y^+ $ }
\includegraphics[width=7.5cm]{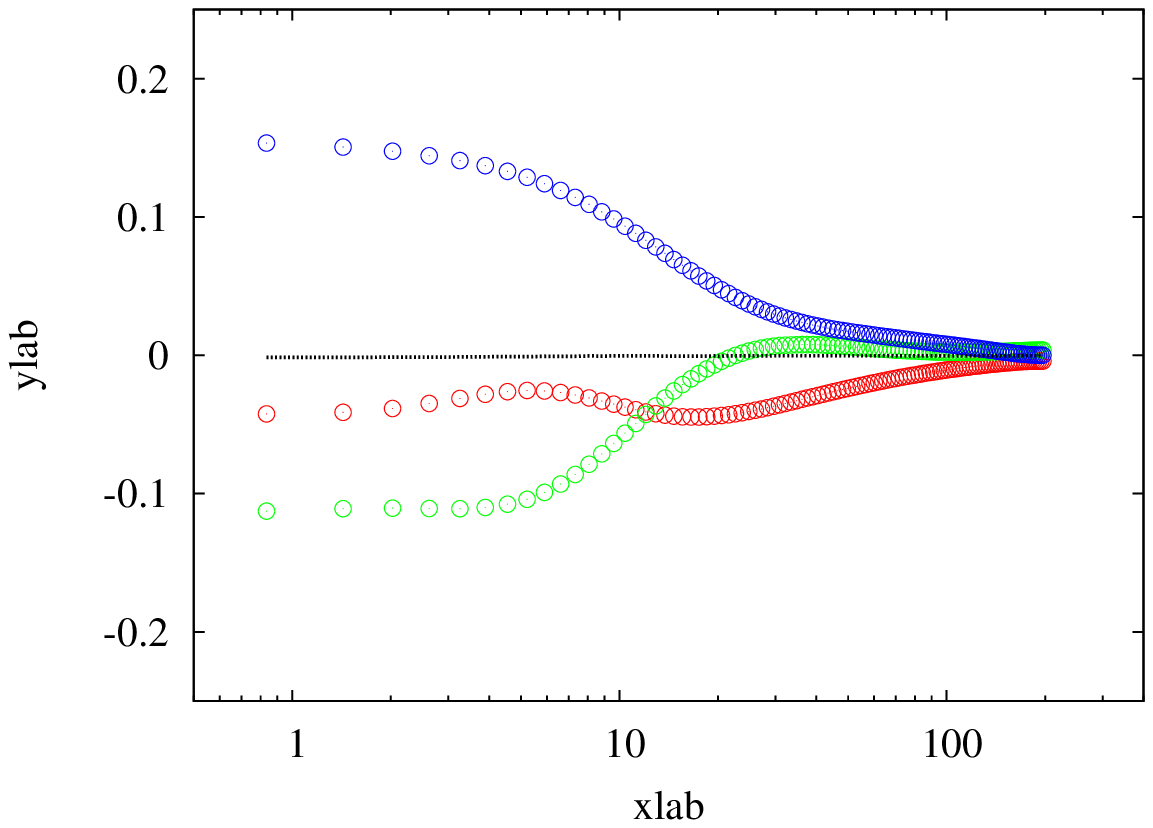}
\hskip -0.35cm
\psfrag{ylab}{\large $ $}
\psfrag{xlab}{\large $ y^+ $ }
\includegraphics[width=7.5cm]{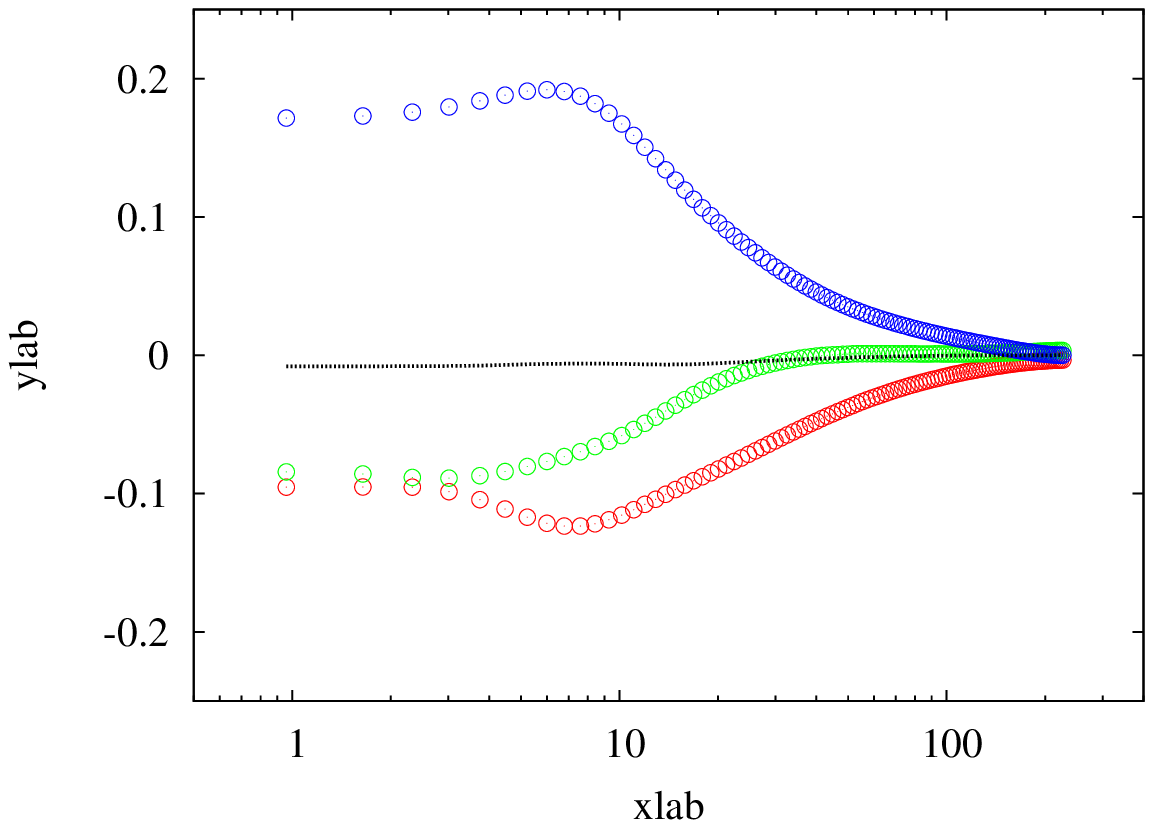}
\vskip -0.5cm \hskip 5cm e) \hskip 7cm f)
\vskip -0.2cm
\hskip -2.0cm
\psfrag{ylab}{ \hskip -0.5cm $BUDG^+ $}
\psfrag{xlab}{\large $ y^+ $ }
\includegraphics[width=7.5cm]{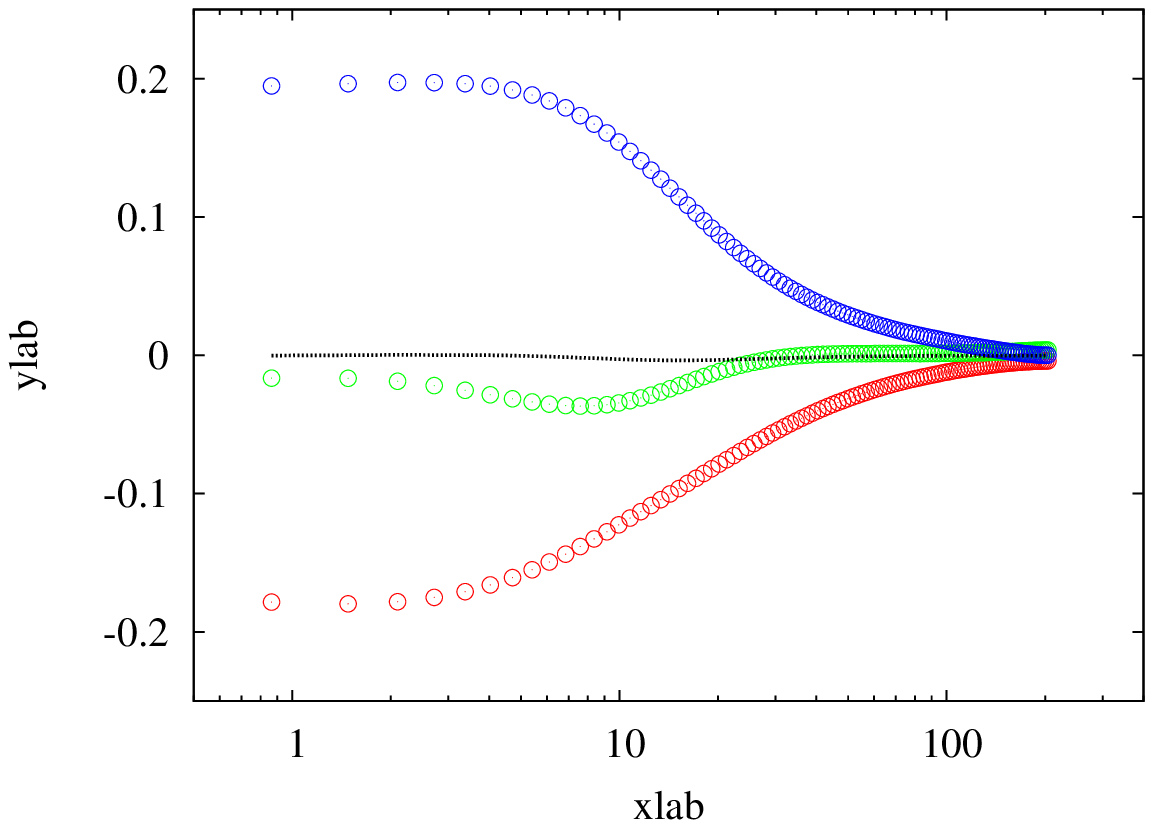}
\hskip -0.35cm
\psfrag{ylab}{\large $ $}
\psfrag{xlab}{\large $ y^+ $ }
\includegraphics[width=7.5cm]{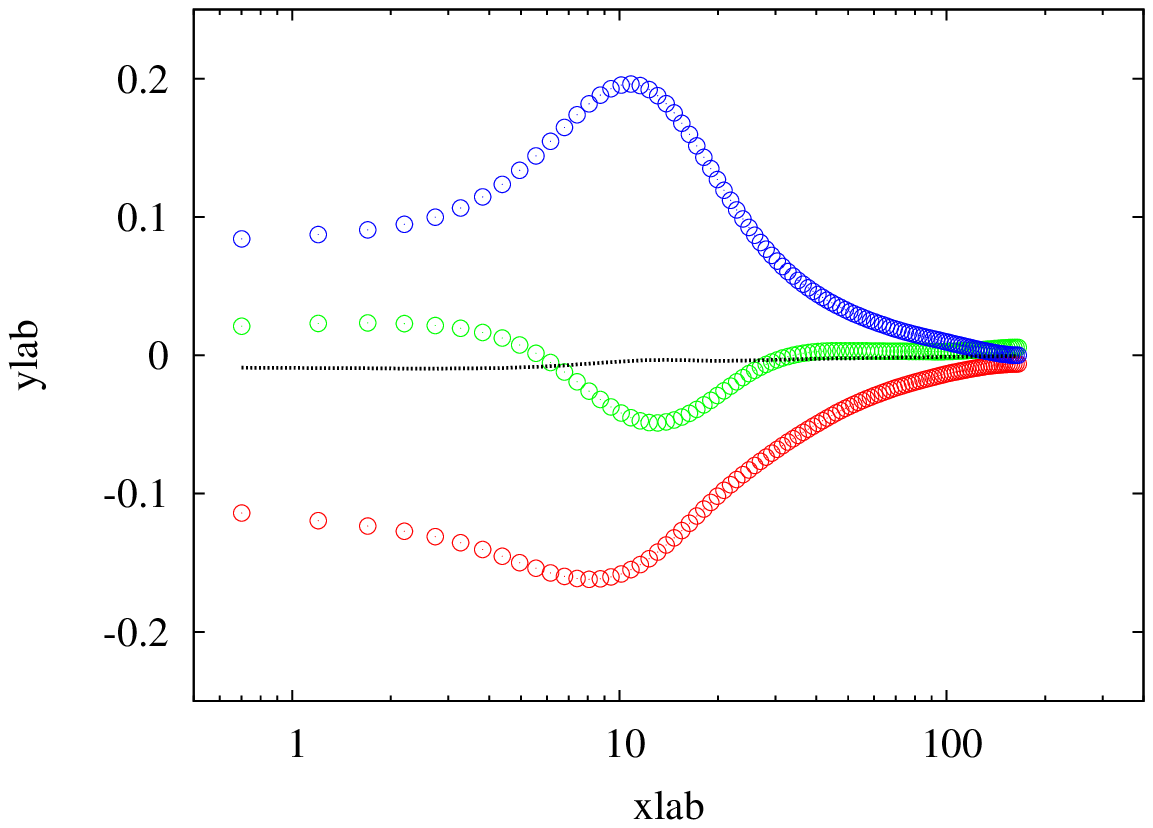}
\vskip -0.5cm  \hskip 5cm g) \hskip 7cm h)
\caption{Profiles in wall  units of the simplified budgets:
$D_k=\nu \langle u_i\nabla^2 u_i \rangle$ , $P_k=-2\langle u_2u_1 \rangle S$,
$T_k=-(\ded{ \langle u_2u_i^2 \rangle}{ x_2}
+\langle u_i\frac{\partial p}{\partial x_i} \rangle)$
a) $SM $ , b) $TS $ , c) $TT $, d) $CS$ ,
e)  $LLS $, f) $LS $ , g) $LT $, h) $LTS$;
in a) line \cite{lee_15} at $R_\tau=180$
symbols present $R_\tau=204$.
}
\label{fig11}
\end{figure}

\subsubsection{ Budgets of turbulent kinetic energy}

To emphasise the differences in the turbulent kinetic energy
budgets in presence of rough surfaces with respect to that
of smooth walls  the simplified budget is considered.
In these circumstances the production is balanced by 
$D_k$ and by the turbulent diffusion due to the non-linear terms
and to the correlations between velocity and pressure gradients. 
For smooth walls all the terms 
are equal to zero at the wall and grow with a different trend; $|D_k|$ and $T_k$ 
proportionally to $y^2$ and $P_k$ to $y^3$. The $T_k$ is positive
in the region with $-\aQ>0$ meaning that the sheet-like structures
loose energy towards the region with $-\aQ<0$ where the tubular-like
structures prevail. This is depicted in figure \ref{fig11}a
with the present data compared with those of \cite{lee_15} at
$R_\tau=180$. The agreement is rather good,  the small
differences in $D_k^+$ and $T_k^+$ are due to the different Reynolds
numbers and to the coarse resolution near the surface in this
simulation.  The transverse square bars $TS$ show a similar trend in
figure \ref{fig11}b with the difference to get the
three terms different from zero at the plane of the crests.
This occurrence is due to the small velocity fluctuations generated inside the 
square cavities. In presence of triangular bars ($TT$) 
the strong fluctuations emerging from the cavities 
produce a high $\langle u_1u_2 \rangle_W$; the maximum production,
in figure \ref{fig11}c, moves at the plane of the crests
with the  consequence to have there a high $D_k^+$. In this flow 
the turbulent transfer is low and negative near the plane of
the crests. This negative contribution is balanced by the positive
contribution at the center of the channel. It
is worth to recall that for smooth walls the total contribution
of the turbulent transfer is null, for rough surfaces it is
smaller than the total production and full rate of dissipation,
but it could be different from zero. 
For the $CS$ surface figure \ref{fig11}d shows a
reduction of $P_k^+$ and $D_k^+$ near the plane of the crests.
The large disturbances emerging from the interior of the surfaces make 
$T_K^+$  a sink of energy comparable to the total rate of dissipation.
This is corroborated by the large values $-\aQ<0$ in figure \ref{fig8}a
implying the prevalence of tubular-like structures in this layer.
For the $LLS$ corrugation the wide cavity
generates large $u_2$ fluctuations, therefore the turbulent
transfer is a sink of turbulent kinetic energy and greater
than $D_k^+$ (figure \ref{fig11}e), implying the formation of 
tubular-like structures near the plane of the crests, as it was
depicted in figure \ref{fig8}a. Figure \ref{fig9}a shows that
the $\langle u^2_2 \rangle$ for $LS$ does not change too much
with respect to that for $LLS$ and therefore the $T_k^+$
profile in figure \ref{fig11}f is similar to that in 
figure \ref{fig11}e. For $LS$ the increase of solid at the plane of the crests
leads to an increase of $S$ greater than 
the reduction of $\langle u_2 u_1 \rangle$,
as it is shown in figure \ref{fig6}a and figure \ref{fig6}b.
This occurrence
explain the increase of $P_k^+$ and $D_k^+$ in figure \ref{fig11}f
with respect to the quantities calculated near the plane of the crests for $LLS$.
For the triangular ($LT$)  as well as for the square ($LS$) longitudinal bars, 
$T_K^+$ is negative in a large part of the channel. In figure \ref{fig11}g the
values of $T_k^+$ are small, therefore the energy produced is directly
dissipated. For the $LTS$ the $u_2$ fluctuations reduce with respect
to those generated in the $LT$ corrugations (figure \ref{fig9}a),
in addition the profiles of $-\aQ$ for $LTS$ in figure \ref{fig8}a are similar to those
for $TS$ and consequently the budgets in figure \ref{fig11}h do not
differ too much from those in figure \ref{fig11}b. 

\begin{figure}
\centering
\vskip 0.0cm
\hskip -1.8cm
\psfrag{ylab} { $\nu_T^+$}
\psfrag{xlab}{ $ y^+ $}
\includegraphics[width=7.5cm]{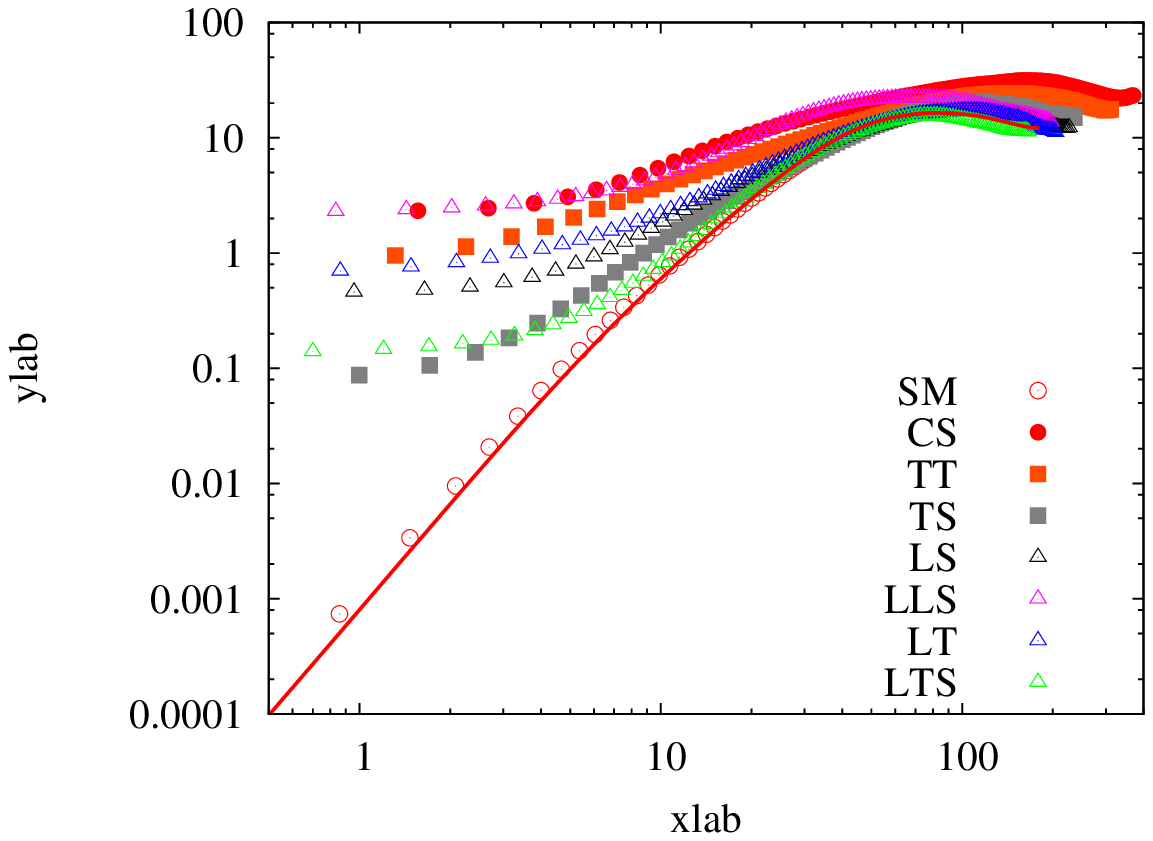}
\psfrag{ylab} {\hskip -0.0cm $\nu_T|^+_W $}
\psfrag{xlab}{ $ v^{\prime +}_W   $}
\includegraphics[width=7.5cm]{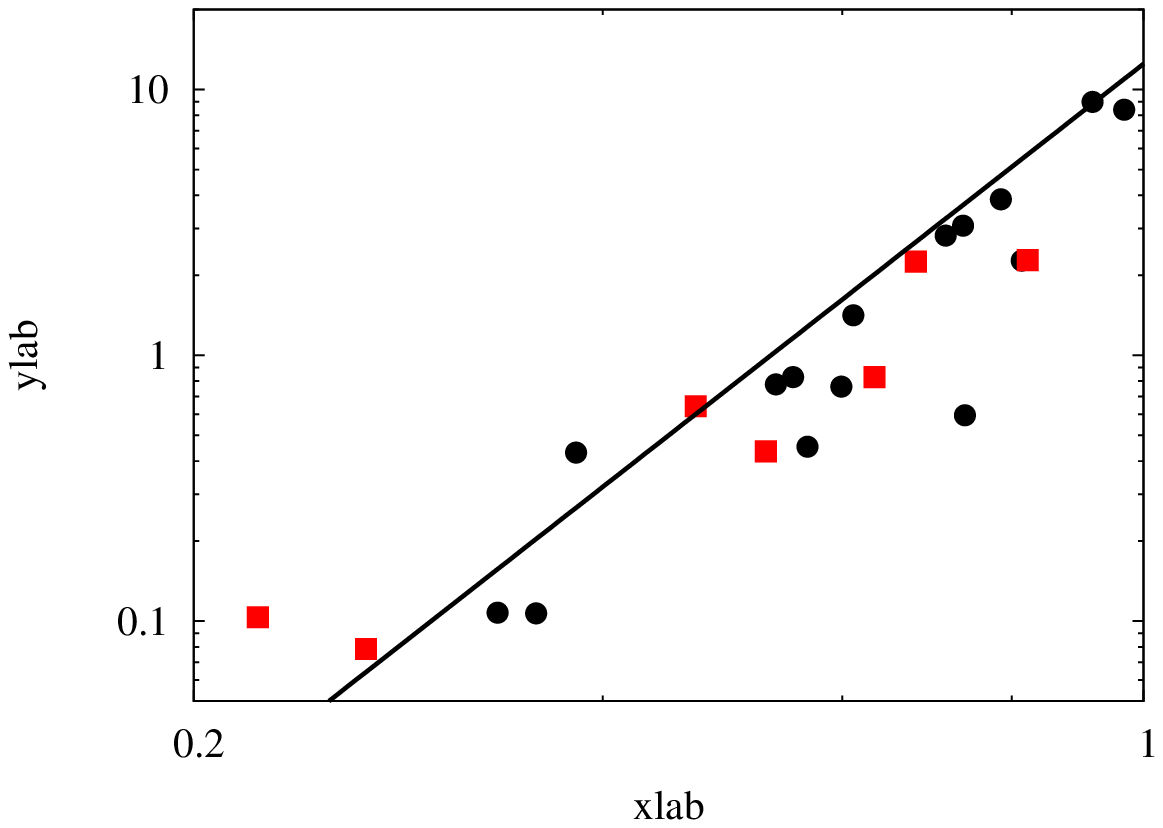}
\vskip -0.5cm \hskip 5cm a) \hskip 7cm b)
\vskip -0.2cm
\hskip -1.8cm
\psfrag{ylab}{ $\Delta U^+$}
\psfrag{xlab}{ $v^{\prime + }_W$}
\includegraphics[width=7.5cm]{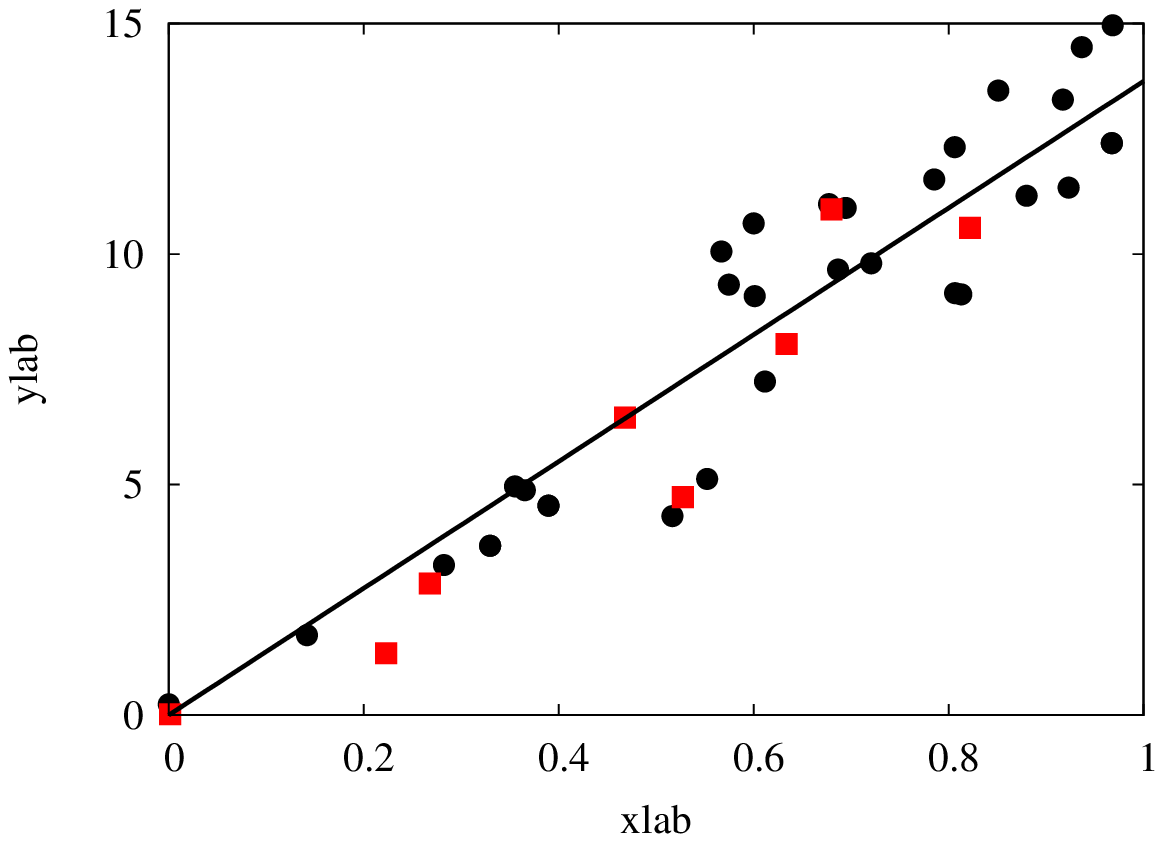}
\psfrag{ylab} { $K_S^+$}
\psfrag{xlab}{ $ v^{\prime +}_W   $}
\includegraphics[width=7.5cm]{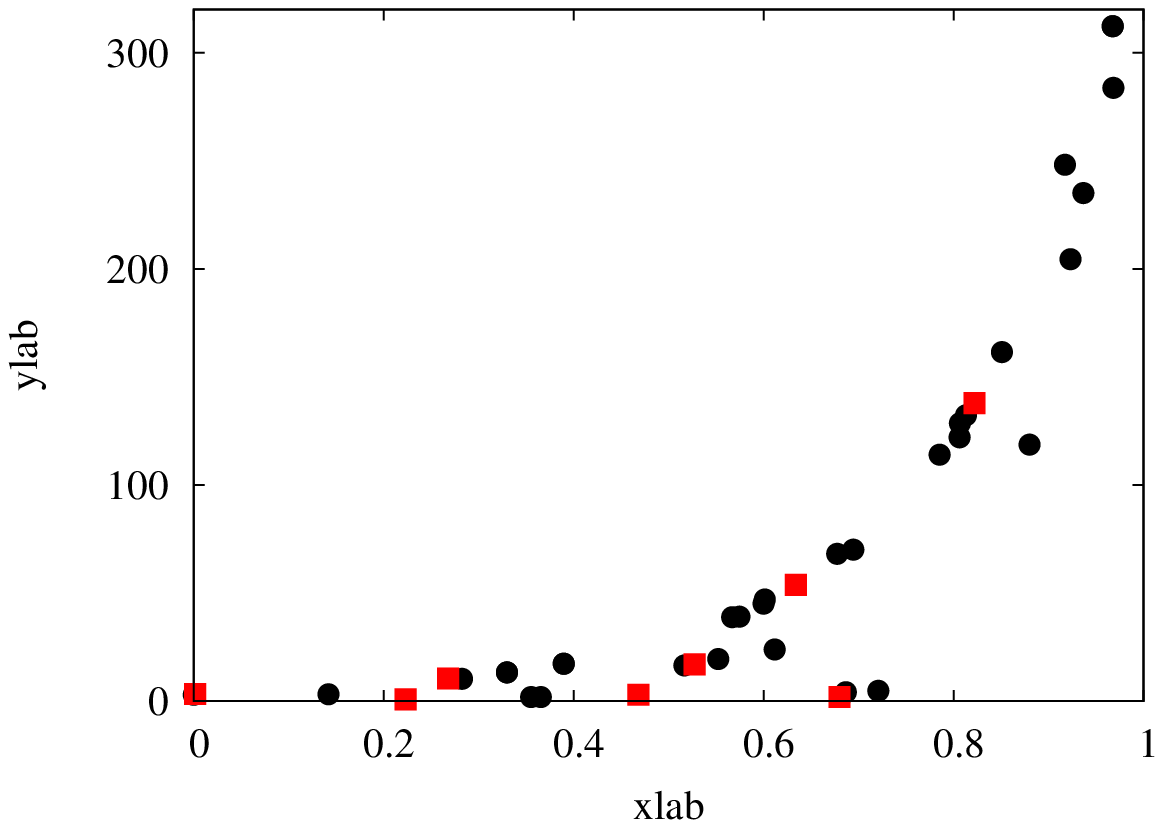}
\vskip -0.5cm \hskip 5cm c) \hskip 7cm d)
\caption{Profiles in wall units of:
a)$\nu_T^+ $
for the flows with rough surfaces (symbols) listed in the inset,
compared with
those in presence of smooth walls  
(open circle present at $R_\tau=204$,
lines \cite{lee_15} at $R_\tau=180$),
b) $\nu_T|_W^+ $,
c) $\Delta U^+ $,
d) $K_S^+$ versus $v^{\prime +}_W$, the black symbols from
the simulations in \cite {Orlandi2013}
the red symbols present results.
}
\label{fig12}
\end{figure}

\subsubsection{ Suggestions for RANS closures }

The simplified budgets in figure \ref{fig11} can be useful to
give directions to the turbulence modellers to improve the Reynolds averaged
closures for simulations of flows past rough surfaces at high
Reynolds numbers. For instance the modification of the Spalart-Allmaras closure
proposed by \cite{Aupoix} requires the modification
of the turbulent viscosity in the near-wall region, as they
reported in their figure 10. The turbulent viscosity profiles obtained by
the present simulations  in figure \ref{fig12}a 
qualitatively agree with the experimental profiles in \cite{Aupoix}.
The correction for the roughness could be achieved by assigning
the value of $\nu_T|_W^+ $ at the plane of the crests, that
depends on the type of surfaces. However, in 
figure \ref{fig12}a it is clear that different surfaces
give the save value of $\nu_T^+$. Therefore a parametrisation
based on the geometrical properties of the rough surface should
be rather difficult. As previously mentioned the parametrization
based on  $v^{\prime +}_W=\sqrt{\langle u^{2 +}_2 \rangle_W}$
could be useful. The arguments in \cite {Orlandi2013}
on the importance of the normal to the wall stress
have been qualitatively reported in commenting
the proportionality between the roughness function 
$\Delta U^+$ and $v^{\prime +}_W$. From
simulations of flows past rough surfaces it was possible
to get the analytical expression $\Delta U^+= B\frac{v^{\prime +}_W}{\kappa}$,
with $B=5.5$ the constant in the expression of the 
$\log$ law for smooth walls, and $\kappa=0.4$ the von Karman constant.
This expression was derived by fitting the data with the black solid symbols
in figure \ref{fig12}c obtained by simulations with one wall
rough and the other smooth.  The present data (red squares in figure \ref{fig12}c)
fit this expression. From the profiles of $\nu_T^+ $ in 
figure \ref{fig12}a and from the profiles calculated
by the simulation in \cite {Orlandi2013} the 
values of $\nu_T|_W^+ $ are given in figure \ref{fig12}b
fitting  rather well the expression $\nu_T|_W^+=12.5{v^{\prime +}_W}^4 $.
\cite{NikuradseEN} from a large number of measurements
of flow past rough surfaces, made by sand grain of different size,
in which the corrugation can not be exactly characterised,
derived the expression for the mean
velocity in wall units $U^+=8.48+1.\log(y^+/K_S^+)/\kappa$
where $K_S^+$ is an equivalent roughness height.
From the present results and from those in \cite {Orlandi2013}
the values of $K_S^+$ are plotted in 
figure \ref{fig12}d versus the corresponding value of
$v^{\prime +}_W$, showing a good collapse of the data,
with the exception of the simulations having
high values of $U_W$ at the plane of the crests.
In \cite{NikuradseEN} an equivalent roughness height
was introduced and was not linked to the shape of
the corrugations, the present results suggest to introduce
a value $v^{\prime +}_W$ equivalent to a roughness height.
The passage from an equivalent roughness height to
a normal to the wall stress should be useful 
in RANS simulation  requiring boundary conditions
at $y=0$ for turbulent statistics.

\begin{table}
 \centering
 \begin{tabular*}{1.\textwidth}{@{\extracolsep{\fill}}cccccccc}
  \hline
   Flow case &$y^+$&$Q_1$  & $Q_2 $ & $ Q_3 $ & $ Q_4 $ & $C_{u_1u_2}$ & \\
  \hline
$SM $    & 0.2073E+01  & 0.6611E-01 & -0.1424E+00  & 0.5680E-01 & -0.2855E+00 & -0.3049E+00\\
$CS $    & 0.3778E+01  & 0.6282E-01 & -0.1907E+00  & 0.2461E-01 & -0.4008E+00 & -0.5040E+00\\
$TT $    & 0.3181E+01  & 0.7204E-01 & -0.2170E+00  & 0.3026E-01 & -0.4115E+00 & -0.5262E+00\\
$TS $    & 0.2413E+01  & 0.7024E-01 & -0.1234E+00  & 0.4393E-01 & -0.2537E+00 & -0.2630E+00\\
$LS $    & 0.2317E+01  & 0.1630E+00 & -0.5222E-01  & 0.1965E-01 & -0.3436E+00 & -0.2132E+00\\
$LLS $   & 0.2016E+01  & 0.1078E+00 & -0.8808E-01  & 0.1652E-01 & -0.2409E+00 & -0.2047E+00\\
$LT $    & 0.2088E+01  & 0.8311E-01 & -0.2295E+00  & 0.4908E-01 & -0.2925E+00 & -0.3898E+00\\
$LTS $   & 0.1693E+01  & 0.9135E-01 & -0.1893E+00  & 0.8727E-01 & -0.2684E+00 & -0.2790E+00\\
  \hline
 \end{tabular*}
\caption{Values of the quadrant contribution to the
$C_{u_1u_2}$ correlation coefficients at $y^+\approx 7$ for the cases
indicated as in table \ref{table1}}
\label{table2}
\end{table}

\subsubsection{ Quadrant analysis  }

\begin{figure}
\centering
\vskip 0.0cm
\hskip -1.8cm
\psfrag{ylab} {\hskip -1.0cm \large $ -C_{u_1u_2}$}
\psfrag{xlab}{ $y^+ $}
\includegraphics[width=7.5cm]{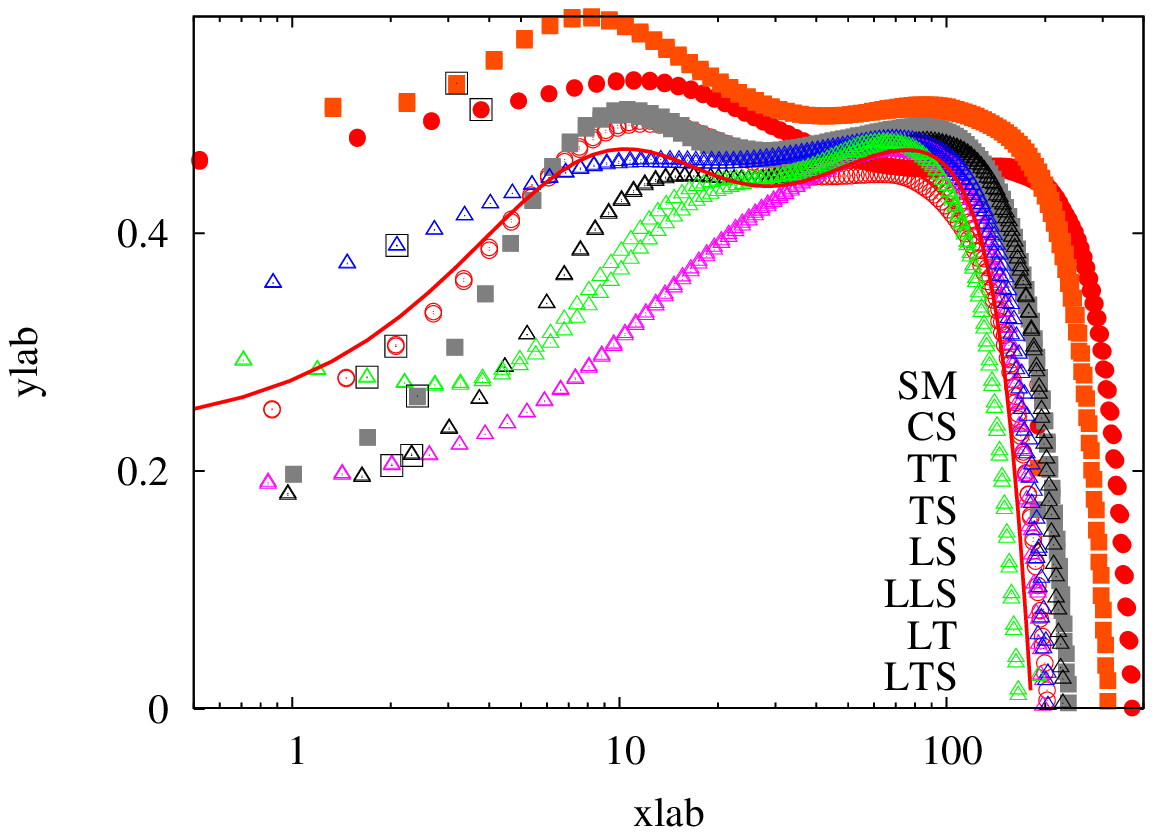}
\psfrag{ylab} {\hskip -1.0cm \large $ Q_i$}
\psfrag{xlab}{ $y^+ $}
\includegraphics[width=7.5cm]{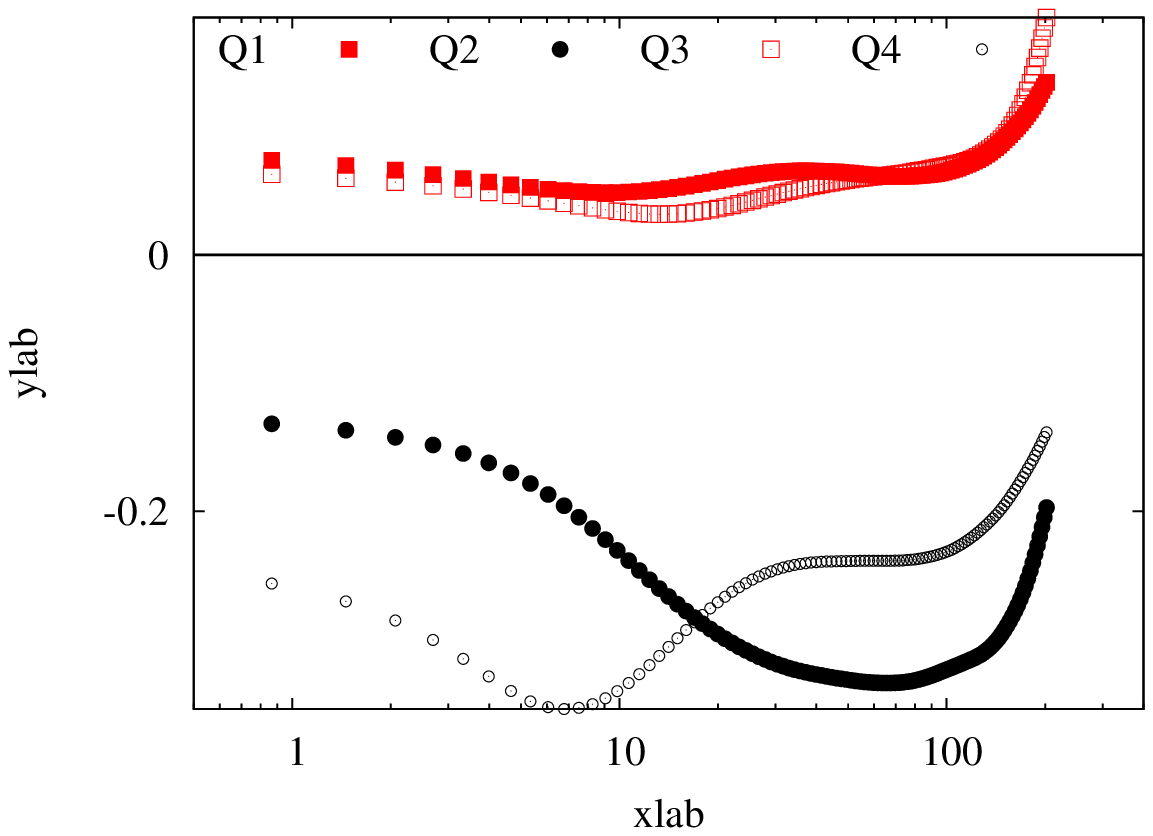}
\vskip -0.5cm \hskip 5cm a) \hskip 7cm b) 
\caption{Profiles in computational  units of:
a) correlation coefficient between $u_1$ and $u_2$,
for the flows with rough surfaces  listed in the inset;
b) quadrant contribution to $ -C_{u_1u_2}$for the $SM$ flow.
}
\label{fig13}
\end{figure}

The statistics previously discussed depicted large variations near
the plane of the crests, that depend on the shape of the surface.
The correlation affecting more the turbulent kinetic energy 
production is $\langle u_1 u_2 \rangle$, therefore it is 
worth analyse the  profiles of the correlation coefficient
$C_{u_1 u_2}=\langle \sigma_1 \sigma_2 \rangle$ 
in figure \ref{fig13}a ($\sigma_i=u_i/\langle u_i^2 \rangle^{1/2}$)
The interesting feature of this figure consists
in a large influence  of the type of surface on the values of
$-C_{u_1 u_2}$ in the near-wall region. The 
satisfactory independence in the outer region, corroborating
the Townsend similarity hypothesis, can be better appreciated by
plotting $-C_{u_1 u_2}$ versus $y$. In presence of smooth walls
from the \cite{lee_15} data it can be observed that $-C_{u_1 u_2}$
is almost independent on the Reynolds number for $R_\tau > 1000$, 
it grows in the near-wall region from a value
equal to $0.2$ to $0.40$ at the location of maximum production.
This correlation coefficient is linked to the flow structures,
the contribution from the different kind of structures can
be derived through the quadrant analysis  described by \cite{Wallace2016}.
This contribution varies across the channel accordingly to the           
kind of flow structures. For instance by plotting the contribution
of the four quadrants $Q_1(+\sigma_1,+\sigma_2)$,$Q_2(-\sigma_1,+\sigma_2)$
,$Q_3(-\sigma_1,-\sigma_2)$,
$Q_4(+\sigma_1,-\sigma_2)$ to $C_{u_1 u_2}$ across the channel
it can be observed that the second and fourth quadrants
prevail on the first and third. The ejection and sweeps events contribute
to $Q_2$ and $Q_4$, and for their relevance have been deeply studied. 
\cite{Wallace2016} defined the events in the first and third quadrants
as outward and inward interactions and their contributions
is constant moving far from the wall, as it is shown in figure \ref{fig13}b.
In the near-wall region $Q_4$ prevails on $Q_2$,  
the location where the two are equal coincides with the location
of the first change of sign of
$(\dedq{ \langle{u_2^2}\rangle}{ x^2_2})^+$
separating the sheet- by the tubular-dominated regions.       

\begin{figure}
\centering
\vskip 0.0cm
\hskip -1.8cm $\sigma_2$
\includegraphics[width=4.0cm]{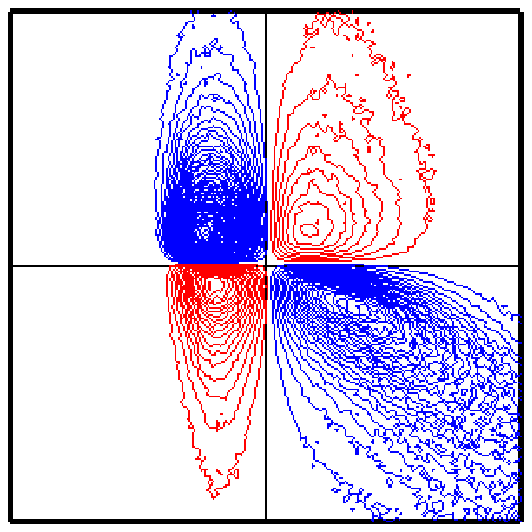}
\hskip -.6cm
\includegraphics[width=4.0cm]{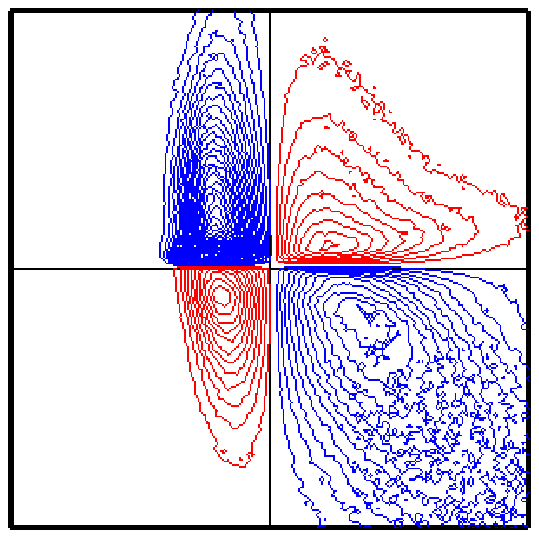}
\hskip -.6cm
\includegraphics[width=4.0cm]{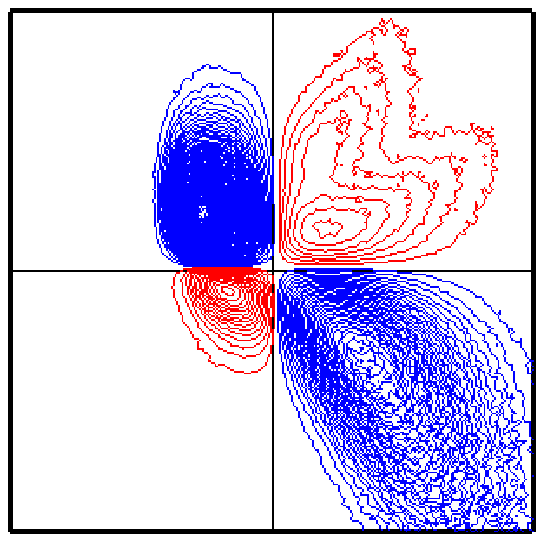}
\hskip -.6cm
\includegraphics[width=4.0cm]{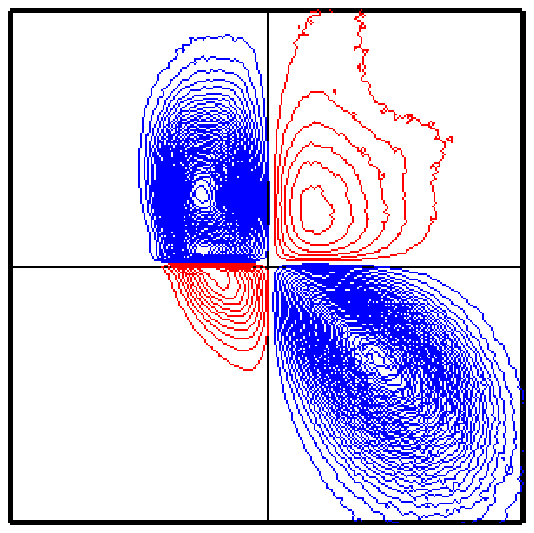}
\vskip -0.1cm \hskip -1cm $\sigma_1$ \hskip 0.5cm a) \hskip 3.5cm b) 
\hskip 3.5cm c) \hskip 3.5cm d)
\vskip -.1cm
\hskip -1.8cm $\sigma_2$
\includegraphics[width=4.0cm]{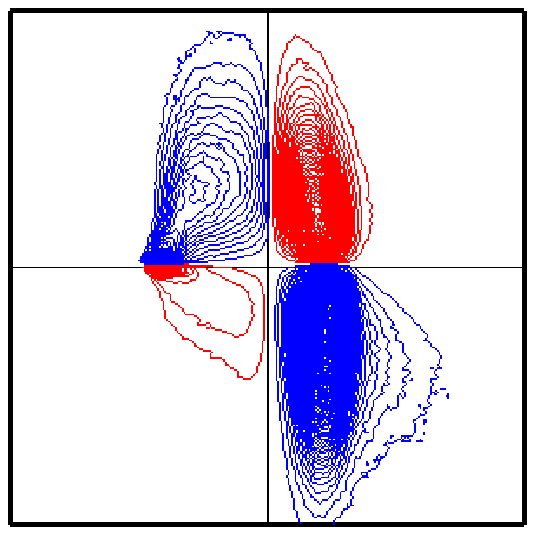}
\hskip -.6cm
\includegraphics[width=4.0cm]{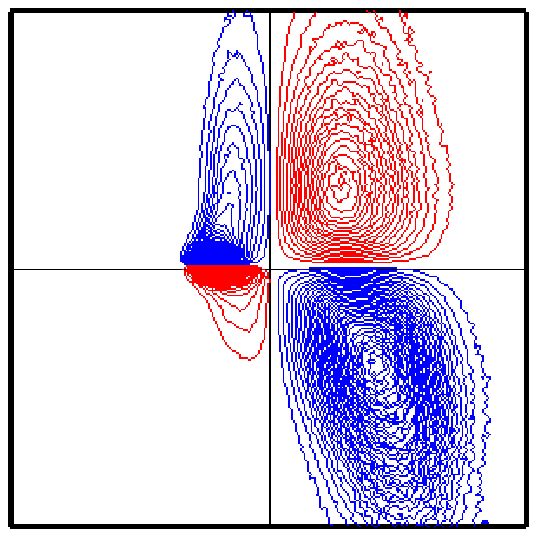}
\hskip -.6cm
\includegraphics[width=4.0cm]{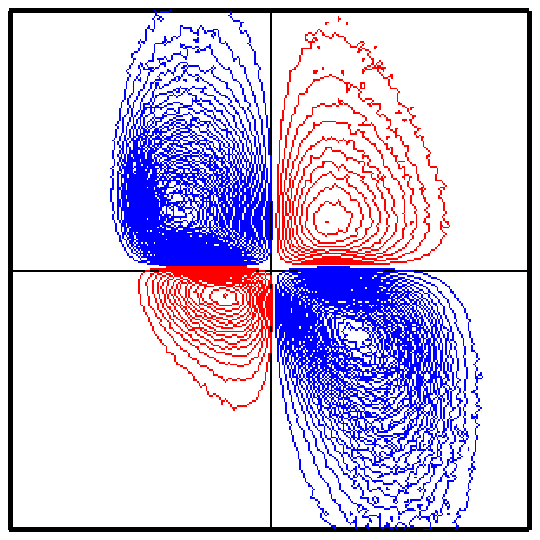}
\hskip -.6cm
\includegraphics[width=4.0cm]{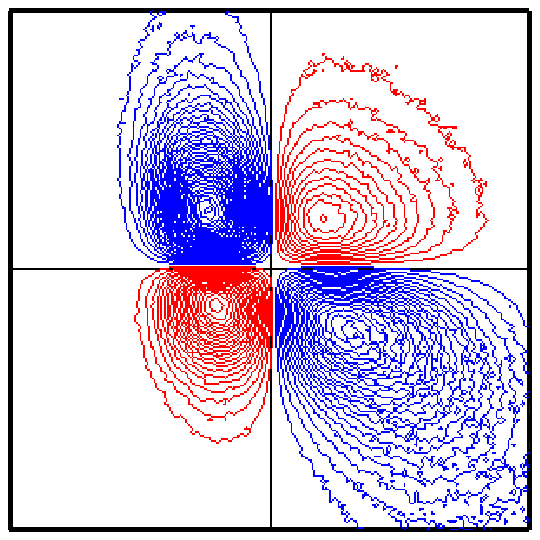}
\vskip -0.1cm \hskip -1cm $\sigma_1$ \hskip 0.5cm e) \hskip 3.5cm f) 
\hskip 3.5cm g) \hskip 3.5cm h)
\vskip -.0cm
\caption{The covariance integrated $q_1q_2P(q_1,q_2)$
at $y^+\approx 5$ between $u_1$ and $u_2$
with contours increments $\Delta=.00001$
a) $SM $ , b) $TS $ , c) $TT $, d) $CS$ ,
e)  $LLS $, f) $LS $ , g) $LT $, h) $LTS$
}
\label{fig14}
\end{figure}

Figure \ref{fig10}e shows that the production of turbulent kinetic
energy for smooth walls  grows in the region dominated by the
sweep events.  The joint pdf $P(q_1,q_2)$
or the covariance integrated $q_1q_2P(q_1,q_2)$ of more interest
have been calculated at $y^+\approx 2$ for any surface,
the distance at which figure \ref{fig13}a shows large variations
due to the different type of corrugation.
The values of $C_{u_1u_2}$ together with the contribution of
the four quadrants are given in table \ref{table2} and are indicated
by the black open squares in figure \ref{fig13}a. The greater values of
$C_{u_1u_2}$ are obtained by the $CS$ and $TT$ flows for the large
increase of the $Q_4$ contribution. The comparison between the covariance
integrated plots for $SM$ (figure \ref{fig14}a) and the $CS$ (figure \ref{fig14}d) 
surfaces depicts minor changes
for the quadrant with $\sigma_1<0$ than for those with $\sigma_1>0$.
\noindent The same behavior is observed in figure \ref{fig14}c for the $TT$ flow.
The contours in the first quadrant have a complex shape,              
due to the form of the surface affecting the ejections of high intensity in
the $CS$ and $TT$ surfaces. From visualizations of $\sigma_2$ 
it can be appreciated that for the corrugations with a large solid region,
at the plane of the crests, the shape of the surfaces is visible up to distances 
$y^+\approx 10$. On the other hand, the surfaces are not  appreciated  by 
the $\sigma_1$ contours, however 
the elongation of the longitudinal structures is strongly reduced
in particular for the $CS$ and $TT$ surfaces. 
In the sheet-dominated region the pdf profiles, evaluated
by the joint pdf, for the $CS$ and $TT$ surfaces are symmetric for $\sigma_2$ 
and positive skewed for the $\sigma_1$.
Due to the weak $\sigma_2$ disturbance in the $TS$ flow the profiles of the 
quadrant contributions, the covariance integrated contours in figure \ref{fig14}b,
and the relative pdf do not change much with respect to those of the smooth 
surface.
The $-C_{u_1 u_2}$ for the surfaces with longitudinal bars are smaller in
particular for the $LLS$ and the $LS$ surfaces, due to the reduction 
of $Q_4$ and the increase  of $Q_1$.
This is clearly depicted in figure \ref{fig14}e and figure \ref{fig14}f.
The corresponding visualizations, not shown, emphasise the formation
of very long streamwise structures with the positive streaks 
over the cavities and the negative over the solid. For $LLS$ the magnitude
of the peaks in the layers with $\sigma_1>0$ is smaller than that for $LS$. 
The symmetric pdf of $\sigma_2$ do not change, instead  the pdf of $\sigma_1$
present, for both surfaces, a sharp decrease leading to negative values of the
skewness coefficient. For the $LT$ and for the $LTS$ surfaces 
$-C_{u_1 u_2}$ increases with respect to the flows past longitudinal
square bars due to the increase of the $Q_2$ contribution. For $LT$ the
highest value of $Q_2$ is given in table \ref{table2};  it is
confirmed by the contours in
figure \ref{fig14}g.

\begin{figure}
\centering
\vskip 0.0cm
\hskip -1.8cm
\includegraphics[width=4.2cm]{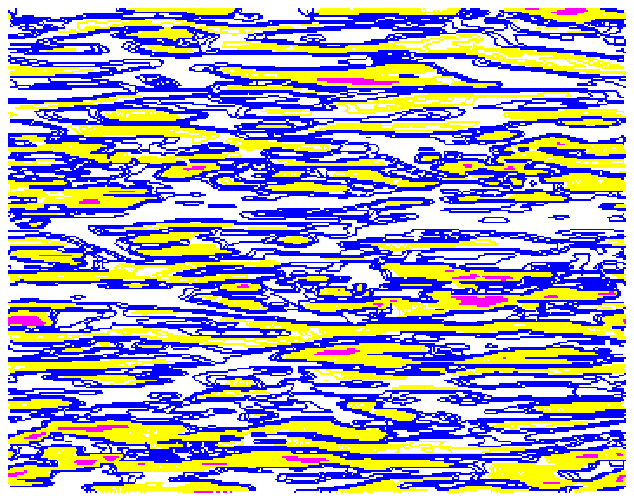}
\includegraphics[width=4.2cm]{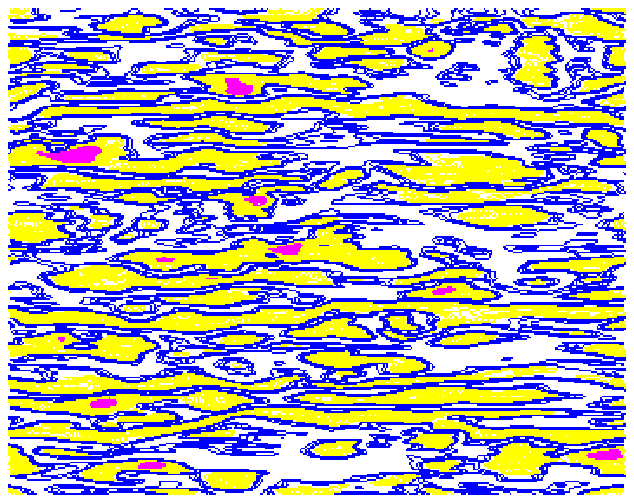}
\includegraphics[width=4.2cm]{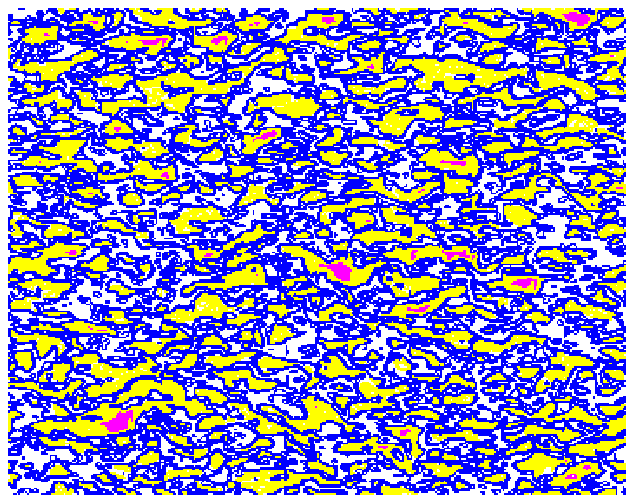}
\vskip -0.3cm \hskip 0cm a1) \hskip 3.9cm a2) \hskip 3.9cm a3) 
\vskip 0.0cm
\hskip -1.8cm
\includegraphics[width=4.2cm]{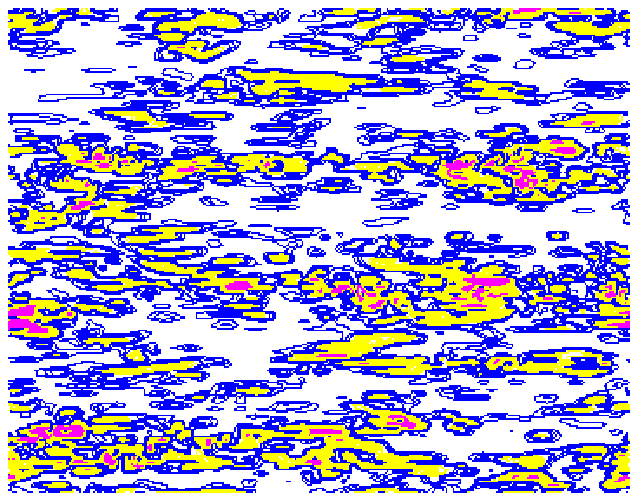}
\includegraphics[width=4.2cm]{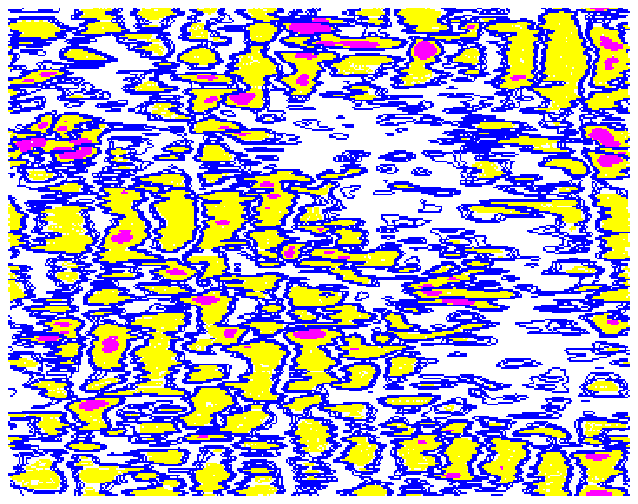}
\includegraphics[width=4.2cm]{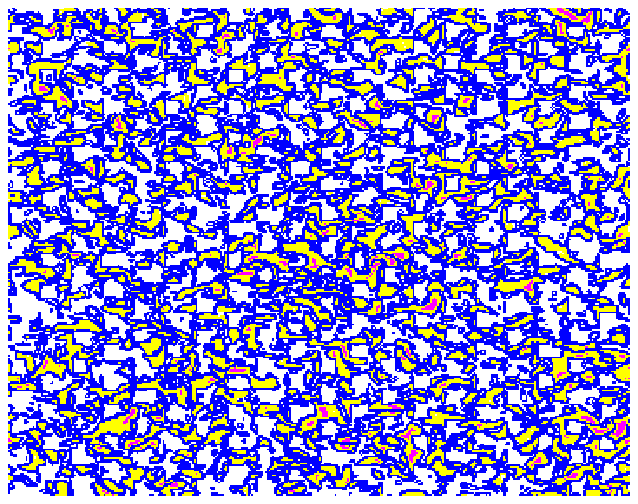}
\vskip -0.3cm \hskip 0cm b1) \hskip 3.9cm b2) \hskip 3.9cm b3) 
\vskip 0.0cm
\hskip -1.8cm
\includegraphics[width=4.2cm]{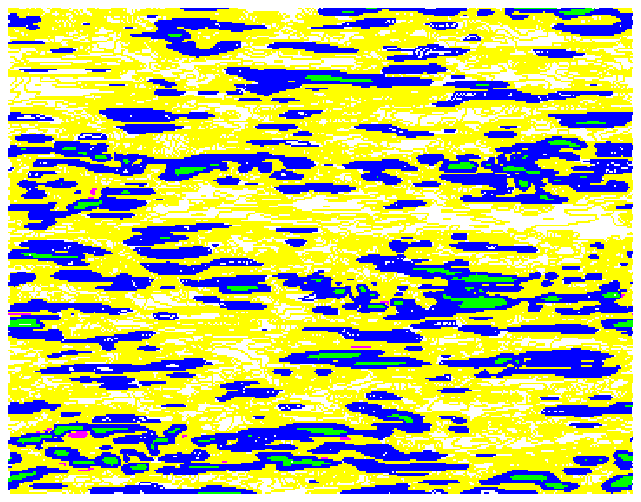}
\includegraphics[width=4.2cm]{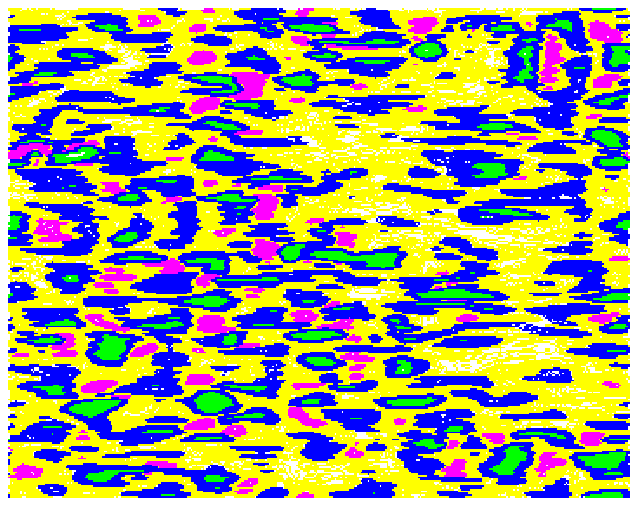}
\includegraphics[width=4.2cm]{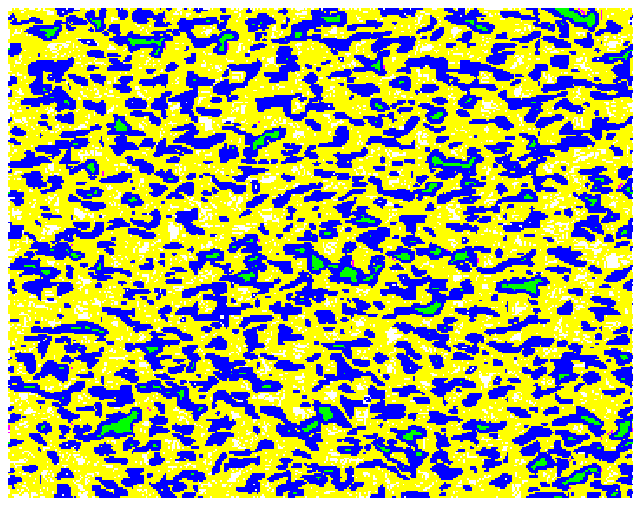}
\vskip -0.3cm \hskip 0cm c1) \hskip 3.9cm c2) \hskip 3.9cm c3) 
\vskip 0.0cm
\hskip -1.8cm
\includegraphics[width=4.2cm]{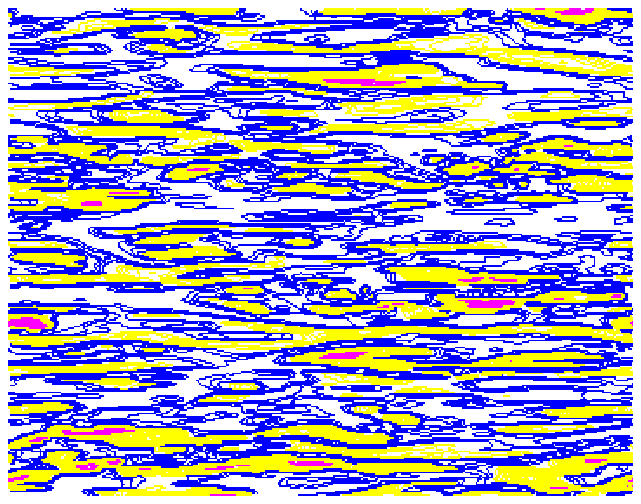}
\includegraphics[width=4.2cm]{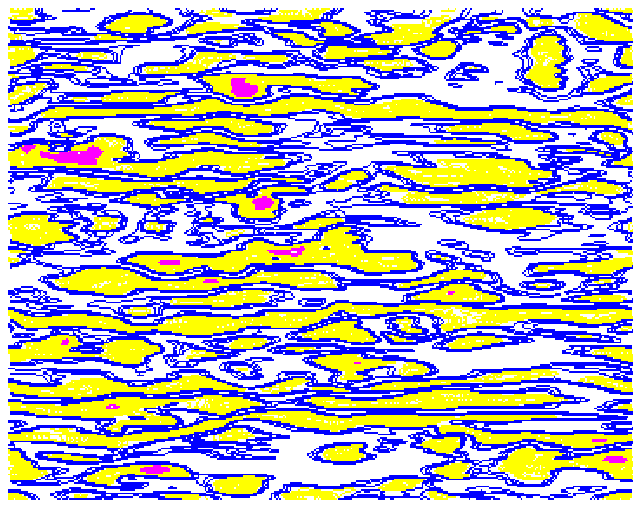}
\includegraphics[width=4.2cm]{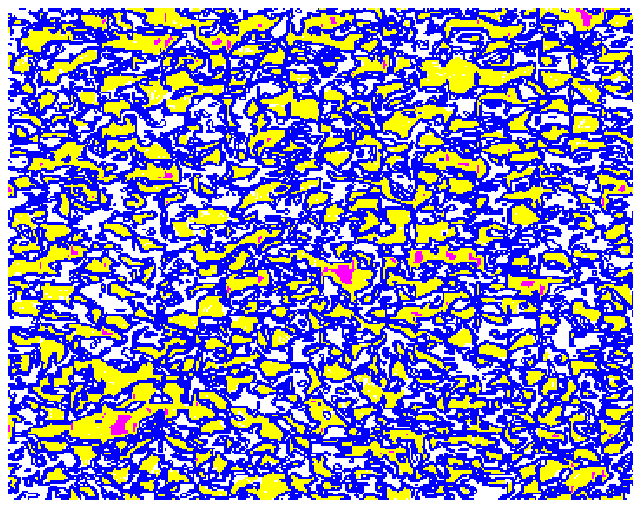}
\vskip -0.3cm \hskip 0cm d1) \hskip 3.9cm d2) \hskip 3.9cm d3) 
\vskip 0.0cm
\hskip -1.8cm
\includegraphics[width=4.2cm]{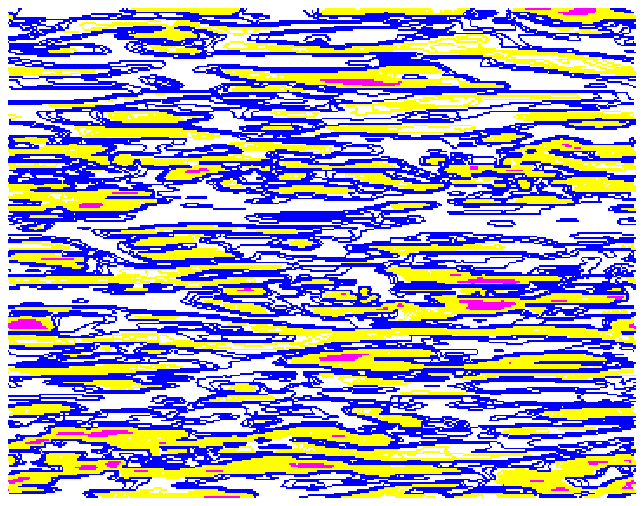}
\includegraphics[width=4.2cm]{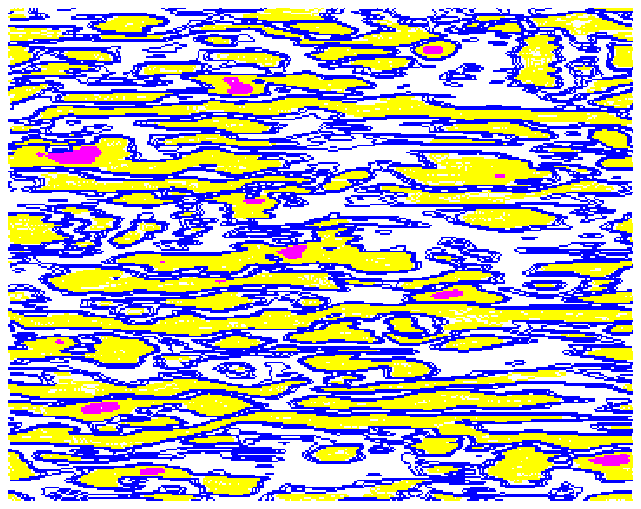}
\includegraphics[width=4.2cm]{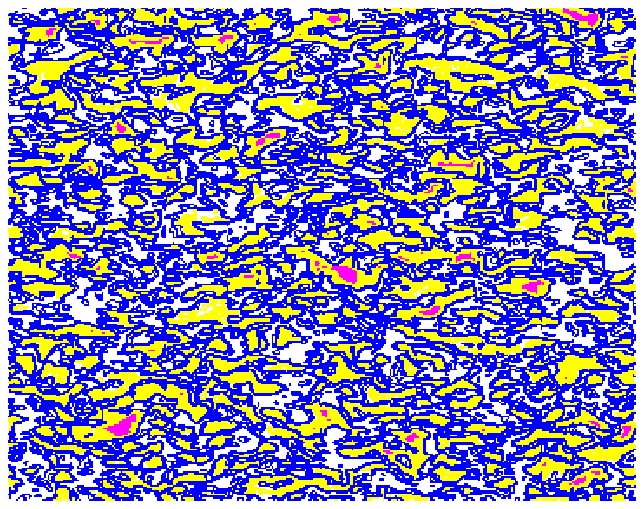}
\vskip -0.3cm \hskip 0cm e1) \hskip 3.9cm e2) \hskip 3.9cm e3) 
\vskip -.1cm
\caption{
Contours of 
al) $\rho{1 1}$, 
bl) $\rho{2 2}$, 
cl) $\rho{1 2}$, 
dl) $\rho{\alpha \alpha}$, 
el) $\rho{\gamma \gamma}$
at $y^+\approx 5$ 
with increments $\Delta=.25$, yellow positive, blue negative
for $< 5$ magenta positive, 
for $> 5$ green negative; left $SM$, center  $LTS$, right $CS$.
}
\label{fig15}
\end{figure}

\subsubsection{Turbulent stresses visualisations}

For the flows past these surfaces it is interesting
to look at the flow visualizations of the stresses $u_lu_m$
in planes $x_1-x_3$ parallel to the plane of the crests.
In these circumstances the stresses are
evaluated by one realisation.
The subscript $l$ and $m$ may indicate either the components in the Cartesian
reference system or those in the frame aligned with the eigenvalues
of the strain tensor $S_{ij}$. The contours in figure \ref{fig15}
are done for $\rho_{l m}(x_1,x_3)=(u_lu_m(x_1,x_3)-R_{l m})/R_{l m}$,
with $R_{l m}=\frac {1}{N1 N3}\Sigma_{N1}\Sigma_{N3} u_l u_m(x_1,x_3)$,
at the distance $y^+\approx 5$ from the plane of the crests. 
Usually the streaks are visualised through contours of  $\sigma_1$
producing a picture with elongated positive and negative regions
similar to those in figure \ref{fig15}a1 for $\rho_{1 1}$.
The yellow positive layers have few peaks with high values (magenta coloured)
the less intense blue negative are located in wider elongated structures.
The contours of $\rho_{2 2}$ in figure \ref{fig15}b1 
depict regions of small size
with a large number of intense positive values. This imply
for $u_2$  a large flatness factor, and agrees with the
covariance integrated distribution  in figure \ref{fig14}a.
In correspondence of the high values of $\rho_{2 2}$ 
high values of negative $\rho_{1 2}$ (green coloured)
can be detected in  figure \ref{fig15}c1. 
The contours of the $\rho_{\alpha \alpha}$
in figure \ref{fig15}d1 and of $\rho_{\gamma \gamma}$ in figure \ref{fig15}e1
are similar. In figure \ref{fig10}b $R_{\gamma \gamma}$
for $SM$ was greater than $R_{\alpha \alpha}$ in figure \ref{fig10}a,
this difference in the visualizations can not be appreciated 
due to the normalisation in the
expression of $\rho_{l m}$. The anisotropy of the near wall region in
the Cartesian reference frame, is clearly drawn by comparing
figure \ref{fig15}a1 and figure \ref{fig15}b1. In the
strain rate reference system the anisotropy 
is visually appreciated by a comparison between  
the contours of $\rho_{\beta \beta}$, equal to those of $\rho_{3 3}$,
with those in in figure \ref{fig15}d1 and in  figure \ref{fig15}e1.

The figures in the central column
for the $LTS$ flow of $\rho_{1 1}$ (figure \ref{fig15}a2),  
$\rho_{\alpha \alpha}$ (figure \ref{fig15}d2) and
$\rho_{\gamma \gamma}$ (figure \ref{fig15}e2)
are similar to those for $SM$, with
more elongated positive regions due to the effects of
the underlying surface, barely visible.
On the other hand,  large differences can be appreciated between the
contours of $\rho_{2 2}$ (figure \ref{fig15}b2) and $\rho_{1 2}$
(figure \ref{fig15}c2) and the corresponding 
figure for $SM$ in the left column.
For $LTS$ is clear  the formation of
spanwise coherent structures with intense positive peaks in
correspondence of which strong negative $\rho_{1 2}$ appear. 
The common features of the $LTS$ and $SM$ surfaces is the
strong influence of the $u_2$ fluctuations on the turbulent
stress $\langle u_1 u_2 \rangle$ and therefore on the
production of turbulence.  This is a further
prof that the $u_2$ fluctuations are those characterising wall
turbulence.  In presence of smooth walls
the streaks do not form in particular locations. 
For the  $LLS$ corrugations the streaks are linked
to the underlying surfaces as it can be observed in visualizations,
not shown for sake of brevity. The influence of the underlying surface
can be appreciated in the visualizations for the $CS$ 
flow in the right column of figure \ref{fig15}. In this case
the disturbances generated within the roughness layer
are strong enough to destroy the near-wall anisotropy. 
The contours of $\rho_{1 1}$ (figure \ref{fig15}a3),  
$\rho_{\alpha \alpha}$ (figure \ref{fig15}d3) and
$\rho_{\gamma \gamma}$ (figure \ref{fig15}e3)
show that the elongated streamwise structures are not
any more visible, and that their size is approximately
the same as that of $\rho_{2 2}$. Therefore the tendency towards the 
isotropy in the near-wall layer is clearly depicted.
In presence of strong $u_2$ and $u_1$ disturbances 
is found that the intense negative
values of $\rho_{1 2}$ in figure \ref{fig15}c3 are
strongly correlated with those of $\rho_{2 2}$ in
figure \ref{fig15}b3 and also with the $\rho_{1 1}$.
To conclude the stress distribution in the near-wall
layer is strictly linked to the staggered distribution of the cubes
in the corrugation.
 
\section{Concluding remarks}

This paper is focused on the connection between turbulent structures and
production of turbulent kinetic energy. Emphasis has been directed towards 
statistics seldom considered in the analysis of wall bounded flows.
Namely the full dissipation rate, the shear parameter and different
expression for the production of turbulent kinetic energy. 
The canonical two-dimensional turbulent channel
has been investigated by taking the data from DNS at high and low
friction velocity Reynolds numbers. In a recent review paper 
\cite{jimenez2018} reported the debate about the eventual universality
of wall bounded flows by increasing the Reynolds number. He shortly
discussed the shear parameter $S^*$ without discussing the universality
of this parameter in the near-wall region. Since $S^+$ does not vary with 
the Reynolds number, in the present paper the eddy turnover time in 
wall units has been evaluated by the DNS data by
concluding that there is a good universality. The 
eddy turnover time  can also be defined as the ratio 
between $q^{2 +}$ and the full rate of 
dissipation $D_k^+$, in this case it has been found that it grows linearly both
in the near-wall region and in the outer region, with two different
constants of proportionally. Therefore there is a small layer
connecting the two regions with linear growth.
This result can be a first indication
that the flow structures near the wall and those in the
outer region are of the same kind. Those near the wall move fast and
those in the outer layer slow. The linear growth near the
wall is greater than that in the outer region, the passage
between one and the other occurs in the layer where the $P_k^+$
sharply grows. 
From these data it can be, also observed that at $R_\tau=5200$
there is a tendency to the linear growth in the outer region. However, our
view is that it will be indeed achieved by
simulations at a slightly higher $Re$.  From the data it
was also possible to conclude that the maximum turbulent kinetic energy
production scales at high Reynolds numbers and that the maximum is located
at a distance from the wall where there is the transition between layers
sheet-dominated and rods-dominated. Namely in
the region where the ribbon unstable structures roll-up to become
tubular structures. Finally it was found that the rate of
isotropic dissipation largely depends on the Reynolds number,
and that the full rate dissipation does not.

Flows past smooth walls have well defined  boundary conditions
for the velocity fields. These boundary conditions 
can be varied by changing the shape
of the walls. Through the DNS of flows past different kind of
corrugations it was observed that it is easy to increase the resistance,
and rather difficult to reduce it. Drag reduction 
is obtained when the viscous stress at the plane of the crests reduces
more than the increase of the turbulent stress $\langle u_1 u_2 \rangle$.
In this regard it is interesting to look at the
results of \cite{arenasleonardi2018} where it is possible to get 
a large drag reduction by imposing $u_2=0$ at the plane of the crests of any 
kind of corrugation.  In real applications this result can be  achieved
if someone is able to find the way to reproduce this boundary condition.
Perhaps this is a very difficult task to reach, but from a mathematical
point of view is important. The simulations of flows past
several types of corrugations allowed to reach the conclusion that a universal
behavior can not be found. However the parametrization of
rough walls can be obtained through the normal to the wall stress
at the plane of the crests. It was reported that the results of
the DNS can give insight on the improvement of turbulence
RANS closures, for instance to the Spalart-Almaras model.
It was also observed that in RANS the reproduction of the turbulent
kinetic energy budget is simpler by considering the full
rate of dissipation instead of the isotropic rate of dissipation.
The flow structures in the near-wall region for corrugations
generating intense $u_2$ fluctuations tend to become more isotropic.
For the drag reducing corrugations spanwise coherent structures
forms which are easily detected by the $u_2$ contours and
even better by pressure contours. These structures were
observed at high Reynolds number by \cite{Raupach1996}
in flows past canopies. They claimed that these structures
were generated by inflectional velocity profiles similar
to those occurring in mixing layers. In this experiment 
the drag is greater than that in presence of smooth walls. Similar conclusions were
reached by \cite{GarciaMayoral} by simulations of flows
past square bars in the case of breakdown of drag reduction
and the spanwise structures were barely visualised.
In the present simulations the spanwise structures were
observed only in the drag reducing corrugations and it has
been observed that a large role should be ascribed to the pressure.
More simulations are currently performed to investigate
how important are these structures.

\section{Acknowledgements}
We acknowledge that some of the results reported in this paper have been
achieved using the PRACE Research Infrastructure resource MARCONI
based at CINECA, Casalecchio di Reno, Italy. A particular thank to
David Sassun  helping me in the implementation of the
immersed boundary method, and to Sergio Pirozzoli through
a large number of discussions on wall turbulence and for
the correction of the draft.
%
%\bibliography{references}

\end{document}